\documentclass[aps, pra,twocolumn,english]{revtex4-2}
\usepackage[colorlinks=true,citecolor=blue,linkcolor=blue,urlcolor=blue, breaklinks]{hyperref}
\usepackage{amsmath, amssymb, mathtools, booktabs, bookmark, braket,bm, float, parskip, nicefrac, xcolor, graphicx, enumitem, bbold, mdframed}
\usepackage[caption=false]{subfig}
\usepackage[mathscr]{eucal}
\newenvironment{aside}{\vspace{\parskip}\begin{mdframed}[linewidth=1.5,linecolor=lightgray, topline=false,rightline=false,bottomline=false, leftmargin=1em]}{\end{mdframed}\vspace{\parskip}\ignorespacesafterend}
\begin{document}

\title{Quantum Extremal Learning}


\author{Savvas Varsamopoulos}
\author{Evan Philip}
\author{Vincent E. Elfving}
\affiliation{Pasqal SAS, 2 av. Augustin Fresnel, 91120 Palaiseau, France}

\author{Herman W. T. van Vlijmen}
\author{Sairam Menon}
\author{Ann Vos}
\author{Natalia Dyubankova}
\author{Bert Torfs}
\author{Anthony Rowe}
\affiliation{Discovery Sciences, Janssen Research \& Development, a division of Janssen Pharmaceutica N.V., Turnhoutseweg 30, B-2340 Beerse, Belgium}

\date{\today}

\begin{abstract}
We propose a quantum algorithm for `extremal learning', which is the process of finding the input to a hidden function that extremizes the function output, without having direct access to the hidden function, given only partial input-output (training) data. The algorithm, called \emph{quantum extremal learning} (QEL), consists of a parametric quantum circuit that is variationally trained to model data input-output relationships and where a trainable quantum feature map, that encodes the input data, is analytically differentiated in order to find the coordinate that extremizes the model. This enables the combination of established quantum machine learning modelling with established quantum optimization, on a single circuit/quantum computer. We have tested our algorithm on a range of classical datasets based on either discrete or continuous input variables, both of which are compatible with the algorithm. In case of discrete variables, we test our algorithm on synthetic problems formulated based on Max-Cut problem generators and also considering higher order correlations in the input-output relationships. In case of the continuous variables, we test our algorithm on synthetic datasets in 1D and simple ordinary differential functions. We find that the algorithm is able to successfully find the extremal value of such problems, even when the training dataset is sparse or a small fraction of the input configuration space. We additionally show how the algorithm can be used for much more general cases of higher dimensionality, complex differential equations, and with full flexibility in the choice of both modeling and optimization ansatz. We envision that due to its general framework and simple construction, the QEL algorithm will be able to solve a wide variety of applications in different fields, opening up areas of further research.
\end{abstract}

\maketitle

\section{Introduction}
\label{section:Introduction}

Optimization problems are ubiquitous in a wide range of scientific and engineering fields, including chemistry, logistics, bio-informatics, finance, mechanical engineering, and mathematics. In its most general form, an optimization problem searches for the best element, according to some criterion, among many candidates, possibly in the presence of some constraints. While many standardized optimization problems are known from mathematical modeling, which can capture the correlations of real-world systems often quite well, there are also a range of complex real-world systems which cannot easily be formulated as a standard optimization problem, due to their inherent hidden complexity or dynamics. In recent years, machine learning has gained a tremendous resurgence since its initial proposals decades ago, finding applications in many learning tasks including classification, clustering and also regression. This begs the question; could one also find algorithms which optimize the output of a machine-learned regressor/surrogate model?

In this paper, we consider the following  combined learning/optimization task: suppose we have access to a set of observations on a system, such as molecular structure descriptors and an associated figure of merit value. Our task is to find/suggest a new structure which would yield an even better figure of merit value. This is a common task in, for example, computer-aided drug design, where a new molecule/compound is sought, and there exists no complete ab-initio model describing the relationship between structure and figure of merit. In other words, such model first needs to be learned, whereafter the model needs to be `extremized' (minimized or maximized), i.e. the input structure is sought which maximizes the figure of merit of the model and hence the expected quality of the drug. 

Typically, optimization occurs by extremizing  a real function by choosing elements from the available set of inputs (independent variable) and computing the associated value of the function for that element. One challenge of solving such an optimization task is that it can be combinatorially expensive in the number of variables of the system, if each can be adjusted independently without major constraints. Indeed, in biochemistry optimal structures can not be found by an exhaustive search \cite{Robert2021, Fox2021, Mulligan2020} due to the large sample space. In material science, structure exploration has been aided by recent classical computational techniques that provide a vast selection of candidates to explore. Finding optimal structures is a very challenging outstanding task, because it typically requires many expensive example simulations or experiments to be performed in order to feed sufficient suggestions to the structure selector~\cite{Kitai2020}.

Although there exist various relatively efficient optimization schemes using classical computers, such methods still either suffer from large computational overhead or unacceptable levels of approximation in heuristics. Quantum computers, a nascent orthogonal computational paradigm, can potentially offer ways to speed up such calculations and various schemes have already been proposed to tackle optimization problems deemed previously intractable. However, these very particular algorithms can only target very specific optimization problems, such as quadratic unconstrained binary optimization problems~\cite{Farhi2014, Kitai2020}.

It would therefore be highly beneficial of quantum optimization algorithms if they could target general trained \textit{quantum} models as well. We call the combination of modeling data and optimizing to find the extremal value of the trained model ``extremal learning"; in this paper, we consider a fully-quantum variant of this which we call ``quantum extremal learning" (QEL).

\subsubsection*{Background}

Recently, extremal learning has been suggested in the context of classical Neural Networks (NNs). A neural network allows us to obtain a prediction function of the form $y = f({\bm{\theta}}; \bm{x})$ that models the relationship between an independent variable (input) $\Set{ \bm{x}_i }$ and a dependent variable (output) $\Set{ y_i }$, by optimizing model parameters ${\bm{\theta}}$. In the setting of optimization, it is desirable to not only find this function $f$ that models data, but also the value of the independent variable $\bm{x}$ that extremizes the value of this function and hence the dependent variable~\cite{Patel2021}. If $f({\bm{\theta}}; \bm{x})$ were a simple function, for example a convex one, this problem would be trivial~\cite{Hiriart-Urruty2001, Niculescu2016}, but conventional techniques for finding the extrema of a function are not suitable in this case since Neural Networks are usually used when the nature of $f({\bm{\theta}}; \bm{x})$ is complex. Although the problem of extremal learning itself was only recently proposed in the classical setting~\cite{Patel2021}, modeling in general with quantum modules~\cite{Mitarai2018} and modeling with classical modules followed by optimization with hybrid quantum/classical modules~\cite{Kitai2020} has received significant interest recently. 
    
Combinatorial optimization (CO) refers to problems where one is interested in finding the minima (maxima) of an objective function whose set of feasible solutions is extremely large and discrete (or can be reduced to a discrete one). In the classical realm, CO problems are typically solved by greedy algorithms~\cite{Kruskal1956, Prim1957}, heuristic algorithms~\cite{Pham2011}, or by relaxing the problem from non-convex to convex. In the quantum realm, CO problems are targeted because of the potential quantum speed-ups that are promised by fault-tolerant quantum computing (FTQC)~\cite{Sanders2020}. However, since FTQC is not yet available in the current Noisy intermediate-scale quantum (NISQ) era~\cite{Preskill2018}, Variational Quantum Algorithms (VQAs) are used, typically reaching comparable performance to the classical optimization algorithms~\cite{Dalzell2020}. 
The most common examples of VQAs are the Variational Quantum Eigensolver (VQE)~\cite{Peruzzo2014} and the Quantum Approximate Optimization Algorithm (QAOA)~\cite{Farhi2014}. A different quantum approach to the VQAs is Quantum Annealing (QA), which has also been successfully applied in CO problems. The main difference between VQAs and QA is that the former are used in gate based quantum computing, whereas the latter is used in special-purpose quantum annealing hardware like DWAVE~\cite{King2022}.

Regression refers to the task of predicting a continuous quantity, while using a set of statistical processes for estimating the relationships between a dependent variable and one or more independent variables. One of the most robust classical regressors are Support Vector Machines (SVMs)~\cite{Boser1992, Vapnik1998}. A SVM constructs a hyperplane or set of hyperplanes in a high- or infinite-dimensional space, focusing on achieving good separation by the hyperplane that has the largest distance to the nearest training-data point of any class, since in general the larger the margin, the lower the generalization error of the classifier~\cite{Nayak2015, Burbidge2001}. Due to the generalization principle, SVMs have the potential to capture very large feature spaces \cite{cristianini_shawe-taylor_2000, Cervantes2020, Joachims2002, Zhan2005}. 

Another classical regression model that can capture highly complex relationships between variables, when the underlying dependence between these variables is unknown, are Neural Networks. The simplest NN type is the Feed-Forward Neural Network (FFNN), which allows the information to move only from input to output nodes without any feedback loops. The hidden layers of a FFNN provide a universal function approximation framework~\cite{Hornik1989, Goodfellow2016}, that can achieve any degree of accuracy depended on the size of the hidden layer. FFNNs have been extensively used in data modeling, due to the universal approximation capabilities and high expressivity of NNs.

Other types of deep NNs that have recently gained attention, are Physics-Informed Neural Networks (PINNs)~\cite{Owhadi2015, Raissi2017, Raissi2019, Kim2021, Cuomo2022}. PINNs are a type of universal function approximator that embed the knowledge of physical laws in the learning process of a given dataset, which acts as a regularization agent to the learning. Furthermore, PINNs can be described by Partial Differential Equations (PDEs), which provides a framework for the physical information to be provided for a given problem~\cite{Yang2019, Meng2020, Karniadakis2021}. Training of PINNs must satisfy both the training data as well as the equations that describe the problem. The combination of the two allows PINNs to find accurate solutions without the need for large training sets or boundary conditions~\cite{Raissi2018, Blechschmidt2021, Kollmannsberger2021, Lu2021}.

In an effort to address scalability issues and sub-optimal performance with many of these classical algorithms, researchers developed quantum algorithms that are inspired from their classical counterparts. For example, quantum SVMs have been developed to combat the computationally expensive step to estimate kernel functions, as the size of the feature space increases. Then, the exponentially large quantum state can be manipulated through controllable entanglement and interference~\cite{Havlicek2019, Paine2022}.

Quantum Neural Networks (QNNs)~\cite{Kak1995, Chrisley1995, Menneer1995, Chrisley1996, Behrman1996, Narayanan1996, Menneer1999, Jeswal2019, Jia2019, Zhao2021a} derive their name from classical NNs as they are universal function approximators, but otherwise they are not a direct translation of classical NNs. Typically, QNNs are referring to variational or parametric quantum circuits (PQCs) that use quantum gates and similar training techniques as the classical NNs to perform optimization. VQAs or PQCs involve quantum circuits that depend on free parameters that can be tuned during training of the QNN. Typically, such quantum algorithms follow a few basic steps: i) Preparation of a fixed (known) quantum state $\ket{\psi}$, ii) Followed by a quantum circuit $U({\bm{\theta}})$ that has trainable ${\bm{\theta}}$ parameters, and iii) Measurement of the desired observable $\hat{A}$ at the output of a quantum circuit in a fixed measurement basis. The expectation values of the function $f({\bm{\theta}}) = \braket{\psi | {U^{\dagger}({\bm{\theta}}) \hat{A} U({\bm{\theta}})} | \psi}$ define a scalar cost for a given task and the free parameters ${\bm{\theta}}$ are tuned to optimize the cost function. The training process of the ${\bm{\theta}}$ parameters typically occurs via classical optimization algorithms that take as input the measurement values of the quantum circuit. This is an iterative process, since the classical optimizer tries to find the best ${\bm{\theta}}$ parameters that will provide such quantum measurements that optimize the cost function.

A particularly compelling use-case for QNNs is regression, for example in the setting of Quantum Circuit Learning (QCL)~\cite{Mitarai2018} or related work on modelling stochastic processes, including protocols for Differentiable Quantum Generative Modeling (DQGM)~\cite{Kyriienko2022}.\\
In the QCL framework, a quantum circuit learns a given task by tuning variational parameters and efficiently inserting classical data using quantum feature map circuits, which is a strategy compatible with NISQ and requiring no QRAM or state tomography.
Another interesting class of algorithms, which are a type of quantum algorithms that solve systems governed by (non-)linear differential equations is proposed by Kyriienko et al. More information regarding this class of algorithms can be found in \textcite{Kyriienko2021, Paine2021, Heim2021, Kyriienko2022}. The functions are defined as expectation values of PQCs, and the function derivatives are obtained through automatic differentiation and are represented by differentiable quantum circuits (DQCs). DQCs are trained to satisfy differential equations and specified boundary conditions, while solving differential equations in a high-dimensional feature space.

The models discussed so far can either perform (combinatorial) optimization or regression; however, to the best of our knowledge, there is no quantum algorithm that combines these two. Quantum extremal learning (QEL) aims to bridge this gap \cite{patent}.

\subsubsection*{Quantum extremal learning}
\begin{figure}[!htb]
	{
		\scalebox{0.85}{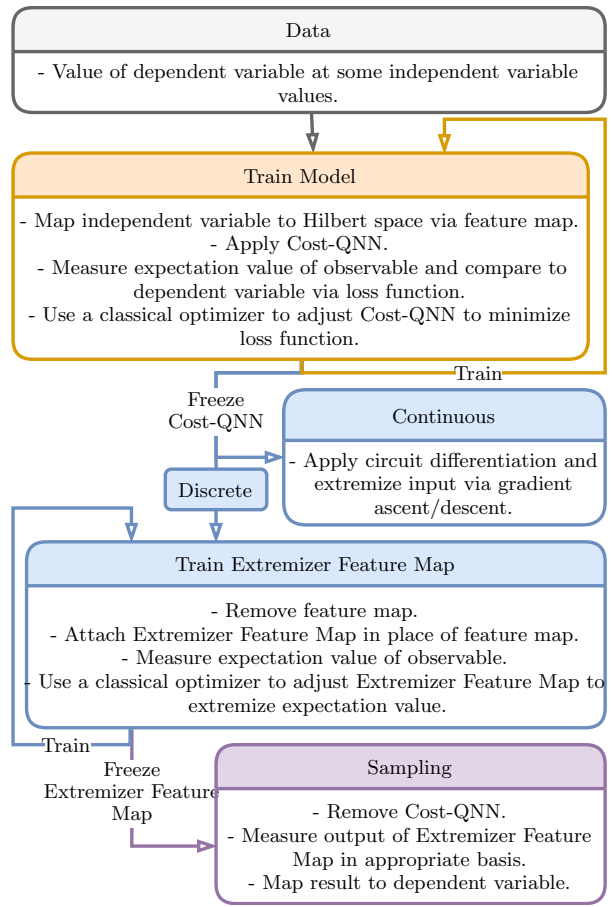}
		\caption{The workflow of QEL\cite{patent} algorithm, showing both the discrete and continuous case.} 
		\label{fig:QEL_workflow}}
\end{figure}
Our proposed quantum algorithm consists of a quantum feature map that encodes the input data to quantum states and a parametrized quantum circuit that is variationally trained to model data input-output relationships. When the model has converged, we analytically differentiate the quantum feature map that encodes the input data with respect to its input, in order to be able to use gradient descent to find the coordinate that extremizes the trained model. The algorithm is able to accept as input either discrete or continuous variables. In case of discrete variables, an additional trainable quantum feature map enables the differentiation of ``interpolated" input data. 

The extremal learning algorithm we propose is a hybrid classical/quantum algorithm, since we use a classical optimizer to find the best parameters for our parametric quantum circuits (PQCs). However, it is crucial to note that we bridge modeling and optimization without losing the ability to leverage quantum superposition. This, combined with the extension to accept discrete input, allows us to solve problems that are intractable with the classical framework of extremal learning in \textcite{Patel2021}. 

There are various hybrid quantum algorithms that focus solely on modeling. For simple continuous input, the modeling step of QEL could be performed as suggested in e.g. \textcite{Mitarai2018}. \textcite{Kyriienko2021, Paine2021, Heim2021, Kyriienko2022}  provides a collection of algorithms to model and solve systems governed by PDEs and generalization is discussed from the perspective of using continuous quantum feature map circuits. These algorithms did not yet discuss optimizing the found solutions with respect to inputs or design parameters for design optimization tasks, which is the extension we make in this work. The innovation of QEL is the incorporation of optimization with modeling without an intermediate classical conversion. 

A hybrid quantum algorithm involving modeling and optimization that handles discrete input is proposed in \textcite{Kitai2020}. A classical machine learning model called a Factorization Machine is trained on available training data. The correlation matrix of the trained model is then given as input to a Quantum Approximate Optimization Algorithm (QAOA), which is a hybrid quantum/classical optimization algorithm to find the optimal input value according to a cost function. The Factorization Machine is used here to produce a quadratic unconstrained binary optimization (QUBO) model, which is the default type of input to the QAOA. It is non-trivial how general models can be mapped to a QUBO, and this places a stringent limit on the format of the original machine learning model and aspects such as the order of correlations that can be incorporated. QEL, on the other hand, can utilize general established quantum machine learning modeling techniques~\cite{Mitarai2018, Kyriienko2021} and is not constrained by these limitations. Using a quantum module for modeling also, allows the direct use of the model for subsequent quantum optimization.\\


We envision that our algorithm can be applied in a wide variety of optimization applications such as computational chemistry and material science. The contributions of this paper can be summarized as follows:

\begin{enumerate}
    \item We propose a quantum algorithm for extremal learning.
    \item The algorithm enables the combination of established quantum machine learning modeling with established quantum optimization, on a single circuit/quantum computer.
    \item We propose strategies to handle continuous and discrete input variables.
    \item We rely on measurement of expectation value of observables rather than quantum state tomography which is important for, and compatible with, NISQ devices.
    \item A pre-trained model is not required; rather our algorithm learns a model based on provided data and optimizes it directly.
\end{enumerate} 

The rest of the paper is organized as follows: in section ~\ref{section:methodology}, we explain how our QEL algorithm works and provide details about the implementation. In section ~\ref{section:results}, we provide the results of the numerical simulations for the tested optimization problems, as well as the analysis of the results. In section ~\ref{section:discussion}, we speculate on how future Fault-Tolerant Quantum Computing (FTQC) will enhance the performance of our algorithm and draw our conclusions about this research. 

\section{Methodology}
\label{section:methodology}

Our algorithm assumes an unknown function that maps a discrete or continuous independent variable (input) $\bm{x} \in X$ to an associated dependent variable (output) $y = f(\bm{x}) \in \mathbb{R}$. A finite set of such pairs $\Set{(\bm{x}_i, y_i)}$ is known and is the training data for our algorithm. The goal of our algorithm is to find $\bm{x}_\mathrm{opt} \in X$ such that $f(\bm{x}_\mathrm{opt})$ is the extremal value of $f(x)$. 

Our quantum algorithm draws inspiration from the classical extremal learning algorithm \cite{Patel2021}, since the goal of both algorithms is to find the extremization input of a learned model. The main difference lies in that we are using Quantum Neural Networks. Training of Quantum Neural Networks in our model is based on Quantum Circuit Learning \cite{Mitarai2018}. Note that QCL is exclusively a means of training classification and regression models and does not attempt to find the extremizing input, which is the goal of our work. Both extremal learning~\cite{Patel2021} and QCL~\cite{Mitarai2018} restricts itself to the the continuous case only. We not only extend extremal learning to using QNNs for continuous input-variables, but also discrete input-variables.

We discuss the quantum extremal learning (QEL) algorithm separately for discrete and continuous input variables. A generalization of the QEL algorithm, in which discrete and continuous input are edge cases, is presented in Appendix~\ref{app:generalization}.  

\subsection*{Step~I: Training a Quantum Model}

\begin{figure}[!htb]
    {
    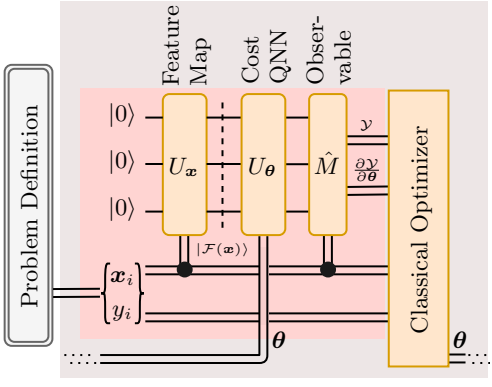
    \caption{One epoch of Step~I is shown in the gray rectangle. The part of the circuit in the inner pink box is executed for each element of the training data ${(\bm{x}_i, y_i)}$ and the results are used by a classical optimizer to suggest $\bm{\theta}$ for the next epoch.} 
    \label{fig:StepI}}
\end{figure}

A quantum model is trained on the basis of the classical dataset that is provided. The quantum model consists of a variational quantum circuit defined on a qubit register. The qubit register is initialized in the zero state $\ket{0}^{\otimes N}$ and a quantum feature map $\mathcal{F}$ is used to map coordinates from input values (independent variable) $\bm{x} \in X$ to a distinct place in the Hilbert space of the qubit register. 
\begin{equation}
\begin{aligned}
    \mathcal{F}\vcentcolon  X &\rightarrow \mathcal{H}_{1/2}^{\otimes N}\\
    \bm{x} &\mapsto \ket{\mathcal{F}(\bm{x})}.
\end{aligned}
\end{equation}
Note that $\mathcal{F}$ is continuous and it is physically accomplished by means of a unitary operator acting on $\ket{0}^{\otimes N}$ whose parameters depend on $\bm{x}$.
\begin{equation}
    U_{\bm{x}} \ket{0}^{\otimes N} \vcentcolon= \ket{\mathcal{F}(\bm{x})}
\end{equation}
The Cost-QNN, a variational unitary circuit (analog or digital) $U_{\bm{\theta}}$, is applied to this and a measurement of the output state in the form of the measurement of an observable $\hat{M}$ is made
\begin{equation}
\begin{aligned}
    \braket{\mathcal{F}(\cdot) | U_{{\bm{\theta}}}^{\dagger} \hat{M} U_{\bm{\theta}} | \mathcal{F}(\cdot)}\vcentcolon X &\rightarrow \mathbb{R}\\
    \bm{x} &\mapsto \mathcal{Y}.
\end{aligned}
\end{equation}

Training of the Cost-QNN is similar to training any classical Neural Network, as shown in Fig.~\ref{fig:StepI}. For each input $\bm{x}_i$, the corresponding output $\mathcal{Y}_i$ (dependent variable) is recorded, where the output is obtained by the above measurement. A distance metric between the true values $\Set{y_i}$ and the predicted values $\Set{\mathcal{Y}_i}$, such as the mean square error (MSE), acts as the loss function $L$. The loss function $L$ is a function of the parameters ${\bm{\theta}}$ of the variational circuit $U_{{\bm{\theta}}}$ and is real-valued. 
\begin{equation}
    \bm{\theta} \mapsto L\left({\bm{\theta}}; \Set{(\bm{x}_i, y_i)}\right) \in \mathbb{R}
\end{equation}
It can be minimized using standard optimization algorithms based on gradient-based techniques. If the training is successful, 
\begin{equation}\label{eq:approx_f}
    \bra{\mathcal{F}(\bm{x})} U_{{\bm{\theta}}}^{\dagger} \hat{M} U_{\bm{\theta}} \ket{\mathcal{F}(\bm{x})} \approx f(\bm{x}).
\end{equation}
     
\subsection*{Step~II: Extremizing the Quantum Model}
\subsubsection*{Continuous case} 
\begin{figure}[!htb]
    {
    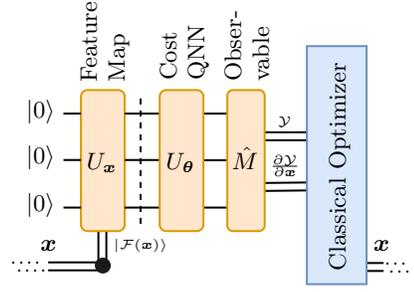
    \caption{Step~II in the continuous case. The part of the circuit retained from Step~I is shown in orange and the new elements are shown in blue. Notice that $\bm{\theta}$ is no longer updated; the optimization is performed on $\bm{x}$.} 
    \label{fig:StepIIcont}}
\end{figure}
At this stage, the parameters ${\bm{\theta}}$ of Cost-QNN are frozen/fixed at the optimal values that were obtained in the previous stage. In this case, $\mathcal{F}$ is continuous and the approximation of $f(\bm{x})$ given in Eq.~\eqref{eq:approx_f} 
\begin{equation}\label{eq:con_out}
    \bm{x} \mapsto \mathcal{Y} \vcentcolon = \bra{\mathcal{F}(\bm{x})} U_{{\bm{\theta}}}^{\dagger} \hat{M} U_{\bm{\theta}} \ket{\mathcal{F}(\bm{x})}\in \mathbb{R}
\end{equation}
is continuous. Since our trainable feature map is analytically differentiable by employing circuit differentiation methods such as parameter shift rules~\cite{Mitarai2018, Schuld2019a}, we can calculate $\partial_{\bm{x}}\mathcal{Y}$. We can now use the gradient ascent / descent to aid in finding the input value $\bm{x}_\mathrm{opt} \in X$ that extremizes the value of the observable $\hat{M}$ on the circuit output state representing $f(\bm{x})$ in Eq.~\eqref{eq:con_out}, as shown in Fig.~\ref{fig:StepIIcont}

\subsubsection*{Discrete case}
\begin{figure}[!htb]
    {
    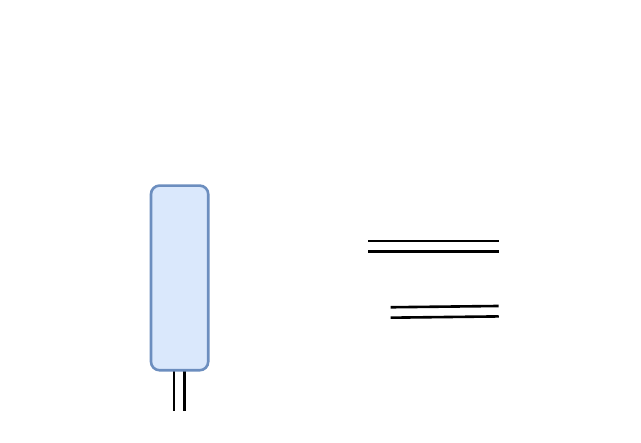
    \caption{First part of Step~II in the discrete case. The part of the circuit retained from Step~I is shown in orange and the new elements are shown in blue. Notice that $\bm{\theta}$ is no longer updated; the optimization is performed on $\bm{\mathscr{X}}$.} 
    \label{fig:StepIIa}}
\end{figure}
For discrete input data, one possible feature map $\mathcal{F}$ is a simple digital encoding, where a binary representation of input $\bm{x}$ is mapped to binary bitstrings. After initializing the qubit register in all-zeros, the injection $\mathcal{F}$ transforms that state to a product state of zeros and ones (bitstrings), by applying an X-gate to each qubit on indices where a 1 is present in the  binary representation of the input $\bm{x}$.
\begin{equation}
\begin{aligned}
    \mathcal{F}\vcentcolon X &\rightarrow \Set{0,1}^{\otimes N}\\
    &\rightarrow \set{ \ket{0},\ket{1} }^{\otimes N} \subset \mathcal{H}_{1/2}^{\otimes N}.
\end{aligned}
\end{equation}

Recollect that the function $f$ is unknown and our goal is to find $\bm{x}_\mathrm{opt} \in X$ such that $f(\bm{x}_\mathrm{opt})$ is the extrema of $f$. Unlike the continuous case, $X$ is discrete and is not path connected, so the continuity of the approximation of $f$ in Eq.~\eqref{eq:approx_f} cannot be exploited to find $\bm{x}_\mathrm{opt}$. Instead, we make use of quantum machine learning to search for $\bm{x}_\mathrm{opt}$.

Initially, we remove the feature map $\mathcal{F}$ from the circuit that trained the Cost-QNN in Step~I. Then, we insert another variational unitary quantum circuit $U_{{ \bm{\mathscr{X}}}}$ (analog/digital) that is connected to the input of the Cost-QNN as shown in Fig.~\ref{fig:StepIIa}. We call the new unitary circuit Extremizer Feature Map, with variational parameters ${ \bm{\mathscr{X}}}$. The state $\ket{0}^{\otimes N}$ is used as input to the Extremizer Feature Map. We define
\begin{equation}
    U_{ \bm{\mathscr{X}}} \ket{0}^{\otimes N} \vcentcolon = \ket{{ \bm{\mathscr{X}}}}  \in \mathcal{H}_{1/2}^{\otimes N}.
\end{equation}
Note that during Extremizer Feature Map training, the parameters ${\bm{\theta}}$ of Cost-QNN remain fixed with values obtained during its own training. The output of the circuit is obtained from the expectation value of the observable $\hat{M}$ 
\begin{equation}\label{eq:dsc_out}
    { \bm{\mathscr{X}}} \mapsto \mathscr{Y} \vcentcolon = \bra{{ \bm{\mathscr{X}}}} U_{{\bm{\theta}}}^{\dagger} \hat{M} U_{\bm{\theta}} \ket{{ \bm{\mathscr{X}}}}\in \mathbb{R}.
\end{equation}
and is maximized (minimized) by variationally optimizing parameters ${ \bm{\mathscr{X}}}$. The optimization of ${ \bm{\mathscr{X}}}$ can be accomplished with gradient ascent (descent), using analytical circuit differentiation rules to compute $\partial_{ \bm{\mathscr{X}}} \mathscr{Y}$. 

The Extremizer Feature Map can be thought of as a trainable feature map, which is trained with respect to ${ \bm{\mathscr{X}}}$. The goal of this training is to learn an intermediate wavefunction that has maximum contribution from states (bitstrings, if we are using digital encoding) that result in a high (low) value of the model to be maximized (minimized). 

\begin{figure}[!htb]
    {
    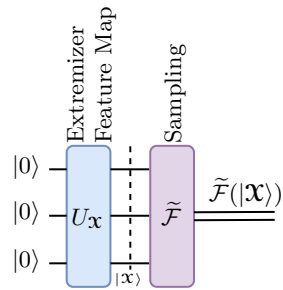
    \caption{The final sampling in Step~II of the discrete case. Due to the previous steps of the algorithm, the bitstring $\widetilde{\mathcal{F}}(\ket{{ \bm{\mathscr{X}}}})$ we measure has an enhanced probability of being mapped to $\bm{x}_\mathrm{opt}$.} 
    \label{fig:StepIIb}}
\end{figure}        
Once the model has converged to an optimal ${ \bm{\mathscr{X}}}$, the Cost-QNN $U_{{\bm{\theta}}}$ is removed from the circuit as shown in Fig.~\ref{fig:StepIIb}. In contrast to $\bm{x}$ in Eq.~\eqref{eq:con_out}, $\bm{\mathscr{X}}$ cannot be directly associated with an input since the input is discrete. Instead, we sample the output of the Extremizer Feature Map circuit to ascertain $\bm{x}_\mathrm{opt}$.
If we are using digital encoding, we do this by making a measurement in the $\hat{\sigma}_Z^{\otimes N}$ basis, which collapses the wavefunction to the space $\set{ \ket{0},\ket{1} }^{\otimes N}$, giving us the map
\begin{equation}
\begin{aligned}
    \widetilde{\mathcal{F}}\vcentcolon \mathcal{H}_{1/2}^{\otimes N} &\rightarrow \set{\ket{0},\ket{1}}^{\otimes N}\\
    &\rightarrow \Set{0,1}^{\otimes N}\\ 
    &\rightarrow X' \supseteq X\\
    \ket{{ \bm{\mathscr{X}}}} &\mapsto \widetilde{\mathcal{F}}(\ket{{ \bm{\mathscr{X}}}}) \ni X' \supseteq X
\end{aligned}
\end{equation}
respecting $\widetilde{\mathcal{F}}^{-1}\mathcal{F} = \mathbb{1}$ (also $\mathcal{F}^{-1}\widetilde{\mathcal{F}}=\mathbb{1}$ if $X'=X$). The previous steps of the algorithm enhance the probabilty that the bitstring $\widetilde{\mathcal{F}}(\ket{{ \bm{\mathscr{X}}}})$ we measure is mapped to $\bm{x}_\mathrm{opt}$.

\subsection*{}

Our algorithm maps the input to the output by means of only quantum gates and a final quantum measurement without intermediate conversion to classical data. Hence, it can leverage quantum superposition for a computational advantage. 

For specific datasets, QNNs have been shown to have fast convergence rates and low generalization error, due to quantum parallelism, which depending on the task might overwhelm the classical NN performance for the same problem. The reasoning is that the cost function of the QNN is encoded as a superposition state in the Hilbert space of the QNN's parameters. This quantum mechanism exploits the hidden structure of QNN to converge to the optimal parameters faster~\cite{Liao2021}. Furthermore, due to various techniques that are designed to avoid the barren plateau problem~\cite{McClean2018, Grant2019, Cerezo2021b, Wang2021, Cerezo2021c, Arrasmith2021, Holmes2021, Ortiz-Marrero2021, Patti2021, Zhao2021b}, the variational circuits are able to provide very good candidate solutions to optimization problems. Additionally, they exhibit a larger model capacity compared to NNs for various tasks~\cite{Abbas2021, Coles2021, Lewenstein2021, Wright2019}.

A critical aspect in any optimization problem setting is the choice of the observable $\hat{M}$. In our case, we decide to use total magnetization, which is the sum of magnetization on all qubits $\sum_i \hat{\sigma}_Z^i$, to retain the large expressibility in our ansatz. Magnetization in such Ising formulated problems reflects the average value of the spin (state of the qubit), thus total magnetization provides information about the state of the system. Also, due to the fact that many of the problems that we are interested in have diagonal Hamiltonians, total magnetization is a simple and easy to measure metric (calculation of the expectation value involves only measurements in one basis, the Z-basis on all qubits), while generalizations can be trained with the variational ansatz before measurement.

\section{Numerical Simulations and Results}
\label{section:results}

In order to evaluate the performance of the QEL algorithm, we performed numerical simulations concerning both the continuous and the discrete input variable case. We examine the modeling accomplished by the training procedure of the QNN and the ability to find the optimal solution to the optimization problem through gradient based techniques.

\subsection{Continuous variables: data fitting}
Continuous variable modeling and optimization concerns the task of fitting a model to a dataset consisting of coordinates $x_i$ and corresponding values $f(x_i)$ of a function $f$ defined on a continuous domain. We demonstrate our method here for the case of a simple function, 
\begin{equation}
    f(x) = \sin(5x), 
\end{equation}
with the aim of maximizing it in the domain $[0,1]$. As a training dataset, we give only values away from the maximum, to avoid potential biases during our modeling and optimization process. 

First, to fit the data, the QCL-type model \cite{Mitarai2018} we use is a QNN circuit with 3 qubits; we initialize in the all-zero state, and apply a feature map with $\hat{R}_y(2j\arccos(x))$ where $j$ is the qubit index starting from 1 and $\hat{R}_y$ is a single-qubit Pauli-Y rotation. Next, we apply a Hardware Efficient Ansatz (HEA)~\cite{Kandala2017}, which is a variational ansatz of depth typically equal to the number of qubits (in this example 3). Finally, we measure the total magnetization $\hat{M}=\sum_j \hat{\sigma}_Z^j$ such that the quantum model is $f(x)=\langle\hat{M}\rangle$.

\begin{figure}[!htb]
    
    \includegraphics[width=0.9\linewidth]{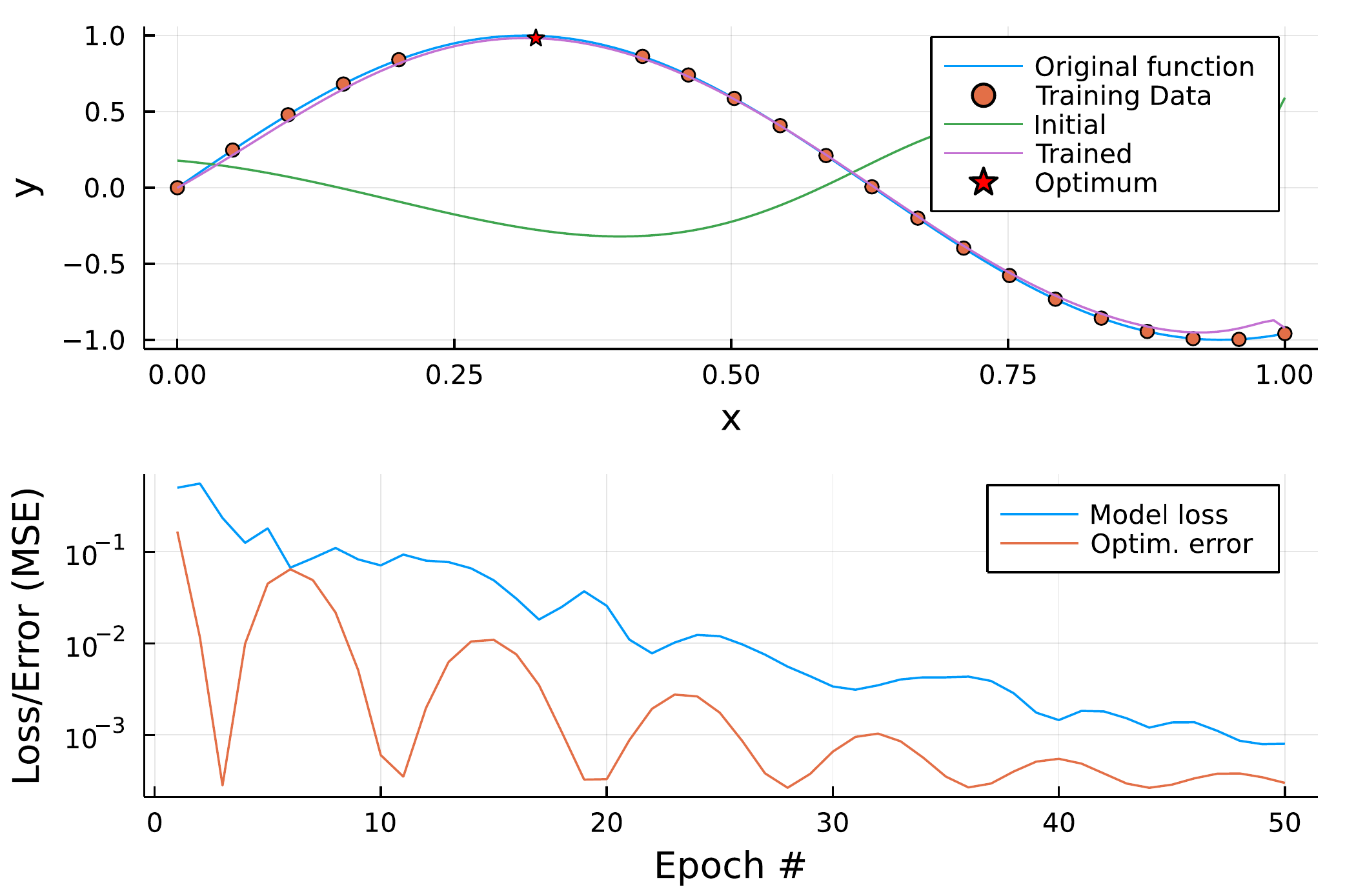}
    \caption{(top) plot showing the original function as a function of x, the training data selected from this function, the initial QNN model, the trained QNN model, and the optimum value of the model, which closely resembles the true optimum. (bottom) plot showing the model loss (MSE) as a function of epoch number in the ADAM optimization. We used a learning rate of $0.5$ and 50 epochs to train the model. Furthermore, we plot the extremizer's MSE as a function of optimization iteration number.}
    \label{fig:qcl_optim}
\end{figure}

The results are found in Figure \ref{fig:qcl_optim}. The optimization shows oscillations that originate from circling around the optimum value of the model function, due to momentum in the ADAM optimizer~\cite{Kingma2014}.

Although this is just a toy example, it shows the simplicity of the approach in the continuous case. Next, we consider how the continuous case extends to differential equations as well.

\subsection{Continuous variables: differential equation solving and extremizing}
We showcase here how a differential equation can be solved by the generalization of our method described in Appendix~\ref{app:generalization}, following \textcite{Kyriienko2021}. Subsequently, the extremal value of the solution within its domain is found.

\begin{figure}[!htb]
    
    \begin{tabular}{c}
     \includegraphics[width=0.9\linewidth]{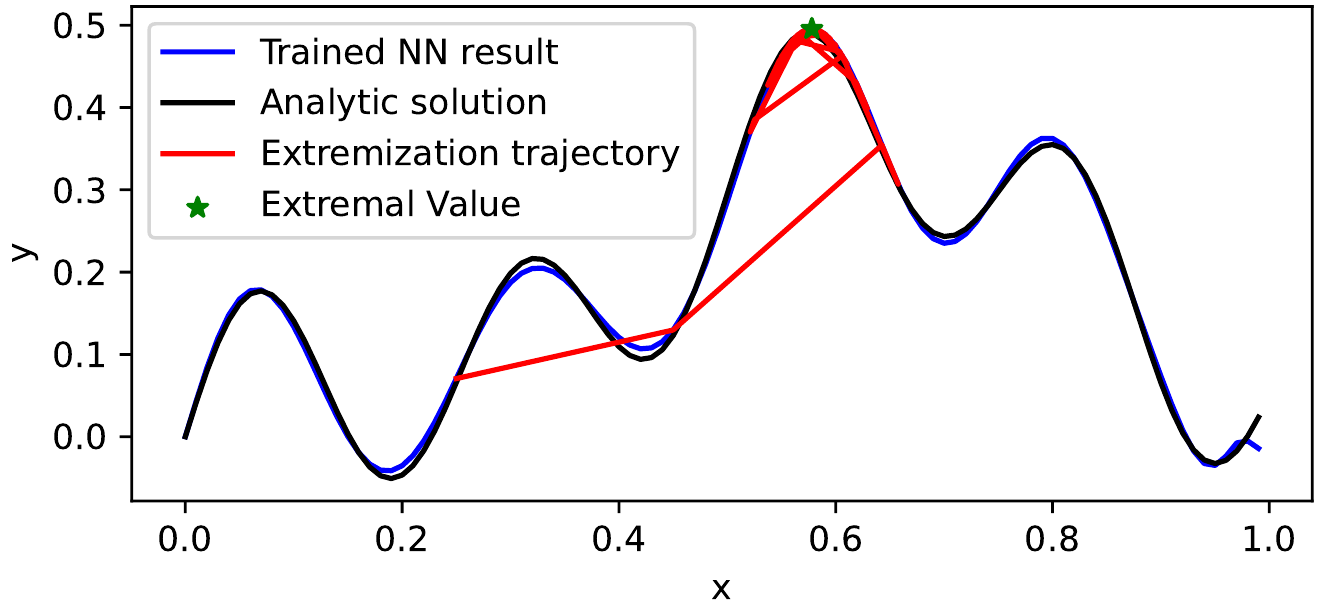}      \\
     \includegraphics[width=0.9\linewidth]{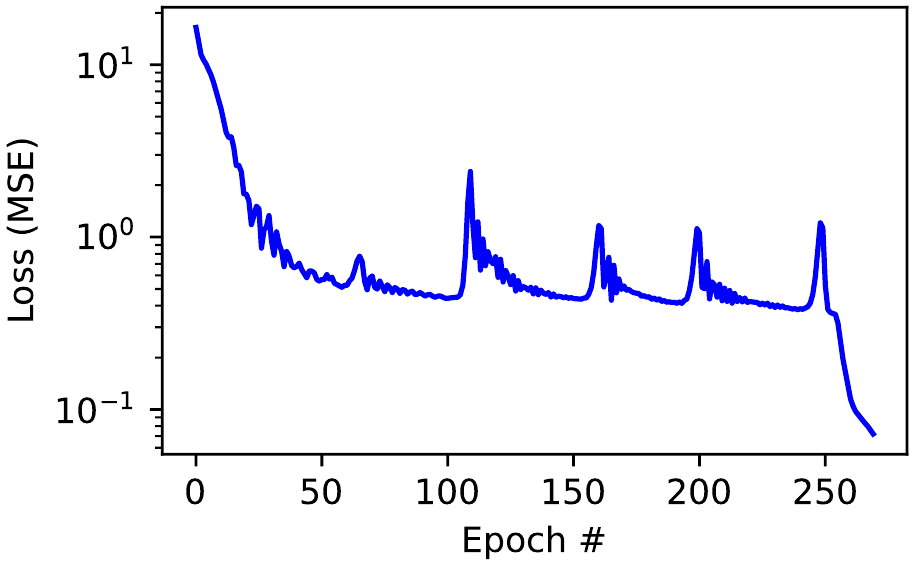}      \\
     \includegraphics[width=0.9\linewidth]{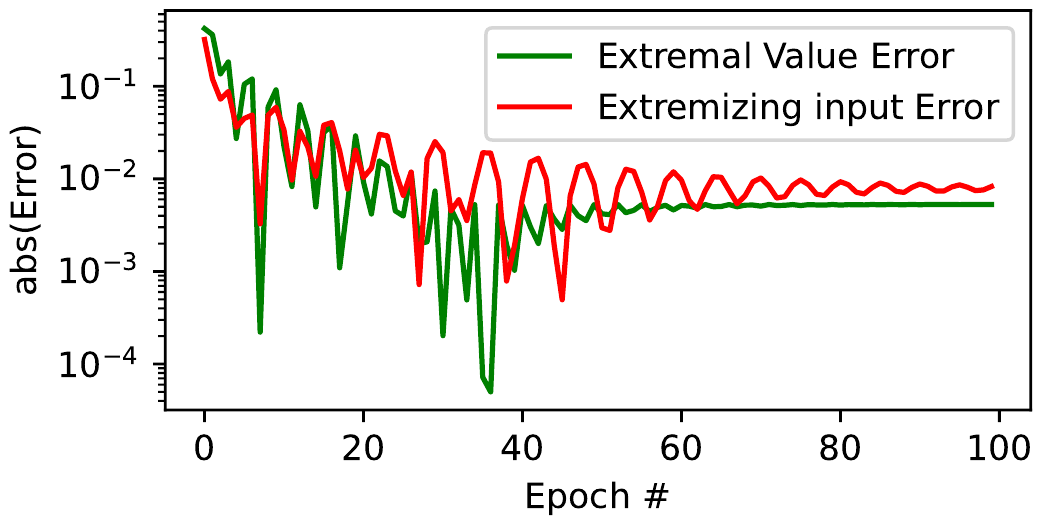}      \\
    \end{tabular}
    \caption{(top) plot showing the analytical solution to Eq.~\eqref{eq:diff_eq}, its approximate solution found by DQC using a trained quantum neural network (NN), and overlaid the extremization trajectory of an initial guess starting at $x=0.25$. (middle) model training loss trajectory with 250 ADAM epochs at a learning rate of 0.1 and 20 L-BFGS epochs at a learning rate of 0.05. (bottom) absolute deviation between the extremizer's suggestions versus exact extremal value and extremizing inputs, respectively, versus extremizer epoch number; a learning rate of 0.2 was chosen for 100 epochs using ADAM.}
    \label{fig:dqc_optim}
\end{figure}

We consider the following 1D ordinary differential equation (ODE):
\begin{align}\label{eq:diff_eq}
    \frac{\partial f}{\partial x} &= -\sin(10x) +3\cos(25x)-2x + 5/4, \\
    f(0) &= 0.
\end{align}
The solution can be computed analytically, and it shows non-trivial oscillatory behavior as shown in Fig.~\ref{fig:dqc_optim}. As a universal function approximator (UFA) we use a 6-qubit QNN with Chebyshev Tower feature map \cite{Kyriienko2021} and train the UFA on 50 points spread uniformly over the domain $x\in \Set{0,1}$. We get good agreement between trained solution and analytical solution, in a loss optimization trajectory of less than 300 epochs (see details in figure caption).

Next, we maximize the value of the trained model with respect to the continuous input $x$, and find a good agreement between exact values and extremizer suggestions of less than $1\%$ errors.

While this example showed the solving of a simple ODE in 1D, the method can be used for any general (set of) PDE(s) \cite{Kyriienko2021} or even SDEs \cite{Paine2021}, as well as extremizing solutions found by means of model discovery \cite{Heim2021}.\\
Furthermore, in this section and the previous, we showcased the specific extremization case of maximizing a 1D function. More complex target `constraints' may be considered as well, such as maximizing the sum of two functions, the product, or minimizing some design constraint which is a function of the model itself. Such generalizations are furthermore included in the description in Appendix~\ref{app:generalization}.

In the next section, we show how the extremal learning paradigm can be applied also to the case where input variables are discrete-valued, where derivatives are not straightforwardly available.

\subsection{Discrete variables}
Discrete variable modeling concerns the task of fitting a model to a dataset consisting of coordinates $x_i$ and the corresponding values $f(x_i)$ of a function $f$ defined on a discrete domain. The optimization part cannot be obtained by explicit gradient based techniques as in the continuous case, since the independent variable $x$ is not continuous-valued, a necessity for differentiation. In this work, the optimization part in the discrete variable case occurs by training a new feature map with respect to ``interpolated" discrete input $x_i$ through gradient based methods. 

We demonstrate our method here for the case of Max-Cut clustering, and a problem based on randomly generated correlations up to a fixed order between nearest neighbors, and an application in computational chemistry. We perform numerical simulations on a wavefunction simulator without the presence of noise.

\subsubsection*{Max-Cut clustering}

Max-Cut problems~\cite{Garey1976, Cook1995, Berman1999} are well-suited here for testing purposes for two reasons: i) to test our quantum algorithm with a synthetic classical dataset, and ii) due to the simple correlations involved (only pairwise correlations, no bias terms). In a clustering setting~\cite{Everitt2011, Altman2017, Nowak2008}, we randomly generate $\nicefrac{N}{2}$ pairs of coordinates $(x,y)$ in 2D Euclidean space that belong to one cluster and another $\nicefrac{N}{2}$ pairs of coordinates that belong to the other cluster, where $N$ is the number of binary variables corresponding also to the number of qubits of the quantum system used. The two clusters by construction are separated by a distance variable, that can be set accordingly. The larger the distance, the more separated the clusters, thus the `easier' the Max-Cut problem is solved. Also, based on this construction of clearly separated clusters, we know the optimal solution before the optimization algorithm is applied, which allows us to assess the performance of the algorithm.

We focus the investigation to a problem with 6 qubits (each representing a binary variable located at particular coordinates) and assess the performance of the QEL algorithm for various sizes of training datasets. As specified in section \ref{section:methodology} concerning the discrete variable case, we use two variational circuits $U_{{\bm{\theta}}}$ and $U_{{ \bm{\mathscr{X}}}}$, each circuit with 6 qubits in the register for the first three use cases and 5 qubits for the chemistry use case. We perform similar training to the continuous case to obtain a trained model via the Cost-QNN. Then, we freeze the parameters of the Cost-QNN and remove its feature map. We attach another trainable quantum feature map (Extremizer Feature Map) in place of the removed feature map and start training the new model. At every iteration, we measure the expectation value of the total magnetization $\hat{M}=\sum_j \hat{\sigma}^Z_j$. The parameters ${{ \bm{\mathscr{X}}}}$ of the Extremizer Feature Map are suggested by a classical optimizer. Finally, when the training has finished and the parameters have been optimized, we remove the Cost-QNN and sample the remaining model. By measuring the output of the Extremizer Feature Map, we are able to map the result to the optimal input variable that is the solution to the optimization problem.

The variational circuits $U_{{\bm{\theta}}}$ and $U_{{ \bm{\mathscr{X}}}}$ are selected with depth $N^2$, so that they are more expressive. Through examination of the numerical simulations, we have found that depth $N$ can still suffice to obtain good results, at least for the small problem sizes that were investigated, although depth $N^2$ provided increased accuracy of finding the optimal solution. Going to larger depths of the variational circuit, did not yield significant increase in accuracy of finding the optimal solution.

We performed numerical simulations involving training the Cost-QNN and the Extremizer Feature Map until convergence based on the provided dataset. Typically, 50 training epochs sufficed for the Cost-QNN and 150 training epochs sufficed for the Extremizer Feature Map. The simulation for each training dataset was repeated 100 times with a new randomly selected group of samples every time (different sampling seed). The classical optimization part that finds the best ${\bm{\theta}}$ and ${ \bm{\mathscr{X}}}$ parameters was done with a gradient based algorithm.

\subsubsection*{Correlations up to higher order terms}
Max-Cut formulated problems only include constant and pairwise correlation terms. We decided to explore the algorithmic performance when a problem with all terms up to a fixed order is provided, and the solution does not involve clustering into complementary groups. Therefore, the problem is not classified as Max-Cut anymore, rather it is an artificially created optimization problem, where one attempts to find the solution with the highest cost.

We create a problem with $N$ points, which is the same number as the qubits that we will use, with nearest-neighbor interactions up to a fixed order. We calculate the cost of each sample in a similar way to the Max-Cut formulation, but the weight of each correlation is randomly generated, thus creating a cost landscape much different from Max-Cut. In such a formulation, there are no symmetries in the cost landscape and the optimal solutions will vary depending on the construction. 

We decided to investigate nearest neighbor correlations up to 2nd order, which enhances the Max-Cut formulation by including linear terms as well, and nearest neighbor correlations up to 3rd order, which provides us with a trend of how the algorithm will perform when higher order correlations are included. However, when higher order correlations between the qubits exist, the problem becomes much harder to solve, thus fine-tuning the parameters of the QEL algorithm might be required. For the purposes of this initial exploration, we kept the same QEL parameters in all of our simulations, regardless of the highest-order included.

\subsubsection*{Molecule optimization}

Finally, we provide an example that involves the optimization of a molecule with five substituents. 
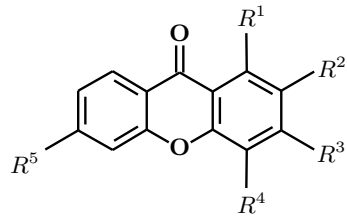
\begin{figure}[!htb]
    {
\begingroup%
  \makeatletter%
  \providecommand\color[2][]{%
    \errmessage{(Inkscape) Color is used for the text in Inkscape, but the package 'color.sty' is not loaded}%
    \renewcommand\color[2][]{}%
  }%
  \providecommand\transparent[1]{%
    \errmessage{(Inkscape) Transparency is used (non-zero) for the text in Inkscape, but the package 'transparent.sty' is not loaded}%
    \renewcommand\transparent[1]{}%
  }%
  \providecommand\rotatebox[2]{#2}%
  \newcommand*\fsize{\dimexpr\f@size pt\relax}%
  \newcommand*\lineheight[1]{\fontsize{\fsize}{#1\fsize}\selectfont}%
  \ifx\svgwidth\undefined%
    \setlength{\unitlength}{184.5bp}%
    \ifx\svgscale\undefined%
      \relax%
    \else%
      \setlength{\unitlength}{\unitlength * \real{\svgscale}}%
    \fi%
  \else%
    \setlength{\unitlength}{\svgwidth}%
  \fi%
  \global\let\svgwidth\undefined%
  \global\let\svgscale\undefined%
  \makeatother%
  \begin{picture}(1,0.44715447)%
    \lineheight{1}%
    \setlength\tabcolsep{0pt}%
    \put(0.4802684,0.12657876){\color[rgb]{0,0,0}\makebox(0,0)[t]{\lineheight{1.25}\smash{\begin{tabular}[t]{c}\textbf{O}\end{tabular}}}}%
    \put(0.4802684,0.35996737){\color[rgb]{0,0,0}\makebox(0,0)[t]{\lineheight{1.25}\smash{\begin{tabular}[t]{c}\textbf{O}\end{tabular}}}}%
    \put(0,0){\includegraphics[width=\unitlength,page=1]{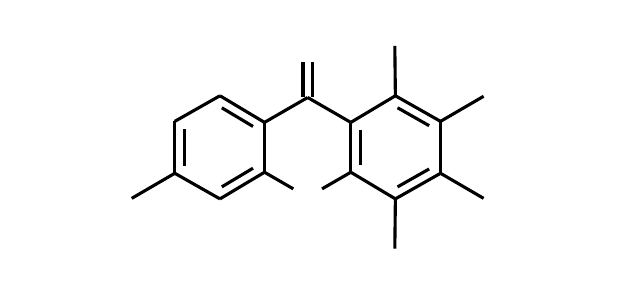}}%
    \put(0.59773589,0.38965853){\makebox(0,0)[lt]{\lineheight{1.25}\smash{\begin{tabular}[t]{l}$R^1$\end{tabular}}}}%
    \put(0.76173333,0.28481604){\makebox(0,0)[lt]{\lineheight{1.25}\smash{\begin{tabular}[t]{l}$R^2$\end{tabular}}}}%
    \put(0.76025161,0.11184928){\makebox(0,0)[lt]{\lineheight{1.25}\smash{\begin{tabular}[t]{l}$R^3$\end{tabular}}}}%
    \put(0.59773267,0.00787493){\makebox(0,0)[lt]{\lineheight{1.25}\smash{\begin{tabular}[t]{l}$R^4$\end{tabular}}}}%
    \put(0.13417886,0.08890264){\makebox(0,0)[lt]{\lineheight{1.25}\smash{\begin{tabular}[t]{l}$R^5$\end{tabular}}}}%
  \end{picture}%
\endgroup%

    \caption{Molecule containing five substituents ($R^{i}$) with two molecular subgroup options at each position.
    \label{fig:molecule_optim}}}
\end{figure}  

The chemical structure of the molecule is shown in Figure~\ref{fig:molecule_optim}, and to further simplify the problem we assume that only two molecular subgroups are allowed per each substituent $R^{i}$. Thus, the input of the problem becomes binary and the size of the possible combinations is $2^5=32$. We avoid any kind of compression of the input data, therefore have one qubit for every substituent. We also assume only nearest neighbor interactions between substituents up to the second order, thus each $R^{i}$ is only correlated to $R^{i-1}$ and $R^{i+1}$. We generate first order correlations according to a normal distribution with standard deviation of 1.0 and mean of 0.0, and second order correlations according to a normal distribution with standard deviation of 2.0 and mean of 0.0. For the Figure-of-Merritt (FOM), we assume that $\text{FOM}_{ij}=\text{FOM}_{ji}$. In physical terms the FOM could represent the calculated binding energy of the molecule to a protein target. The first order correlations would represent the interaction of the substituent with the protein target and the second order correlations the interactions between the substituents. The Figure-of-Merritt that was assumed in this use case was the total energy of the molecule, which was calculated based on the energies of the linear correlations $E_{i}$ and the energies of the quadratic correlations $E_{ij}$:

\begin{equation}
    E_{total} = \sum_{i}^{N} E_{i} + \sum_{ij}^{N} E_{ij} \text{ for } j>i
\end{equation}

Based on such configuration, we can easily increase the complexity of the problem by adding more substituents/qubits or assuming higher order correlations. However, scaling of such an artificial problem is not targeted at the moment, since this is an initial investigation of the QEL algorithm.

\subsubsection*{Results analysis: various thresholds}
\begin{figure}
     
     \subfloat{
         
         \includegraphics[width=\linewidth]{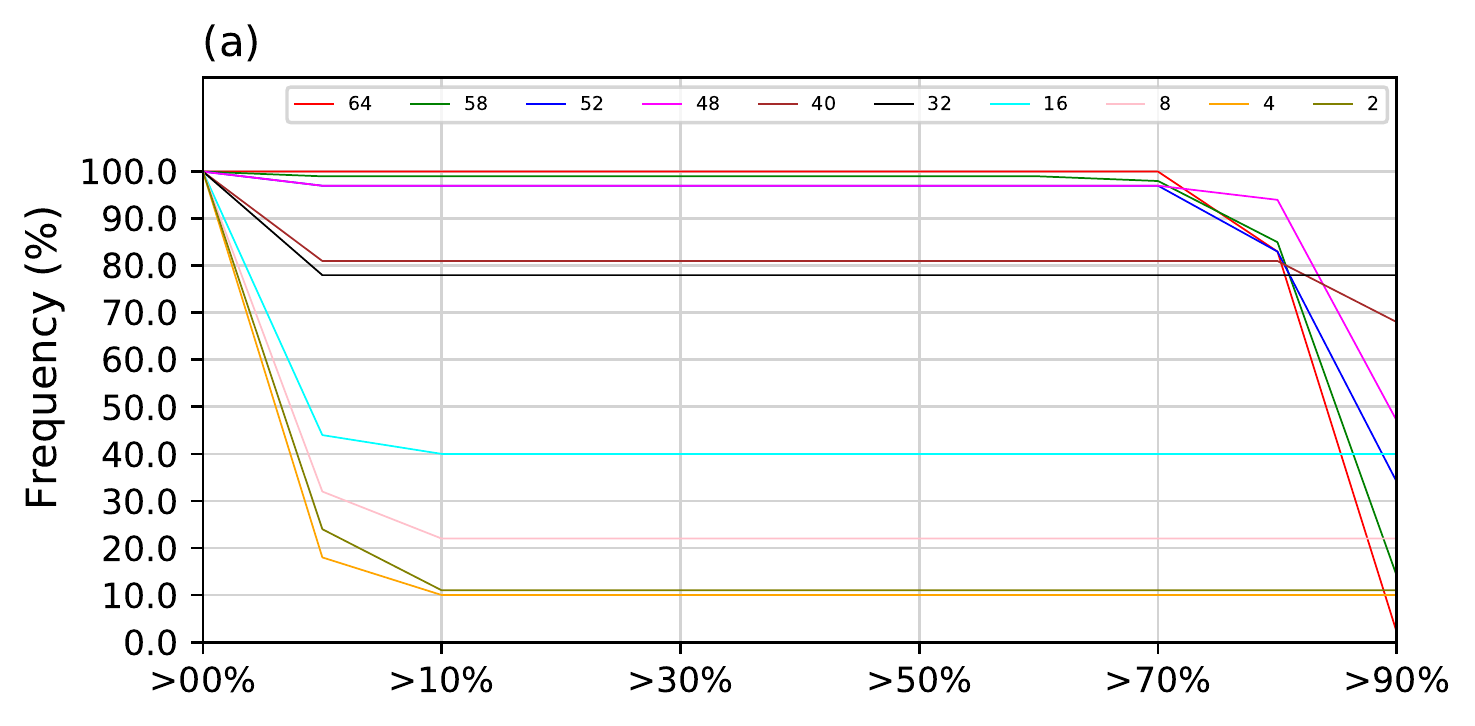}
     }
     \hfill
     \subfloat{
         
         \includegraphics[width=\linewidth]{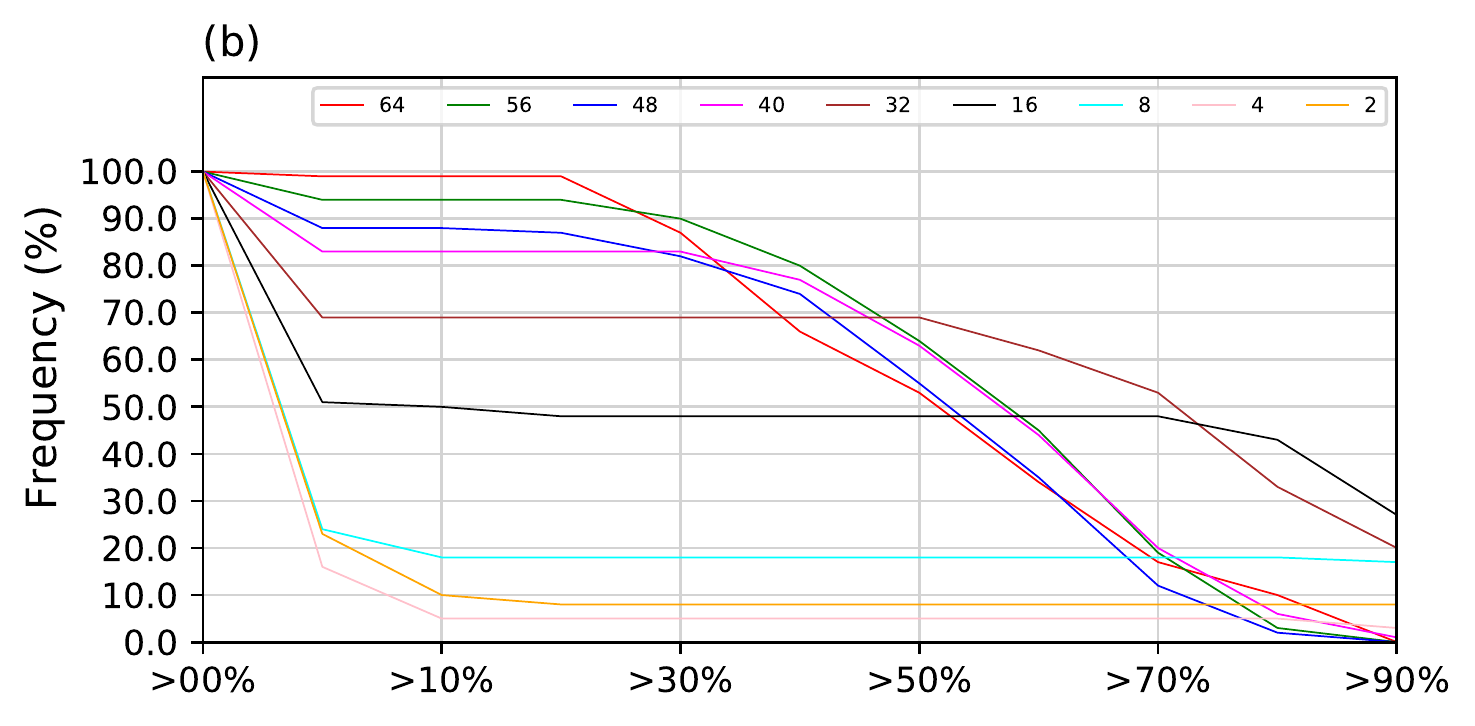}
     }
     \hfill
     \subfloat{
         
         \includegraphics[width=\linewidth]{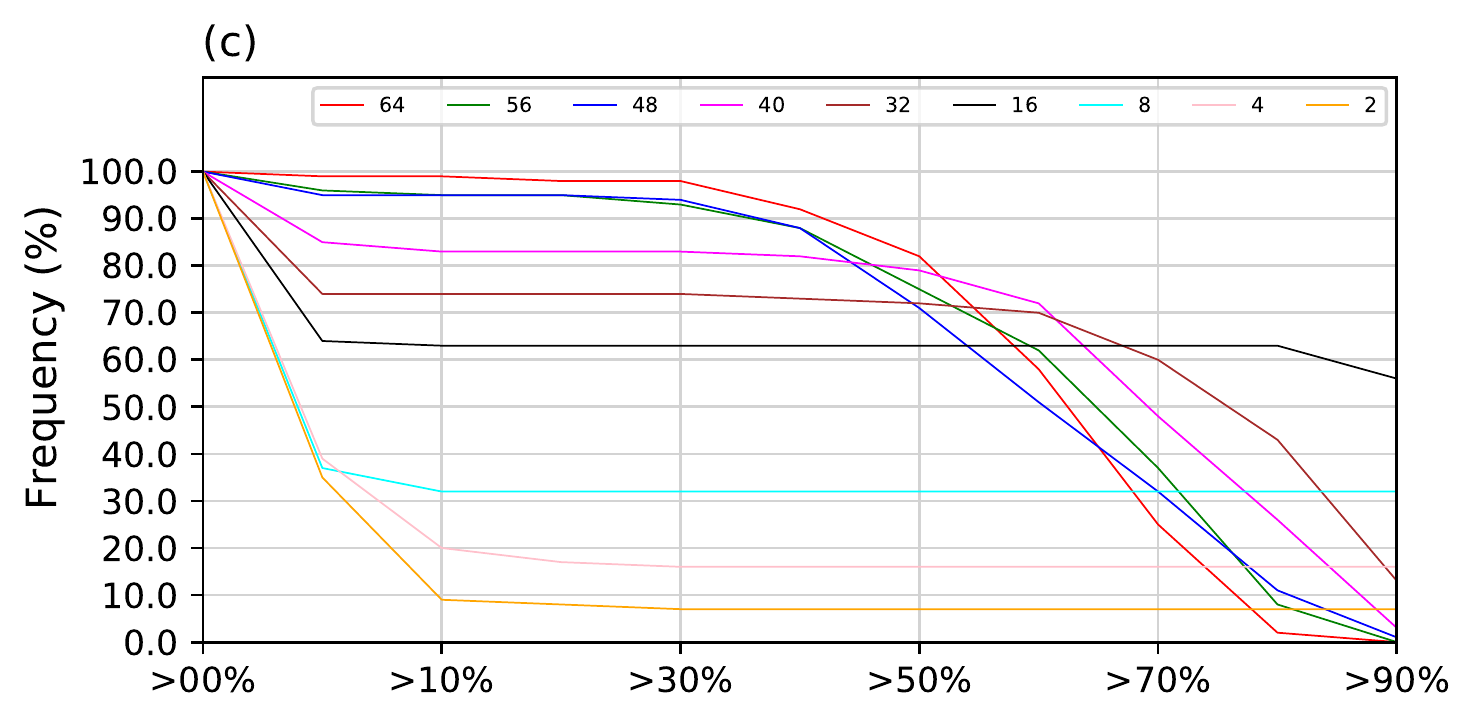}
     }
     \hfill
     \subfloat{
         
         \includegraphics[width=\linewidth]{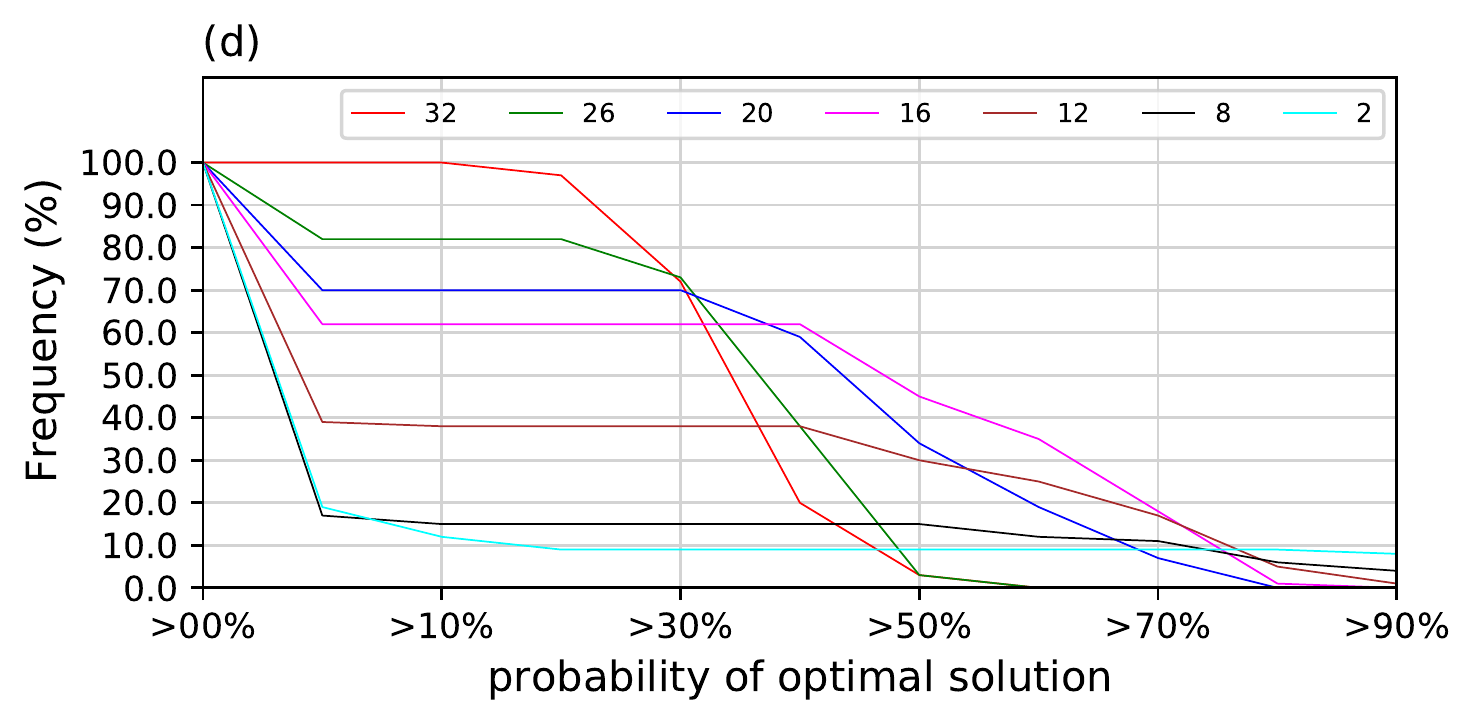}
     }
        \caption{Plot showing the investigation of the total probability of sampling any of the optimal samples, for a variety of training datasets for the problem of (a) Max-Cut clustering in two clusters for 6 qubits, (b) a system with nearest neighbor interactions up to 2nd order correlations for 6 qubits, (c) a system with nearest neighbor interactions up to 3rd order correlations for 6 qubits, (d) a molecule with 5 substituents with two molecular subgroup options per substituent.
        \label{fig:discrete_various_thresholds}}
\end{figure}

In Figure~\ref{fig:discrete_various_thresholds}, we present the results of the numerical simulations for (a) Max-Cut 2D clustering with 6 qubits with the distance between the two clusters being 5, (b) a system with nearest neighbor interactions up to 2nd order correlations for 6 qubits, (c) a system with nearest neighbor interactions up to 3rd order correlations for 6 qubits, (d) a molecule with 5 substituents with two molecular subgroup options per substituent. We provide the total probability to sample any of the optimal bitstrings ($\textit{total prob}$) for each training dataset size, as shown in the legend. On the x-axis, we provide the threshold that the $\textit{total prob}$ exceeds for each training set size. On the y-axis, we provide the frequency with which we obtain the ($\textit{total prob}$) for each threshold after running 100 randomly initialized simulations.

The general trend of the plots provided in Figure~\ref{fig:discrete_various_thresholds} is the following: When the number of training samples is increased reaching values close to the $2^N$, then the $\textit{total prob}$ of the optimal solutions is in general decreased. The reasoning for the decreased probability is that the algorithm is enhancing the probability of all samples, avoiding overfitting. We also observe that when a small amount of training samples is provided, the QEL algorithm is able to suggest at least one of the optimal solutions in most cases, with a moderately high probability ($>10\%$). Suggesting the optimal solution with a probability of $10\%$ is deemed enough for quantum computers, since the measurement to obtain the optimal solution will be repeated multiple times and evaluating the FOM of each obtained solution can be done fast and cheaply. Therefore, suggesting the optimal solution with a $10\%$ probability with a frequency of $10\%$ of the times that the simulation is run, will be enough to find the optimal solution.

We observe in Figure~\ref{fig:discrete_various_thresholds}(a), (b) and (c) that the QEL  algorithm can find the optimal solutions with higher probability in the case of Max-Cut compared to the datasets including higher order correlations. Although the algorithm is still successful in suggesting the optimal solutions with high probability, the addition of the first-order correlations slightly decreases the performance because the problem is now more complex. For the case of the molecule optimization presented in Figure~\ref{fig:discrete_various_thresholds}(d), the solution landscape is completely different to the other use cases, however, the QEL algorithm is still able to successfully find the optimal solution with a high probability.

\subsubsection*{Results analysis: fixed threshold}

\begin{figure}
     
     \subfloat{
         
         \includegraphics[width=\linewidth]{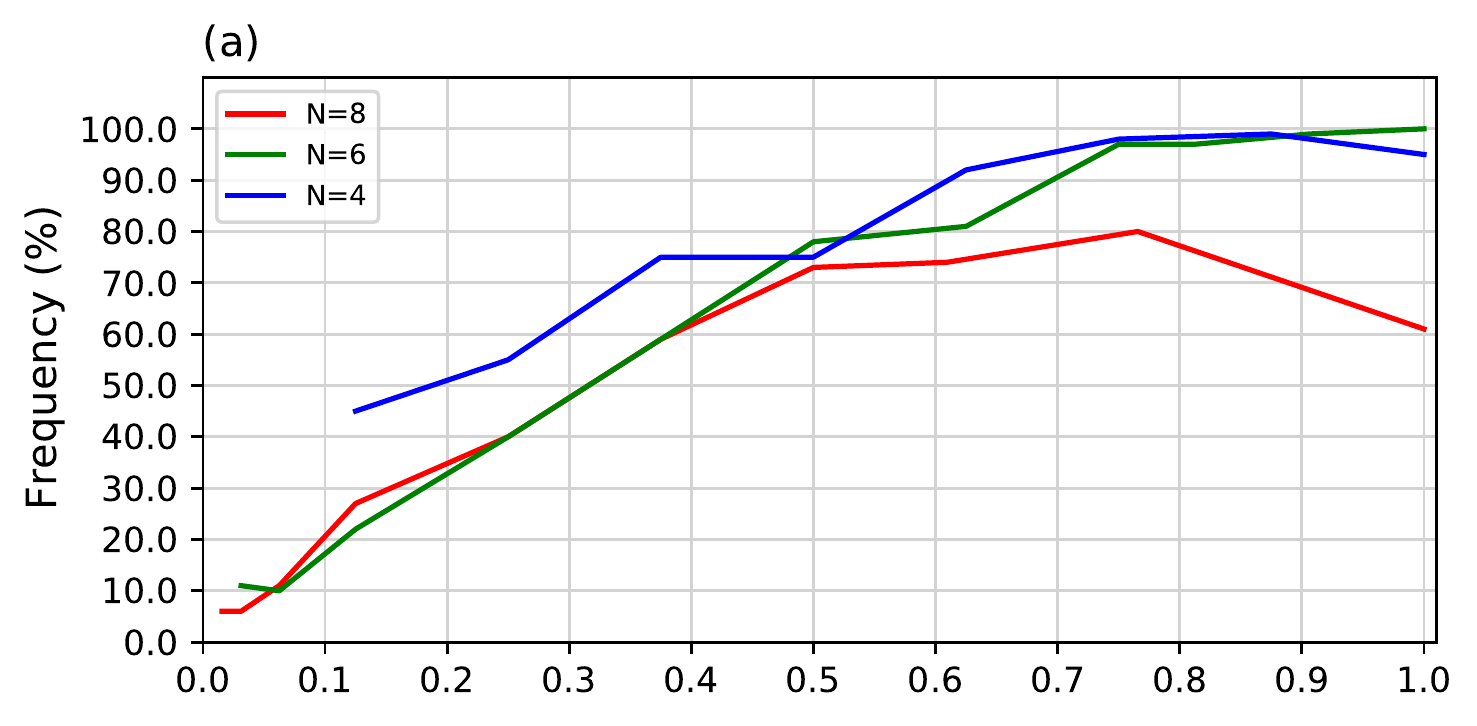}
     }
     \hfill
     \subfloat{
         
         \includegraphics[width=\linewidth]{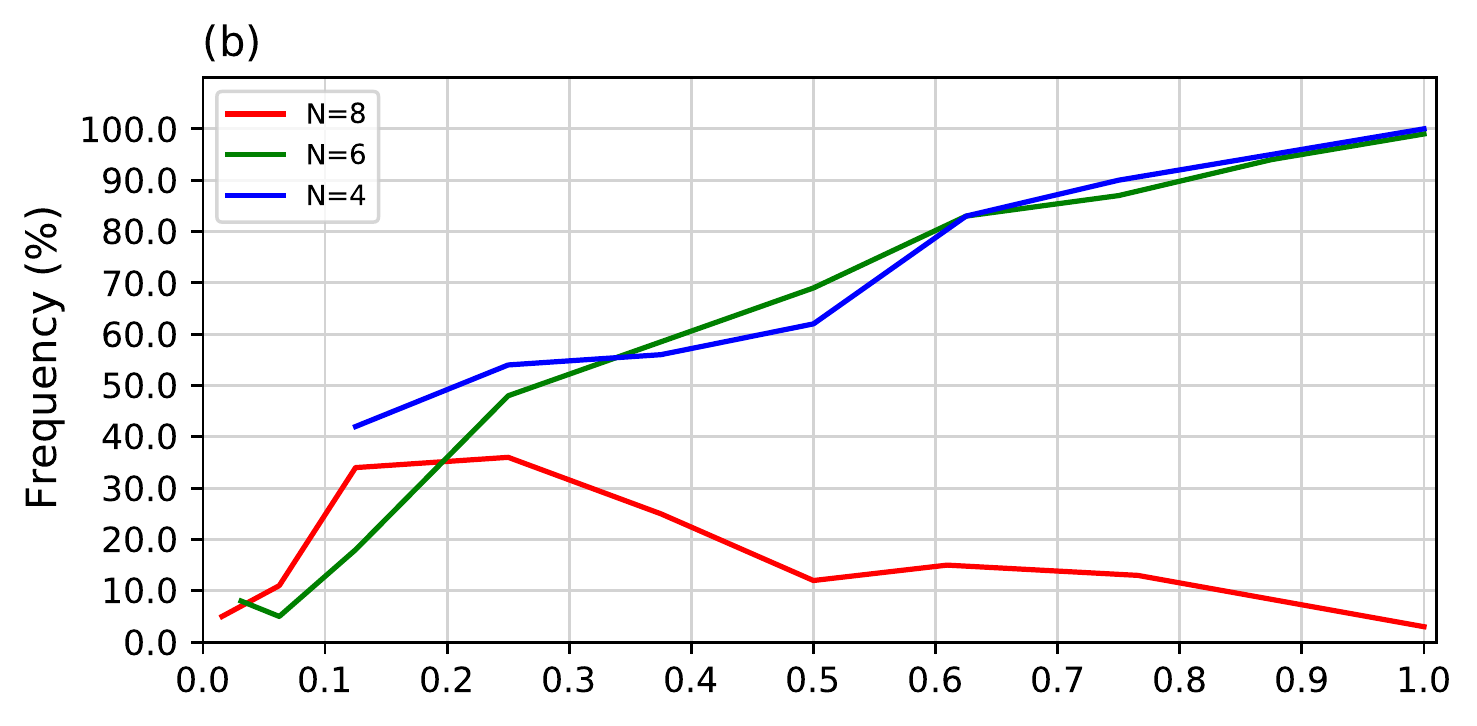}
     }
     \hfill
     \subfloat{
         
         \includegraphics[width=\linewidth]{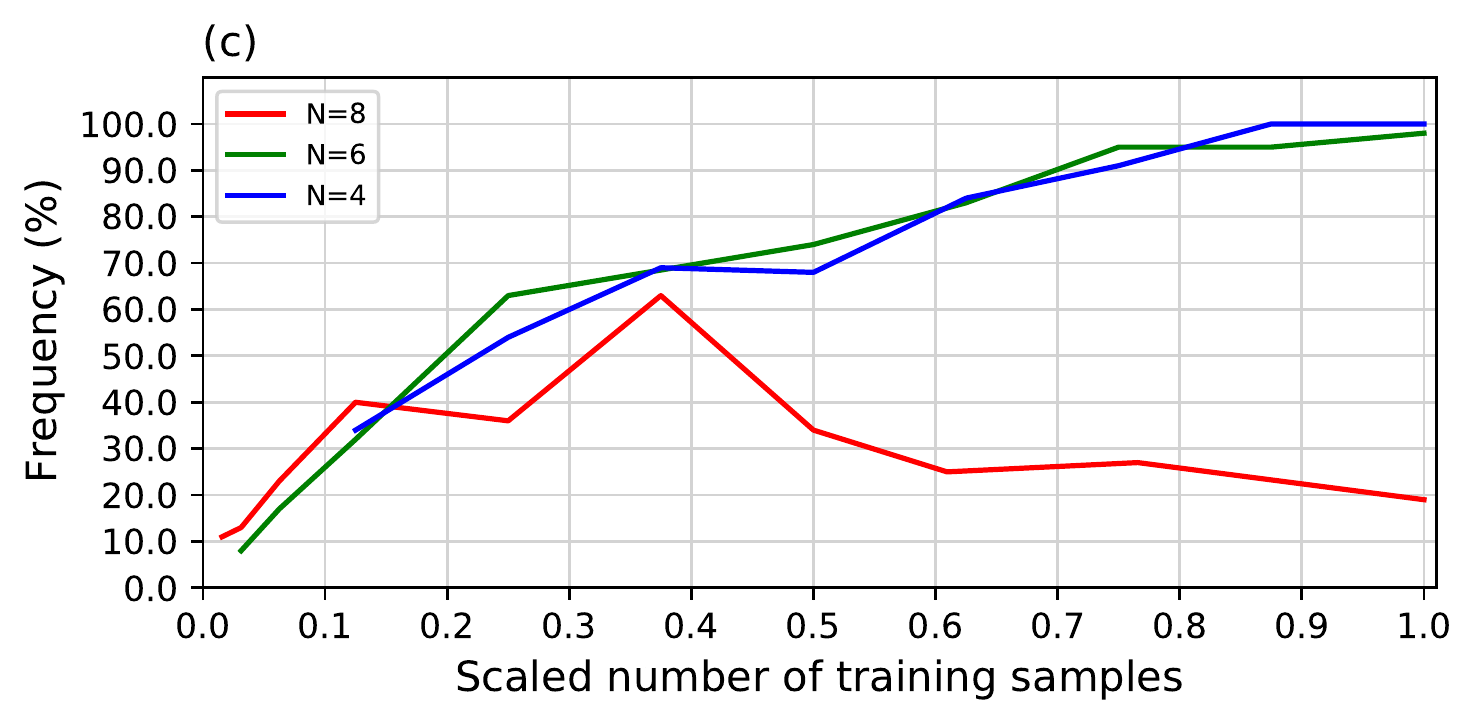}
     }
     \caption{plot showing the investigation for different system sizes (N=4,6,8) for various training datasets sizes. The threshold is set as such so that the total probability of all optimal solutions is larger than 20\%. (a) Max-Cut clustering in two clusters, (b) a system with nearest neighbor interactions up to 2nd order correlations, (c) a system with nearest neighbor interactions up to 3rd order correlations.
        \label{fig:discrete_fixed_thresholds}}
\end{figure}

In Figure~\ref{fig:discrete_fixed_thresholds}, we present the performance of our algorithm once again for (a) the Max-Cut problem with two clusters, (b) a system with nearest neighbor interactions up to 2nd order, and (c) a system with nearest neighbor interactions up to 3rd order; however, here we compare the performance of different problem sizes (N=4,6,8) for a fixed threshold, namely $20\%$, rather than a variety of thresholds. To make a fair comparison, we have scaled the x-axis accordingly, so that we can compare problems of different sizes based on the percentage of the training dataset over the $2^N$ possible samples. We see that even for small training datasets, the QEL algorithm is able to suggest optimal solutions with a probability higher than 20\% with moderate frequency. On the other hand, when the size of the training dataset approaches $2^N$, we observe that the frequency of suggesting optimal solutions with a probability greater than 20\% increases significantly for the smaller system sizes. The explanation of this phenomenon is that when almost all samples are used during training, the algorithm has a better understanding of how the cost landscape is, therefore the probability of suggesting the optimal solutions is increased.

Comparing Figure~\ref{fig:discrete_fixed_thresholds}(a) with Figures ~\ref{fig:discrete_fixed_thresholds}(b) and ~\ref{fig:discrete_fixed_thresholds}(c), we observe that as the size of the problem increases, the algorithm will suggest the optimal solutions with a lower frequency in the nearest neighbor with high-order correlations case. We can attribute that to the fact that the Max-Cut problem is easier to solve than the one that includes all nearest neighbor correlations up to a fixed order.

\subsubsection*{Results analysis: model output scaling}

In an effort to help our algorithm perform more efficient training, we employ the well-known approach in classical machine learning of scaling the output value to be between 0 and 1. Similarly to the classical NNs, QNNs are able to fit the data faster and better when the target values are within a known range. Furthermore, always scaling the output within a known range keeps the range of the target values fixed regardless of the size of the problem, which makes comparison easier. The goal of the training process is to learn a mapping of inputs to outputs based on examples included in a training dataset. The initial parameters of the model, usually take small random values and are updated via the optimization algorithm according to the error estimate on the training dataset. Therefore, unscaled input variables can lead to a slow and unstable learning process, while unscaled target (output) variables can lead to exploding gradients resulting in bad learning process.

In the event that the training samples are incomplete ($\leq 2^N$), the maximum feature value in training will not necessarily be equal to the true maximum feature value. Therefore, the value of the true maximum feature corresponding to the optimal solution will be out of bounds, since it should be higher than the assumed maximum value. To compensate for this issue, we define a variable $\alpha \in [1.0, 100.0]$ and a constant $\beta = 0.5$, as a multiplicative and additive factor to the model output. The new expectation value will be 

\begin{equation}
    \text{expectation value} = \frac{\alpha * \text{model output}}{2*N} + \beta
\end{equation}

The observable that we used in all use cases is the total magnetization, which takes values between $-N$ and $+N$. Therefore, assuming $\alpha=1.0$ and $\beta=0.5$ will create a range between 0 and 1. Using $\alpha>1.0$, the maximum feature value predicted by the model can be increased, so that optimization can suggest candidate solutions with high probability for inputs that might have not been included in the training dataset. As $\alpha$ increases, we find that the variational circuit has more flexibility to improve good candidate solutions; however, a trade-off must be found between a large $\alpha$ value and the probability of suggesting the optimal solution.

\begin{figure}[!htb]
    
    \includegraphics[width=1.0\linewidth]{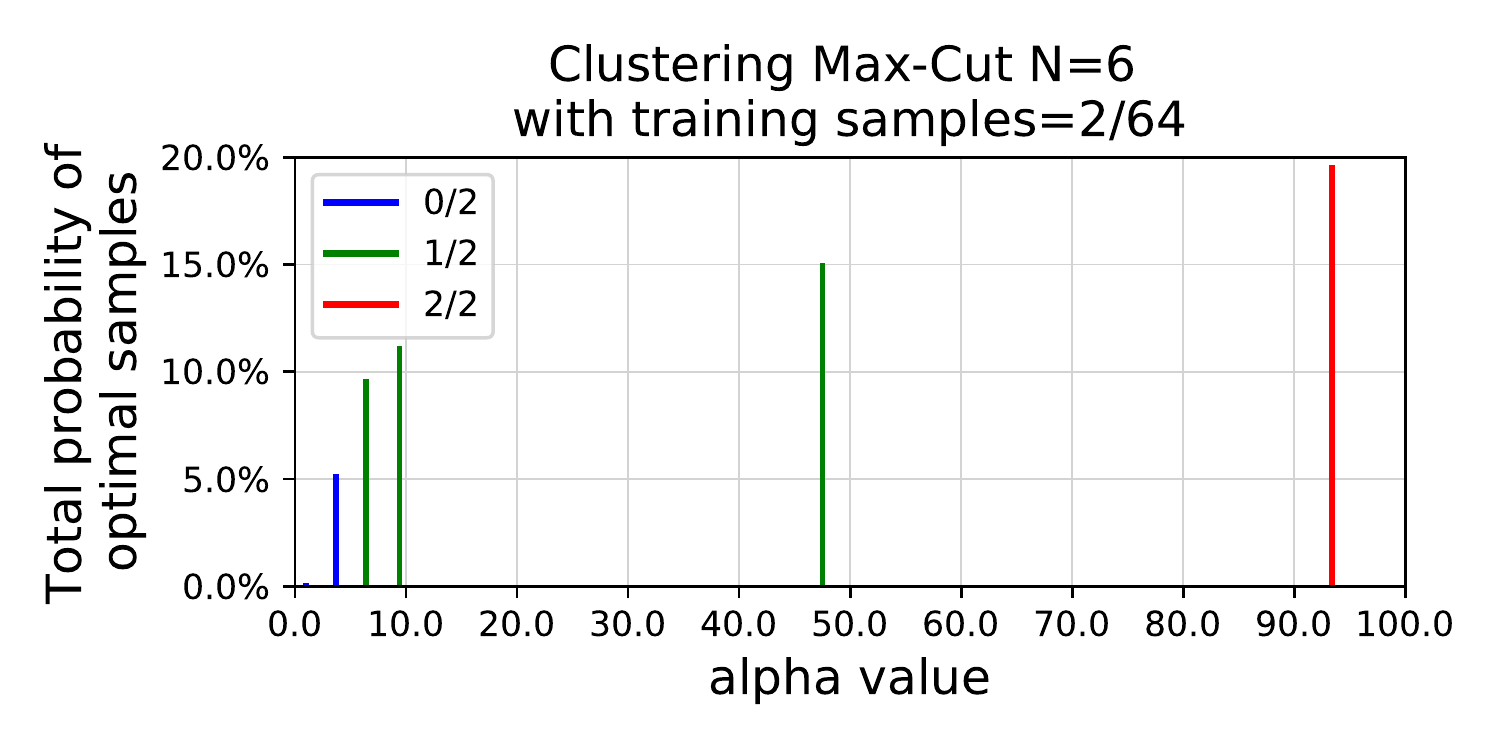}
    \caption{plot showcasing how the scaling of the model output through the alpha variable affects the performance of the algorithm for Max-Cut clustering with 6 qubits and a small training dataset size.}
    \label{fig:alpha_N=6}
\end{figure}

In Figure~\ref{fig:alpha_N=6}, we present how the total probability of the optimal bitstrings increases, while $\alpha$ increases incrementally by $0.1$. In order to keep the graph as clear as possible, we only showcase the $\alpha$ values that provided a significant increase of the total probability of the optimal solutions, as $\alpha$ was increased.

The investigation was performed for the Max-Cut clustering formulation with 6 qubits. By construction, we know that there are two optimal solutions regardless of the problem size in such a Max-Cut formulated problem. Therefore, for each $\alpha$ value, we keep track of whether the two suggested solutions by the algorithm with the highest probabilities, correspond to the optimal solutions. This is presented via the coloring of the bar and is showcased in the legend of the graph.

We observe in Figure~\ref{fig:alpha_N=6} that as $\alpha$ is increased, the total probability of the optimal solutions is increased, and the algorithm is eventually able to find both optimal solutions. In the case of Max-Cut, the two optimal solutions are complementary and have the same target value, but our algorithm is not programmed to search for symmetrical solutions; therefore, finding only one of the two optimal solutions is deemed sufficient. It is important to note that the number of samples used is deliberately small to test the performance of the algorithm in the extreme case of very few training samples available. For this example, we have only used 2 out the possible 64 inputs during training.

Clearly, there is a lot of room for improving understanding of scaling and shifting of training data versus the model output, in order to maximize expressivity, generalizability, and trainability altogether.

\section{Discussion}
\label{section:discussion}

We presented a hybrid quantum/classical algorithm that performs extremal learning. The algorithm uses a quantum feature map that encodes the classical data into quantum states, and then creates a model of the input/output relationship of data with a variational quantum circuit. After the model has been trained, then we analytically differentiate the quantum feature map that encodes the input data in order to find the coordinate that extremizes the trained model. The differentiation can be done directly in the case of continuous input data or indirectly with the assistance of another trainable quantum feature map in the case of discrete input data.

The results of our experimentation indicate that the QEL algorithm is able to successfully model and then optimize the modeled function to find the input value that extremizes the output function. We have proven that for the simple problems that were investigated for both continuous and discrete input data, the algorithmic performance is good, at least assuming a noiseless quantum simulation. The algorithm is formulated in a generic manner, so that it can cover a wide variety of optimization problems and applications based on its formulation and structure. However, the research is still in its early stage and there is much to explore.

In the next steps, we are planning to explore several aspects of the algorithm. One critical aspect is the training process of the feature maps and the variational circuits. We will investigate how the training procedure is evolving through metrics like the loss function (investigating multiple loss functions as well) and how stopping conditions are affecting the algorithmic performance (whether the algorithm is over- or under-fitting). We will also continue the investigation of how scaling the model output can contribute to increasing the performance of our algorithm. Furthermore, assessing how good our algorithm is in avoiding barren plateaus is another challenging issue, since this can be a limitation of other hybrid algorithms that utilize variational circuits. Furthermore, the choice of the feature maps used is also crucial, since a large variety of feature maps have already been suggested in literature that provide good results. Another critical aspect that requires further investigation is the ansatz optimization in terms of the ansatz type, the depth of the ansatz used, etc. We have used a HEA in this research because it is considered a generic and efficient one, however based on the application there might be better options. An important choice is also the observable that is selected. Our choice of the total magnetization was based on the fact that it can be easily obtained from measurements of all qubits in the quantum circuit, and that all qubits contribute to its value; however, other options may outperform it in practice. Finally, time complexity and scalability of accurate performance should also be investigated, since we have only tested small scale problems. 

Currently, different values of discrete variables are mapped by our algorithm to perpendicular subspaces of a Hilbert space. There is no difference in distance between them. The current mapping provides an easy path to access all points at the cost of higher dimensionality. More efficient mapping strategies that consider the topology of the discrete space~\cite{1999Bengio} would be advantageous, but is beyond the scope of this expository work.

We have only demonstrated function approximation using the measurement of the expectation value of a linear operator. More complex methods involving for example multiple nonlinear operator measurements and kernels~\cite{Schuld2019b, Henry2021} could be used to approximate the function, and then the generalization of the modeling step is straightforward. However, the construction of an Extremizer Feature Map that permits efficient exploration of the space of input variables is nontrivial and remains for future examination.

The variational nature of the QEL algorithm is beneficial for the current NISQ era, since variational circuits have been proven to be more resilient against errors. We have not yet tested our algorithm under any noise settings, including error mitigation techniques etc, which would be essential to gain advantages in the near-term and thus should be investigated numerically in simulations as well as on quantum hardware.

We anticipate that our algorithm can be further enhanced in an FTQC setting as well, assuming the use of error correction techniques which allow for deep circuits. In particular, in that case we may make use of Linear Combination of Unitaries (LCU) techniques for parallelizing training. The LCU framework is known to provide a potential exponential speed-up for quantum simulation and matrix inversion, using ancillary registers of qubits~\cite{Long2008, Childs2012, Berry2015, Childs2018}.

\par
\paragraph*{Ethics declaration} A patent application for the method described in this manuscript has been submitted by PASQAL SAS with SV, EP and VEE as inventors~\cite{patent}.

\bibliography{references} 

\begin{thebibliography}{91}%
\makeatletter
\providecommand \@ifxundefined [1]{%
 \@ifx{#1\undefined}
}%
\providecommand \@ifnum [1]{%
 \ifnum #1\expandafter \@firstoftwo
 \else \expandafter \@secondoftwo
 \fi
}%
\providecommand \@ifx [1]{%
 \ifx #1\expandafter \@firstoftwo
 \else \expandafter \@secondoftwo
 \fi
}%
\providecommand \natexlab [1]{#1}%
\providecommand \enquote  [1]{``#1''}%
\providecommand \bibnamefont  [1]{#1}%
\providecommand \bibfnamefont [1]{#1}%
\providecommand \citenamefont [1]{#1}%
\providecommand \href@noop [0]{\@secondoftwo}%
\providecommand \href [0]{\begingroup \@sanitize@url \@href}%
\providecommand \@href[1]{\@@startlink{#1}\@@href}%
\providecommand \@@href[1]{\endgroup#1\@@endlink}%
\providecommand \@sanitize@url [0]{\catcode `\\12\catcode `\$12\catcode
  `\&12\catcode `\#12\catcode `\^12\catcode `\_12\catcode `\%12\relax}%
\providecommand \@@startlink[1]{}%
\providecommand \@@endlink[0]{}%
\providecommand \url  [0]{\begingroup\@sanitize@url \@url }%
\providecommand \@url [1]{\endgroup\@href {#1}{\urlprefix }}%
\providecommand \urlprefix  [0]{URL }%
\providecommand \Eprint [0]{\href }%
\providecommand \doibase [0]{https://doi.org/}%
\providecommand \selectlanguage [0]{\@gobble}%
\providecommand \bibinfo  [0]{\@secondoftwo}%
\providecommand \bibfield  [0]{\@secondoftwo}%
\providecommand \translation [1]{[#1]}%
\providecommand \BibitemOpen [0]{}%
\providecommand \bibitemStop [0]{}%
\providecommand \bibitemNoStop [0]{.\EOS\space}%
\providecommand \EOS [0]{\spacefactor3000\relax}%
\providecommand \BibitemShut  [1]{\csname bibitem#1\endcsname}%
\let\auto@bib@innerbib\@empty
\bibitem [{\citenamefont {Robert}\ \emph {et~al.}(2021)\citenamefont {Robert},
  \citenamefont {Barkoutsos}, \citenamefont {Woerner},\ and\ \citenamefont
  {Tavernelli}}]{Robert2021}%
  \BibitemOpen
  \bibfield  {author} {\bibinfo {author} {\bibfnamefont {A.}~\bibnamefont
  {Robert}}, \bibinfo {author} {\bibfnamefont {P.~K.}\ \bibnamefont
  {Barkoutsos}}, \bibinfo {author} {\bibfnamefont {S.}~\bibnamefont
  {Woerner}},\ and\ \bibinfo {author} {\bibfnamefont {I.}~\bibnamefont
  {Tavernelli}},\ }\href {https://doi.org/10.1038/s41534-021-00368-4}
  {\bibfield  {journal} {\bibinfo  {journal} {npj Quantum Information}\
  }\textbf {\bibinfo {volume} {7}} (\bibinfo {year} {2021})}\BibitemShut
  {NoStop}%
\bibitem [{\citenamefont {Fox}\ \emph {et~al.}(2021)\citenamefont {Fox},
  \citenamefont {Branson},\ and\ \citenamefont {Walker}}]{Fox2021}%
  \BibitemOpen
  \bibfield  {author} {\bibinfo {author} {\bibfnamefont {D.~M.}\ \bibnamefont
  {Fox}}, \bibinfo {author} {\bibfnamefont {K.~M.}\ \bibnamefont {Branson}},\
  and\ \bibinfo {author} {\bibfnamefont {R.~C.}\ \bibnamefont {Walker}},\
  }\href {https://doi.org/10.1371/journal.pone.0259101} {\bibfield  {journal}
  {\bibinfo  {journal} {PLOS ONE}\ }\textbf {\bibinfo {volume} {16}},\ \bibinfo
  {pages} {1} (\bibinfo {year} {2021})}\BibitemShut {NoStop}%
\bibitem [{\citenamefont {Mulligan}\ \emph {et~al.}(2020)\citenamefont
  {Mulligan}, \citenamefont {Melo}, \citenamefont {Merritt}, \citenamefont
  {Slocum}, \citenamefont {Weitzner}, \citenamefont {Watkins}, \citenamefont
  {Renfrew}, \citenamefont {Pelissier}, \citenamefont {Arora},\ and\
  \citenamefont {Bonneau}}]{Mulligan2020}%
  \BibitemOpen
  \bibfield  {author} {\bibinfo {author} {\bibfnamefont {V.~K.}\ \bibnamefont
  {Mulligan}}, \bibinfo {author} {\bibfnamefont {H.}~\bibnamefont {Melo}},
  \bibinfo {author} {\bibfnamefont {H.~I.}\ \bibnamefont {Merritt}}, \bibinfo
  {author} {\bibfnamefont {S.}~\bibnamefont {Slocum}}, \bibinfo {author}
  {\bibfnamefont {B.~D.}\ \bibnamefont {Weitzner}}, \bibinfo {author}
  {\bibfnamefont {A.~M.}\ \bibnamefont {Watkins}}, \bibinfo {author}
  {\bibfnamefont {P.~D.}\ \bibnamefont {Renfrew}}, \bibinfo {author}
  {\bibfnamefont {C.}~\bibnamefont {Pelissier}}, \bibinfo {author}
  {\bibfnamefont {P.~S.}\ \bibnamefont {Arora}},\ and\ \bibinfo {author}
  {\bibfnamefont {R.}~\bibnamefont {Bonneau}},\ }\bibfield  {journal} {\bibinfo
   {journal} {bioRxiv}\ }\href {https://doi.org/10.1101/752485}
  {10.1101/752485} (\bibinfo {year} {2020})\BibitemShut {NoStop}%
\bibitem [{\citenamefont {Kitai}\ \emph {et~al.}(2020)\citenamefont {Kitai},
  \citenamefont {Guo}, \citenamefont {Ju}, \citenamefont {Tanaka},
  \citenamefont {Tsuda}, \citenamefont {Shiomi},\ and\ \citenamefont
  {Tamura}}]{Kitai2020}%
  \BibitemOpen
  \bibfield  {author} {\bibinfo {author} {\bibfnamefont {K.}~\bibnamefont
  {Kitai}}, \bibinfo {author} {\bibfnamefont {J.}~\bibnamefont {Guo}}, \bibinfo
  {author} {\bibfnamefont {S.}~\bibnamefont {Ju}}, \bibinfo {author}
  {\bibfnamefont {S.}~\bibnamefont {Tanaka}}, \bibinfo {author} {\bibfnamefont
  {K.}~\bibnamefont {Tsuda}}, \bibinfo {author} {\bibfnamefont
  {J.}~\bibnamefont {Shiomi}},\ and\ \bibinfo {author} {\bibfnamefont
  {R.}~\bibnamefont {Tamura}},\ }\href
  {https://doi.org/10.1103/PhysRevResearch.2.013319} {\bibfield  {journal}
  {\bibinfo  {journal} {Phys. Rev. Research}\ }\textbf {\bibinfo {volume}
  {2}},\ \bibinfo {pages} {013319} (\bibinfo {year} {2020})}\BibitemShut
  {NoStop}%
\bibitem [{\citenamefont {Farhi}\ and\ \citenamefont
  {Gutmann}(2014)}]{Farhi2014}%
  \BibitemOpen
  \bibfield  {author} {\bibinfo {author} {\bibfnamefont {E.}~\bibnamefont
  {Farhi}}\ and\ \bibinfo {author} {\bibfnamefont {J.~G.~S.}\ \bibnamefont
  {Gutmann}},\ }\href@noop {} {\bibinfo {title} {A quantum approximate
  optimization algorithm}} (\bibinfo {year} {2014}),\ \Eprint
  {https://arxiv.org/abs/1411.4028} {arXiv:1411.4028 [quant-ph]} \BibitemShut
  {NoStop}%
\bibitem [{\citenamefont {Patel}\ and\ \citenamefont
  {Rummel}(2021)}]{Patel2021}%
  \BibitemOpen
  \bibfield  {author} {\bibinfo {author} {\bibfnamefont {Z.}~\bibnamefont
  {Patel}}\ and\ \bibinfo {author} {\bibfnamefont {M.}~\bibnamefont {Rummel}},\
  }\href@noop {} {\bibinfo {title} {Extremal learning: extremizing the output
  of a neural network in regression problems}} (\bibinfo {year} {2021}),\
  \Eprint {https://arxiv.org/abs/2102.03626} {arXiv:2102.03626 [cs.LG]}
  \BibitemShut {NoStop}%
\bibitem [{\citenamefont {Hiriart-Urruty}\ and\ \citenamefont
  {Lemar\'echal}(2001)}]{Hiriart-Urruty2001}%
  \BibitemOpen
  \bibfield  {author} {\bibinfo {author} {\bibfnamefont {J.-B.}\ \bibnamefont
  {Hiriart-Urruty}}\ and\ \bibinfo {author} {\bibfnamefont {C.}~\bibnamefont
  {Lemar\'echal}},\ }\href
  {https://doi.org/https://doi.org/10.1007/978-3-642-56468-0} {\emph {\bibinfo
  {title} {Fundamentals of Convex Analysis}}}\ (\bibinfo  {publisher} {Springer
  Berlin, Heidelberg},\ \bibinfo {year} {2001})\BibitemShut {NoStop}%
\bibitem [{\citenamefont {Niculescu}\ and\ \citenamefont
  {Persson}(2006)}]{Niculescu2016}%
  \BibitemOpen
  \bibfield  {author} {\bibinfo {author} {\bibfnamefont {C.~P.}\ \bibnamefont
  {Niculescu}}\ and\ \bibinfo {author} {\bibfnamefont {L.-E.}\ \bibnamefont
  {Persson}},\ }\href {https://doi.org/https://doi.org/10.1007/0-387-31077-0}
  {\emph {\bibinfo {title} {Convex Functions and their Applications:A
  Contemporary Approach}}}\ (\bibinfo  {publisher} {Springer New York, NY},\
  \bibinfo {year} {2006})\BibitemShut {NoStop}%
\bibitem [{\citenamefont {Mitarai}\ \emph {et~al.}(2018)\citenamefont
  {Mitarai}, \citenamefont {Negoro}, \citenamefont {Kitagawa},\ and\
  \citenamefont {Fujii}}]{Mitarai2018}%
  \BibitemOpen
  \bibfield  {author} {\bibinfo {author} {\bibfnamefont {K.}~\bibnamefont
  {Mitarai}}, \bibinfo {author} {\bibfnamefont {M.}~\bibnamefont {Negoro}},
  \bibinfo {author} {\bibfnamefont {M.}~\bibnamefont {Kitagawa}},\ and\
  \bibinfo {author} {\bibfnamefont {K.}~\bibnamefont {Fujii}},\ }\href
  {https://doi.org/10.1103/PhysRevA.98.032309} {\bibfield  {journal} {\bibinfo
  {journal} {Phys. Rev. A}\ }\textbf {\bibinfo {volume} {98}},\ \bibinfo
  {pages} {032309} (\bibinfo {year} {2018})}\BibitemShut {NoStop}%
\bibitem [{\citenamefont {Kruskal}(1956)}]{Kruskal1956}%
  \BibitemOpen
  \bibfield  {author} {\bibinfo {author} {\bibfnamefont {J.~B.}\ \bibnamefont
  {Kruskal}},\ }\href {https://doi.org/10.2307/2033241} {\bibfield  {journal}
  {\bibinfo  {journal} {American Mathematical Society}\ }\textbf {\bibinfo
  {volume} {7}},\ \bibinfo {pages} {48} (\bibinfo {year} {1956})}\BibitemShut
  {NoStop}%
\bibitem [{\citenamefont {Prim}(1957)}]{Prim1957}%
  \BibitemOpen
  \bibfield  {author} {\bibinfo {author} {\bibfnamefont {R.~C.}\ \bibnamefont
  {Prim}},\ }\href
  {https://doi.org/https://doi.org/10.1002/j.1538-7305.1957.tb01515.x}
  {\bibfield  {journal} {\bibinfo  {journal} {Bell System Technical Journal}\
  }\textbf {\bibinfo {volume} {36}},\ \bibinfo {pages} {1389} (\bibinfo {year}
  {1957})}\BibitemShut {NoStop}%
\bibitem [{\citenamefont {Pham}\ and\ \citenamefont
  {Karaboga}(2011)}]{Pham2011}%
  \BibitemOpen
  \bibfield  {author} {\bibinfo {author} {\bibfnamefont {D.~T.}\ \bibnamefont
  {Pham}}\ and\ \bibinfo {author} {\bibfnamefont {D.}~\bibnamefont
  {Karaboga}},\ }\href@noop {} {\emph {\bibinfo {title} {Intelligent
  Optimisation Techniques: Genetic Algorithms, Tabu Search, Simulated Annealing
  and Neural Networks}}},\ \bibinfo {edition} {1st}\ ed.\ (\bibinfo
  {publisher} {Springer Publishing Company, Incorporated},\ \bibinfo {year}
  {2011})\BibitemShut {NoStop}%
\bibitem [{\citenamefont {Sanders}\ \emph {et~al.}(2020)\citenamefont
  {Sanders}, \citenamefont {Berry}, \citenamefont {Costa}, \citenamefont
  {Tessler}, \citenamefont {Wiebe}, \citenamefont {Gidney}, \citenamefont
  {Neven},\ and\ \citenamefont {Babbush}}]{Sanders2020}%
  \BibitemOpen
  \bibfield  {author} {\bibinfo {author} {\bibfnamefont {Y.~R.}\ \bibnamefont
  {Sanders}}, \bibinfo {author} {\bibfnamefont {D.~W.}\ \bibnamefont {Berry}},
  \bibinfo {author} {\bibfnamefont {P.~C.}\ \bibnamefont {Costa}}, \bibinfo
  {author} {\bibfnamefont {L.~W.}\ \bibnamefont {Tessler}}, \bibinfo {author}
  {\bibfnamefont {N.}~\bibnamefont {Wiebe}}, \bibinfo {author} {\bibfnamefont
  {C.}~\bibnamefont {Gidney}}, \bibinfo {author} {\bibfnamefont
  {H.}~\bibnamefont {Neven}},\ and\ \bibinfo {author} {\bibfnamefont
  {R.}~\bibnamefont {Babbush}},\ }\href
  {https://doi.org/10.1103/PRXQuantum.1.020312} {\bibfield  {journal} {\bibinfo
   {journal} {PRX Quantum}\ }\textbf {\bibinfo {volume} {1}},\ \bibinfo {pages}
  {020312} (\bibinfo {year} {2020})}\BibitemShut {NoStop}%
\bibitem [{\citenamefont {Preskill}(2018)}]{Preskill2018}%
  \BibitemOpen
  \bibfield  {author} {\bibinfo {author} {\bibfnamefont {J.}~\bibnamefont
  {Preskill}},\ }\href {https://doi.org/10.22331/q-2018-08-06-79} {\bibfield
  {journal} {\bibinfo  {journal} {{Quantum}}\ }\textbf {\bibinfo {volume}
  {2}},\ \bibinfo {pages} {79} (\bibinfo {year} {2018})}\BibitemShut {NoStop}%
\bibitem [{\citenamefont {Dalzell}\ \emph {et~al.}(2020)\citenamefont
  {Dalzell}, \citenamefont {Harrow}, \citenamefont {Koh},\ and\ \citenamefont
  {La~Placa}}]{Dalzell2020}%
  \BibitemOpen
  \bibfield  {author} {\bibinfo {author} {\bibfnamefont {A.~M.}\ \bibnamefont
  {Dalzell}}, \bibinfo {author} {\bibfnamefont {A.~W.}\ \bibnamefont {Harrow}},
  \bibinfo {author} {\bibfnamefont {D.~E.}\ \bibnamefont {Koh}},\ and\ \bibinfo
  {author} {\bibfnamefont {R.~L.}\ \bibnamefont {La~Placa}},\ }\href
  {https://doi.org/10.22331/q-2020-05-11-264} {\bibfield  {journal} {\bibinfo
  {journal} {{Quantum}}\ }\textbf {\bibinfo {volume} {4}},\ \bibinfo {pages}
  {264} (\bibinfo {year} {2020})}\BibitemShut {NoStop}%
\bibitem [{\citenamefont {Peruzzo}\ \emph {et~al.}(2014)\citenamefont
  {Peruzzo}, \citenamefont {McClean}, \citenamefont {Shadbolt}, \citenamefont
  {Yung}, \citenamefont {Zhou}, \citenamefont {Love}, \citenamefont
  {Aspuru-Guzik},\ and\ \citenamefont {O’Brien}}]{Peruzzo2014}%
  \BibitemOpen
  \bibfield  {author} {\bibinfo {author} {\bibfnamefont {A.}~\bibnamefont
  {Peruzzo}}, \bibinfo {author} {\bibfnamefont {J.}~\bibnamefont {McClean}},
  \bibinfo {author} {\bibfnamefont {P.}~\bibnamefont {Shadbolt}}, \bibinfo
  {author} {\bibfnamefont {M.-H.}\ \bibnamefont {Yung}}, \bibinfo {author}
  {\bibfnamefont {X.-Q.}\ \bibnamefont {Zhou}}, \bibinfo {author}
  {\bibfnamefont {P.~J.}\ \bibnamefont {Love}}, \bibinfo {author}
  {\bibfnamefont {A.}~\bibnamefont {Aspuru-Guzik}},\ and\ \bibinfo {author}
  {\bibfnamefont {J.~L.}\ \bibnamefont {O’Brien}},\ }\bibfield  {journal}
  {\bibinfo  {journal} {Nature Communications}\ }\textbf {\bibinfo {volume}
  {5}},\ \href {https://doi.org/10.1038/ncomms5213} {10.1038/ncomms5213}
  (\bibinfo {year} {2014})\BibitemShut {NoStop}%
\bibitem [{\citenamefont {King}\ \emph {et~al.}(2022)\citenamefont {King},
  \citenamefont {Suzuki}, \citenamefont {Raymond}, \citenamefont {Zucca},
  \citenamefont {Lanting}, \citenamefont {Altomare}, \citenamefont {Berkley},
  \citenamefont {Ejtemaee}, \citenamefont {Hoskinson}, \citenamefont {Huang},
  \citenamefont {Ladizinsky}, \citenamefont {MacDonald}, \citenamefont
  {Marsden}, \citenamefont {Oh}, \citenamefont {Poulin-Lamarre}, \citenamefont
  {Reis}, \citenamefont {Rich}, \citenamefont {Sato}, \citenamefont
  {Whittaker}, \citenamefont {Yao}, \citenamefont {Harris}, \citenamefont
  {Lidar}, \citenamefont {Nishimori},\ and\ \citenamefont {Amin}}]{King2022}%
  \BibitemOpen
  \bibfield  {author} {\bibinfo {author} {\bibfnamefont {A.~D.}\ \bibnamefont
  {King}}, \bibinfo {author} {\bibfnamefont {S.}~\bibnamefont {Suzuki}},
  \bibinfo {author} {\bibfnamefont {J.}~\bibnamefont {Raymond}}, \bibinfo
  {author} {\bibfnamefont {A.}~\bibnamefont {Zucca}}, \bibinfo {author}
  {\bibfnamefont {T.}~\bibnamefont {Lanting}}, \bibinfo {author} {\bibfnamefont
  {F.}~\bibnamefont {Altomare}}, \bibinfo {author} {\bibfnamefont {A.~J.}\
  \bibnamefont {Berkley}}, \bibinfo {author} {\bibfnamefont {S.}~\bibnamefont
  {Ejtemaee}}, \bibinfo {author} {\bibfnamefont {E.}~\bibnamefont {Hoskinson}},
  \bibinfo {author} {\bibfnamefont {S.}~\bibnamefont {Huang}}, \bibinfo
  {author} {\bibfnamefont {E.}~\bibnamefont {Ladizinsky}}, \bibinfo {author}
  {\bibfnamefont {A.}~\bibnamefont {MacDonald}}, \bibinfo {author}
  {\bibfnamefont {G.}~\bibnamefont {Marsden}}, \bibinfo {author} {\bibfnamefont
  {T.}~\bibnamefont {Oh}}, \bibinfo {author} {\bibfnamefont {G.}~\bibnamefont
  {Poulin-Lamarre}}, \bibinfo {author} {\bibfnamefont {M.}~\bibnamefont
  {Reis}}, \bibinfo {author} {\bibfnamefont {C.}~\bibnamefont {Rich}}, \bibinfo
  {author} {\bibfnamefont {Y.}~\bibnamefont {Sato}}, \bibinfo {author}
  {\bibfnamefont {J.~D.}\ \bibnamefont {Whittaker}}, \bibinfo {author}
  {\bibfnamefont {J.}~\bibnamefont {Yao}}, \bibinfo {author} {\bibfnamefont
  {R.}~\bibnamefont {Harris}}, \bibinfo {author} {\bibfnamefont {D.~A.}\
  \bibnamefont {Lidar}}, \bibinfo {author} {\bibfnamefont {H.}~\bibnamefont
  {Nishimori}},\ and\ \bibinfo {author} {\bibfnamefont {M.~H.}\ \bibnamefont
  {Amin}},\ }\href@noop {} {\bibinfo {title} {Coherent quantum annealing in a
  programmable 2000-qubit ising chain}} (\bibinfo {year} {2022}),\ \Eprint
  {https://arxiv.org/abs/2202.05847} {arXiv:2202.05847 [quant-ph]} \BibitemShut
  {NoStop}%
\bibitem [{\citenamefont {Boser}\ \emph {et~al.}(1992)\citenamefont {Boser},
  \citenamefont {Guyon},\ and\ \citenamefont {Vapnik}}]{Boser1992}%
  \BibitemOpen
  \bibfield  {author} {\bibinfo {author} {\bibfnamefont {B.~E.}\ \bibnamefont
  {Boser}}, \bibinfo {author} {\bibfnamefont {I.}~\bibnamefont {Guyon}},\ and\
  \bibinfo {author} {\bibfnamefont {V.~N.}\ \bibnamefont {Vapnik}},\ }in\
  \href@noop {} {\emph {\bibinfo {booktitle} {COLT '92}}}\ (\bibinfo {year}
  {1992})\BibitemShut {NoStop}%
\bibitem [{\citenamefont {Vapnik}(1998)}]{Vapnik1998}%
  \BibitemOpen
  \bibfield  {author} {\bibinfo {author} {\bibfnamefont {V.~N.}\ \bibnamefont
  {Vapnik}},\ }\href@noop {} {\emph {\bibinfo {title} {Statistical Learning
  Theory}}}\ (\bibinfo  {publisher} {Wiley},\ \bibinfo {year}
  {1998})\BibitemShut {NoStop}%
\bibitem [{\citenamefont {Nayak}\ \emph {et~al.}(2015)\citenamefont {Nayak},
  \citenamefont {Naik},\ and\ \citenamefont {Behera}}]{Nayak2015}%
  \BibitemOpen
  \bibfield  {author} {\bibinfo {author} {\bibfnamefont {J.}~\bibnamefont
  {Nayak}}, \bibinfo {author} {\bibfnamefont {B.}~\bibnamefont {Naik}},\ and\
  \bibinfo {author} {\bibfnamefont {H.~S.}\ \bibnamefont {Behera}},\ }\bibfield
   {journal} {\bibinfo  {journal} {International Journal of Database Theory and
  Application}\ }\textbf {\bibinfo {volume} {8}},\ \href
  {https://doi.org/10.14257/ijdta.2015.8.1.18} {10.14257/ijdta.2015.8.1.18}
  (\bibinfo {year} {2015})\BibitemShut {NoStop}%
\bibitem [{\citenamefont {Burbidge}\ \emph {et~al.}(2001)\citenamefont
  {Burbidge}, \citenamefont {Trotter}, \citenamefont {Buxton},\ and\
  \citenamefont {Holden}}]{Burbidge2001}%
  \BibitemOpen
  \bibfield  {author} {\bibinfo {author} {\bibfnamefont {R.}~\bibnamefont
  {Burbidge}}, \bibinfo {author} {\bibfnamefont {M.}~\bibnamefont {Trotter}},
  \bibinfo {author} {\bibfnamefont {B.}~\bibnamefont {Buxton}},\ and\ \bibinfo
  {author} {\bibfnamefont {S.}~\bibnamefont {Holden}},\ }\bibfield  {journal}
  {\bibinfo  {journal} {Computers \& chemistry}\ }\textbf {\bibinfo {volume}
  {26}},\ \href {https://doi.org/10.1016/s0097-8485(01)00094-8}
  {10.1016/s0097-8485(01)00094-8} (\bibinfo {year} {2001})\BibitemShut
  {NoStop}%
\bibitem [{\citenamefont {Cristianini}\ and\ \citenamefont
  {Shawe-Taylor}(2000)}]{cristianini_shawe-taylor_2000}%
  \BibitemOpen
  \bibfield  {author} {\bibinfo {author} {\bibfnamefont {N.}~\bibnamefont
  {Cristianini}}\ and\ \bibinfo {author} {\bibfnamefont {J.}~\bibnamefont
  {Shawe-Taylor}},\ }\href {https://doi.org/10.1017/CBO9780511801389} {\emph
  {\bibinfo {title} {An Introduction to Support Vector Machines and Other
  Kernel-based Learning Methods}}}\ (\bibinfo  {publisher} {Cambridge
  University Press},\ \bibinfo {year} {2000})\BibitemShut {NoStop}%
\bibitem [{\citenamefont {Cervantes}\ \emph {et~al.}(2020)\citenamefont
  {Cervantes}, \citenamefont {Garcia-Lamont}, \citenamefont
  {Rodríguez-Mazahua},\ and\ \citenamefont {Lopez}}]{Cervantes2020}%
  \BibitemOpen
  \bibfield  {author} {\bibinfo {author} {\bibfnamefont {J.}~\bibnamefont
  {Cervantes}}, \bibinfo {author} {\bibfnamefont {F.}~\bibnamefont
  {Garcia-Lamont}}, \bibinfo {author} {\bibfnamefont {L.}~\bibnamefont
  {Rodríguez-Mazahua}},\ and\ \bibinfo {author} {\bibfnamefont
  {A.}~\bibnamefont {Lopez}},\ }\href
  {https://doi.org/https://doi.org/10.1016/j.neucom.2019.10.118} {\bibfield
  {journal} {\bibinfo  {journal} {Neurocomputing}\ }\textbf {\bibinfo {volume}
  {408}},\ \bibinfo {pages} {189} (\bibinfo {year} {2020})}\BibitemShut
  {NoStop}%
\bibitem [{\citenamefont {Joachims}(2002)}]{Joachims2002}%
  \BibitemOpen
  \bibfield  {author} {\bibinfo {author} {\bibfnamefont {T.}~\bibnamefont
  {Joachims}},\ }\href@noop {} {\emph {\bibinfo {title} {Learning to Classify
  Text Using Support Vector Machines: Methods, Theory and Algorithms}}}\
  (\bibinfo  {publisher} {Kluwer Academic Publishers},\ \bibinfo {year}
  {2002})\BibitemShut {NoStop}%
\bibitem [{\citenamefont {Shen}(2005)}]{Zhan2005}%
  \BibitemOpen
  \bibfield  {author} {\bibinfo {author} {\bibfnamefont {Y.~Z.~D.}\
  \bibnamefont {Shen}},\ }\bibfield  {journal} {\bibinfo  {journal} {Pattern
  Recognition}\ }\textbf {\bibinfo {volume} {38}},\ \href
  {https://doi.org/10.1016/j.patcog.2004.06.001} {10.1016/j.patcog.2004.06.001}
  (\bibinfo {year} {2005})\BibitemShut {NoStop}%
\bibitem [{\citenamefont {Hornik}\ \emph {et~al.}(1989)\citenamefont {Hornik},
  \citenamefont {Stinchcombe},\ and\ \citenamefont {White}}]{Hornik1989}%
  \BibitemOpen
  \bibfield  {author} {\bibinfo {author} {\bibfnamefont {K.}~\bibnamefont
  {Hornik}}, \bibinfo {author} {\bibfnamefont {M.}~\bibnamefont
  {Stinchcombe}},\ and\ \bibinfo {author} {\bibfnamefont {H.}~\bibnamefont
  {White}},\ }\bibfield  {journal} {\bibinfo  {journal} {Neural Networks}\
  }\textbf {\bibinfo {volume} {2}},\ \href
  {https://doi.org/10.1016/0893-6080(89)90020-8} {10.1016/0893-6080(89)90020-8}
  (\bibinfo {year} {1989})\BibitemShut {NoStop}%
\bibitem [{\citenamefont {Goodfellow}\ \emph {et~al.}(2016)\citenamefont
  {Goodfellow}, \citenamefont {Bengio},\ and\ \citenamefont
  {Courville}}]{Goodfellow2016}%
  \BibitemOpen
  \bibfield  {author} {\bibinfo {author} {\bibfnamefont {I.}~\bibnamefont
  {Goodfellow}}, \bibinfo {author} {\bibfnamefont {Y.}~\bibnamefont {Bengio}},\
  and\ \bibinfo {author} {\bibfnamefont {A.}~\bibnamefont {Courville}},\
  }\href@noop {} {\emph {\bibinfo {title} {Deep Learning}}}\ (\bibinfo
  {publisher} {MIT Press},\ \bibinfo {year} {2016})\ \bibinfo {note}
  {\url{http://www.deeplearningbook.org}}\BibitemShut {NoStop}%
\bibitem [{\citenamefont {Owhadi}(2015)}]{Owhadi2015}%
  \BibitemOpen
  \bibfield  {author} {\bibinfo {author} {\bibfnamefont {H.}~\bibnamefont
  {Owhadi}},\ }\href {https://doi.org/10.1137/140974596} {\bibfield  {journal}
  {\bibinfo  {journal} {Multiscale Modeling \& Simulation}\ }\textbf {\bibinfo
  {volume} {13}},\ \bibinfo {pages} {812} (\bibinfo {year} {2015})},\ \Eprint
  {https://arxiv.org/abs/https://doi.org/10.1137/140974596}
  {https://doi.org/10.1137/140974596} \BibitemShut {NoStop}%
\bibitem [{\citenamefont {Raissi}\ \emph {et~al.}(2017)\citenamefont {Raissi},
  \citenamefont {Perdikaris},\ and\ \citenamefont {Karniadakis}}]{Raissi2017}%
  \BibitemOpen
  \bibfield  {author} {\bibinfo {author} {\bibfnamefont {M.}~\bibnamefont
  {Raissi}}, \bibinfo {author} {\bibfnamefont {P.}~\bibnamefont {Perdikaris}},\
  and\ \bibinfo {author} {\bibfnamefont {G.~E.}\ \bibnamefont {Karniadakis}},\
  }\href {https://doi.org/https://doi.org/10.1016/j.jcp.2017.01.060} {\bibfield
   {journal} {\bibinfo  {journal} {Journal of Computational Physics}\ }\textbf
  {\bibinfo {volume} {335}},\ \bibinfo {pages} {736} (\bibinfo {year}
  {2017})}\BibitemShut {NoStop}%
\bibitem [{\citenamefont {Raissi}\ \emph {et~al.}(2019)\citenamefont {Raissi},
  \citenamefont {Perdikaris},\ and\ \citenamefont {Karniadakis}}]{Raissi2019}%
  \BibitemOpen
  \bibfield  {author} {\bibinfo {author} {\bibfnamefont {M.}~\bibnamefont
  {Raissi}}, \bibinfo {author} {\bibfnamefont {P.}~\bibnamefont {Perdikaris}},\
  and\ \bibinfo {author} {\bibfnamefont {G.}~\bibnamefont {Karniadakis}},\
  }\href {https://doi.org/https://doi.org/10.1016/j.jcp.2018.10.045} {\bibfield
   {journal} {\bibinfo  {journal} {Journal of Computational Physics}\ }\textbf
  {\bibinfo {volume} {378}},\ \bibinfo {pages} {686} (\bibinfo {year}
  {2019})}\BibitemShut {NoStop}%
\bibitem [{\citenamefont {Kim}\ \emph {et~al.}(2021)\citenamefont {Kim},
  \citenamefont {Kim}, \citenamefont {Lee},\ and\ \citenamefont
  {Lee}}]{Kim2021}%
  \BibitemOpen
  \bibfield  {author} {\bibinfo {author} {\bibfnamefont {S.~W.}\ \bibnamefont
  {Kim}}, \bibinfo {author} {\bibfnamefont {I.}~\bibnamefont {Kim}}, \bibinfo
  {author} {\bibfnamefont {J.}~\bibnamefont {Lee}},\ and\ \bibinfo {author}
  {\bibfnamefont {S.}~\bibnamefont {Lee}},\ }\bibfield  {journal} {\bibinfo
  {journal} {Journal of Mechanical Science and Technology}\ }\textbf {\bibinfo
  {volume} {35}},\ \href {https://doi.org/10.1007/s12206-021-0342-5}
  {10.1007/s12206-021-0342-5} (\bibinfo {year} {2021})\BibitemShut {NoStop}%
\bibitem [{\citenamefont {Cuomo}\ \emph {et~al.}(2022)\citenamefont {Cuomo},
  \citenamefont {Cola}, \citenamefont {Giampaolo}, \citenamefont {Rozza},
  \citenamefont {Raissi},\ and\ \citenamefont {Piccialli}}]{Cuomo2022}%
  \BibitemOpen
  \bibfield  {author} {\bibinfo {author} {\bibfnamefont {S.}~\bibnamefont
  {Cuomo}}, \bibinfo {author} {\bibfnamefont {V.~S.~D.}\ \bibnamefont {Cola}},
  \bibinfo {author} {\bibfnamefont {F.}~\bibnamefont {Giampaolo}}, \bibinfo
  {author} {\bibfnamefont {G.}~\bibnamefont {Rozza}}, \bibinfo {author}
  {\bibfnamefont {M.}~\bibnamefont {Raissi}},\ and\ \bibinfo {author}
  {\bibfnamefont {F.}~\bibnamefont {Piccialli}},\ }\href@noop {} {\bibinfo
  {title} {Scientific machine learning through physics-informed neural
  networks: Where we are and what’s next}} (\bibinfo {year} {2022}),\ \Eprint
  {https://arxiv.org/abs/2201.05624v3} {arXiv:2201.05624v3 [cs.LG]}
  \BibitemShut {NoStop}%
\bibitem [{\citenamefont {Yang}\ and\ \citenamefont
  {Perdikaris}(2019)}]{Yang2019}%
  \BibitemOpen
  \bibfield  {author} {\bibinfo {author} {\bibfnamefont {Y.}~\bibnamefont
  {Yang}}\ and\ \bibinfo {author} {\bibfnamefont {P.}~\bibnamefont
  {Perdikaris}},\ }\href
  {https://doi.org/https://doi.org/10.1016/j.jcp.2019.05.027} {\bibfield
  {journal} {\bibinfo  {journal} {Journal of Computational Physics}\ }\textbf
  {\bibinfo {volume} {394}},\ \bibinfo {pages} {136} (\bibinfo {year}
  {2019})}\BibitemShut {NoStop}%
\bibitem [{\citenamefont {Meng}\ \emph {et~al.}(2020)\citenamefont {Meng},
  \citenamefont {Li}, \citenamefont {Zhang},\ and\ \citenamefont
  {Karniadakis}}]{Meng2020}%
  \BibitemOpen
  \bibfield  {author} {\bibinfo {author} {\bibfnamefont {X.}~\bibnamefont
  {Meng}}, \bibinfo {author} {\bibfnamefont {Z.}~\bibnamefont {Li}}, \bibinfo
  {author} {\bibfnamefont {D.}~\bibnamefont {Zhang}},\ and\ \bibinfo {author}
  {\bibfnamefont {G.~E.}\ \bibnamefont {Karniadakis}},\ }\href
  {https://doi.org/https://doi.org/10.1016/j.cma.2020.113250} {\bibfield
  {journal} {\bibinfo  {journal} {Computer Methods in Applied Mechanics and
  Engineering}\ }\textbf {\bibinfo {volume} {370}},\ \bibinfo {pages} {113250}
  (\bibinfo {year} {2020})}\BibitemShut {NoStop}%
\bibitem [{\citenamefont {Karniadakis}\ \emph {et~al.}(2021)\citenamefont
  {Karniadakis}, \citenamefont {Kevrekidis}, \citenamefont {Lu}, \citenamefont
  {Perdikaris}, \citenamefont {Wang},\ and\ \citenamefont
  {Yang}}]{Karniadakis2021}%
  \BibitemOpen
  \bibfield  {author} {\bibinfo {author} {\bibfnamefont {G.~E.}\ \bibnamefont
  {Karniadakis}}, \bibinfo {author} {\bibfnamefont {I.~G.}\ \bibnamefont
  {Kevrekidis}}, \bibinfo {author} {\bibfnamefont {L.}~\bibnamefont {Lu}},
  \bibinfo {author} {\bibfnamefont {P.}~\bibnamefont {Perdikaris}}, \bibinfo
  {author} {\bibfnamefont {S.}~\bibnamefont {Wang}},\ and\ \bibinfo {author}
  {\bibfnamefont {L.}~\bibnamefont {Yang}},\ }\bibfield  {journal} {\bibinfo
  {journal} {Nature Reviews Physics}\ }\textbf {\bibinfo {volume} {3}},\ \href
  {https://doi.org/10.1038/s42254-021-00314-5} {10.1038/s42254-021-00314-5}
  (\bibinfo {year} {2021})\BibitemShut {NoStop}%
\bibitem [{\citenamefont {Raissi}(2018)}]{Raissi2018}%
  \BibitemOpen
  \bibfield  {author} {\bibinfo {author} {\bibfnamefont {M.}~\bibnamefont
  {Raissi}},\ }\href {http://jmlr.org/papers/v19/18-046.html} {\bibfield
  {journal} {\bibinfo  {journal} {Journal of Machine Learning Research}\
  }\textbf {\bibinfo {volume} {19}},\ \bibinfo {pages} {1} (\bibinfo {year}
  {2018})}\BibitemShut {NoStop}%
\bibitem [{\citenamefont {Blechschmidt}\ and\ \citenamefont
  {Ernst}(2021)}]{Blechschmidt2021}%
  \BibitemOpen
  \bibfield  {author} {\bibinfo {author} {\bibfnamefont {J.}~\bibnamefont
  {Blechschmidt}}\ and\ \bibinfo {author} {\bibfnamefont {O.~G.}\ \bibnamefont
  {Ernst}},\ }\href {https://doi.org/https://doi.org/10.1002/gamm.202100006}
  {\bibfield  {journal} {\bibinfo  {journal} {GAMM-Mitteilungen}\ }\textbf
  {\bibinfo {volume} {44}},\ \bibinfo {pages} {e202100006} (\bibinfo {year}
  {2021})}\BibitemShut {NoStop}%
\bibitem [{\citenamefont {Kollmannsberger}\ \emph {et~al.}(2021)\citenamefont
  {Kollmannsberger}, \citenamefont {D'Angella}, \citenamefont {Jokeit},\ and\
  \citenamefont {Herrmann}}]{Kollmannsberger2021}%
  \BibitemOpen
  \bibfield  {author} {\bibinfo {author} {\bibfnamefont {S.}~\bibnamefont
  {Kollmannsberger}}, \bibinfo {author} {\bibfnamefont {D.}~\bibnamefont
  {D'Angella}}, \bibinfo {author} {\bibfnamefont {M.}~\bibnamefont {Jokeit}},\
  and\ \bibinfo {author} {\bibfnamefont {L.}~\bibnamefont {Herrmann}},\
  }\bibinfo {title} {Physics-informed neural networks},\ in\ \href
  {https://doi.org/10.1007/978-3-030-76587-3_5} {\emph {\bibinfo {booktitle}
  {Deep Learning in Computational Mechanics: An Introductory Course}}}\
  (\bibinfo  {publisher} {Springer International Publishing},\ \bibinfo
  {address} {Cham},\ \bibinfo {year} {2021})\ pp.\ \bibinfo {pages}
  {55--84}\BibitemShut {NoStop}%
\bibitem [{\citenamefont {Lu}\ \emph {et~al.}(2021)\citenamefont {Lu},
  \citenamefont {Meng}, \citenamefont {Mao},\ and\ \citenamefont
  {Karniadakis}}]{Lu2021}%
  \BibitemOpen
  \bibfield  {author} {\bibinfo {author} {\bibfnamefont {L.}~\bibnamefont
  {Lu}}, \bibinfo {author} {\bibfnamefont {X.}~\bibnamefont {Meng}}, \bibinfo
  {author} {\bibfnamefont {Z.}~\bibnamefont {Mao}},\ and\ \bibinfo {author}
  {\bibfnamefont {G.~E.}\ \bibnamefont {Karniadakis}},\ }\href
  {https://doi.org/10.1137/19M1274067} {\bibfield  {journal} {\bibinfo
  {journal} {SIAM Review}\ }\textbf {\bibinfo {volume} {63}},\ \bibinfo {pages}
  {208} (\bibinfo {year} {2021})},\ \Eprint
  {https://arxiv.org/abs/https://doi.org/10.1137/19M1274067}
  {https://doi.org/10.1137/19M1274067} \BibitemShut {NoStop}%
\bibitem [{\citenamefont {Havlíček}\ \emph {et~al.}(2019)\citenamefont
  {Havlíček}, \citenamefont {Córcoles}, \citenamefont {Temme}, \citenamefont
  {Harrow}, \citenamefont {Kandala}, \citenamefont {Chow},\ and\ \citenamefont
  {Gambetta}}]{Havlicek2019}%
  \BibitemOpen
  \bibfield  {author} {\bibinfo {author} {\bibfnamefont {V.}~\bibnamefont
  {Havlíček}}, \bibinfo {author} {\bibfnamefont {A.~D.}\ \bibnamefont
  {Córcoles}}, \bibinfo {author} {\bibfnamefont {K.}~\bibnamefont {Temme}},
  \bibinfo {author} {\bibfnamefont {A.~W.}\ \bibnamefont {Harrow}}, \bibinfo
  {author} {\bibfnamefont {A.}~\bibnamefont {Kandala}}, \bibinfo {author}
  {\bibfnamefont {J.~M.}\ \bibnamefont {Chow}},\ and\ \bibinfo {author}
  {\bibfnamefont {J.~M.}\ \bibnamefont {Gambetta}},\ }\href
  {https://doi.org/https://doi.org/10.1038/s41586-019-0980-2} {\bibfield
  {journal} {\bibinfo  {journal} {Nature}\ }\textbf {\bibinfo {volume} {567}},\
  \bibinfo {pages} {209} (\bibinfo {year} {2019})}\BibitemShut {NoStop}%
\bibitem [{\citenamefont {Paine}\ \emph {et~al.}(2022)\citenamefont {Paine},
  \citenamefont {Elfving},\ and\ \citenamefont {Kyriienko}}]{Paine2022}%
  \BibitemOpen
  \bibfield  {author} {\bibinfo {author} {\bibfnamefont {A.~E.}\ \bibnamefont
  {Paine}}, \bibinfo {author} {\bibfnamefont {V.~E.}\ \bibnamefont {Elfving}},\
  and\ \bibinfo {author} {\bibfnamefont {O.}~\bibnamefont {Kyriienko}},\ }\href
  {https://doi.org/10.48550/ARXIV.2203.08884} {\bibinfo {title} {Quantum kernel
  methods for solving differential equations}} (\bibinfo {year}
  {2022})\BibitemShut {NoStop}%
\bibitem [{\citenamefont {Kak}(1995)}]{Kak1995}%
  \BibitemOpen
  \bibfield  {author} {\bibinfo {author} {\bibfnamefont {S.}~\bibnamefont
  {Kak}},\ }\href
  {https://doi.org/https://doi.org/10.1016/0020-0255(94)00095-S} {\bibfield
  {journal} {\bibinfo  {journal} {Information Sciences}\ }\textbf {\bibinfo
  {volume} {83}},\ \bibinfo {pages} {143} (\bibinfo {year} {1995})}\BibitemShut
  {NoStop}%
\bibitem [{\citenamefont {Chrisley}(1995)}]{Chrisley1995}%
  \BibitemOpen
  \bibfield  {author} {\bibinfo {author} {\bibfnamefont {R.}~\bibnamefont
  {Chrisley}},\ }\href@noop {} {\bibfield  {journal} {\bibinfo  {journal} {New
  directions in cognitive science: Proceedings of the international symposium,
  Saariselka}\ }\textbf {\bibinfo {volume} {4}},\ \bibinfo {pages} {77}
  (\bibinfo {year} {1995})}\BibitemShut {NoStop}%
\bibitem [{\citenamefont {Menneer}\ and\ \citenamefont
  {Narayanan}(1995)}]{Menneer1995}%
  \BibitemOpen
  \bibfield  {author} {\bibinfo {author} {\bibfnamefont {T.}~\bibnamefont
  {Menneer}}\ and\ \bibinfo {author} {\bibfnamefont {A.}~\bibnamefont
  {Narayanan}},\ }\href@noop {} {\emph {\bibinfo {title} {Quantum-inspired
  Neural Networks}}},\ \bibinfo {type} {Tech. Rep.}\ (\bibinfo {year}
  {1995})\BibitemShut {NoStop}%
\bibitem [{\citenamefont {Chrisley}(1996)}]{Chrisley1996}%
  \BibitemOpen
  \bibfield  {author} {\bibinfo {author} {\bibfnamefont {R.~L.}\ \bibnamefont
  {Chrisley}},\ }in\ \href@noop {} {\emph {\bibinfo {booktitle} {In P.
  Pylkk{\"a}nen \& P. Pylkk{\"o} (Eds.), Brain, Mind and Physics}}}\ (\bibinfo
  {publisher} {IOS Press},\ \bibinfo {year} {1996})\ pp.\ \bibinfo {pages}
  {126--139}\BibitemShut {NoStop}%
\bibitem [{\citenamefont {Behrman}\ \emph {et~al.}(1996)\citenamefont
  {Behrman}, \citenamefont {Niemel}, \citenamefont {Steck},\ and\ \citenamefont
  {Skinner}}]{Behrman1996}%
  \BibitemOpen
  \bibfield  {author} {\bibinfo {author} {\bibfnamefont {E.}~\bibnamefont
  {Behrman}}, \bibinfo {author} {\bibfnamefont {J.}~\bibnamefont {Niemel}},
  \bibinfo {author} {\bibfnamefont {J.}~\bibnamefont {Steck}},\ and\ \bibinfo
  {author} {\bibfnamefont {S.}~\bibnamefont {Skinner}},\ }\href@noop {}
  {\bibinfo {title} {A quantum dot neural network}} (\bibinfo {year}
  {1996})\BibitemShut {NoStop}%
\bibitem [{\citenamefont {Narayanan}\ and\ \citenamefont
  {Moore}(1996)}]{Narayanan1996}%
  \BibitemOpen
  \bibfield  {author} {\bibinfo {author} {\bibfnamefont {A.}~\bibnamefont
  {Narayanan}}\ and\ \bibinfo {author} {\bibfnamefont {M.}~\bibnamefont
  {Moore}},\ }in\ \href {https://doi.org/10.1109/ICEC.1996.542334} {\emph
  {\bibinfo {booktitle} {Proceedings of IEEE International Conference on
  Evolutionary Computation}}}\ (\bibinfo {year} {1996})\ pp.\ \bibinfo {pages}
  {61--66}\BibitemShut {NoStop}%
\bibitem [{\citenamefont {Menneer}(1999)}]{Menneer1999}%
  \BibitemOpen
  \bibfield  {author} {\bibinfo {author} {\bibfnamefont {T.}~\bibnamefont
  {Menneer}}\ }(\bibinfo {year} {1999})\BibitemShut {NoStop}%
\bibitem [{\citenamefont {Jeswal}\ and\ \citenamefont
  {Chakraverty}(2019)}]{Jeswal2019}%
  \BibitemOpen
  \bibfield  {author} {\bibinfo {author} {\bibfnamefont {S.~K.}\ \bibnamefont
  {Jeswal}}\ and\ \bibinfo {author} {\bibfnamefont {S.}~\bibnamefont
  {Chakraverty}},\ }\bibfield  {journal} {\bibinfo  {journal} {Archives of
  Computational Methods in Engineering}\ }\textbf {\bibinfo {volume} {26}},\
  \href {https://doi.org/10.1007/s11831-018-9269-0} {10.1007/s11831-018-9269-0}
  (\bibinfo {year} {2019})\BibitemShut {NoStop}%
\bibitem [{\citenamefont {Jia}\ \emph {et~al.}(2019)\citenamefont {Jia},
  \citenamefont {Yi}, \citenamefont {Zhai}, \citenamefont {Wu}, \citenamefont
  {Guo},\ and\ \citenamefont {Guo}}]{Jia2019}%
  \BibitemOpen
  \bibfield  {author} {\bibinfo {author} {\bibfnamefont {Z.-A.}\ \bibnamefont
  {Jia}}, \bibinfo {author} {\bibfnamefont {B.}~\bibnamefont {Yi}}, \bibinfo
  {author} {\bibfnamefont {R.}~\bibnamefont {Zhai}}, \bibinfo {author}
  {\bibfnamefont {Y.-C.}\ \bibnamefont {Wu}}, \bibinfo {author} {\bibfnamefont
  {G.-C.}\ \bibnamefont {Guo}},\ and\ \bibinfo {author} {\bibfnamefont {G.-P.}\
  \bibnamefont {Guo}},\ }\href
  {https://doi.org/https://doi.org/10.1002/qute.201800077} {\bibfield
  {journal} {\bibinfo  {journal} {Advanced Quantum Technologies}\ }\textbf
  {\bibinfo {volume} {2}},\ \bibinfo {pages} {1800077} (\bibinfo {year}
  {2019})}\BibitemShut {NoStop}%
\bibitem [{\citenamefont {Zhao}\ and\ \citenamefont {Wang}(2021)}]{Zhao2021a}%
  \BibitemOpen
  \bibfield  {author} {\bibinfo {author} {\bibfnamefont {R.}~\bibnamefont
  {Zhao}}\ and\ \bibinfo {author} {\bibfnamefont {S.}~\bibnamefont {Wang}},\
  }\href@noop {} {\bibinfo {title} {A review of quantum neural networks:
  Methods, models, dilemma}} (\bibinfo {year} {2021}),\ \Eprint
  {https://arxiv.org/abs/2109.01840v1} {arXiv:2109.01840v1 [cs.ET]}
  \BibitemShut {NoStop}%
\bibitem [{\citenamefont {Kyriienko}\ \emph {et~al.}(2022)\citenamefont
  {Kyriienko}, \citenamefont {Paine},\ and\ \citenamefont
  {Elfving}}]{Kyriienko2022}%
  \BibitemOpen
  \bibfield  {author} {\bibinfo {author} {\bibfnamefont {O.}~\bibnamefont
  {Kyriienko}}, \bibinfo {author} {\bibfnamefont {A.~E.}\ \bibnamefont
  {Paine}},\ and\ \bibinfo {author} {\bibfnamefont {V.~E.}\ \bibnamefont
  {Elfving}},\ }\href@noop {} {\bibinfo {title} {Protocols for trainable and
  differentiable quantum generative modelling}} (\bibinfo {year} {2022}),\
  \Eprint {https://arxiv.org/abs/2202.08253} {arXiv:2202.08253 [quant-ph]}
  \BibitemShut {NoStop}%
\bibitem [{\citenamefont {{Kyriienko}}\ \emph {et~al.}(2021)\citenamefont
  {{Kyriienko}}, \citenamefont {{Paine}},\ and\ \citenamefont
  {{Elfving}}}]{Kyriienko2021}%
  \BibitemOpen
  \bibfield  {author} {\bibinfo {author} {\bibfnamefont {O.}~\bibnamefont
  {{Kyriienko}}}, \bibinfo {author} {\bibfnamefont {A.~E.}\ \bibnamefont
  {{Paine}}},\ and\ \bibinfo {author} {\bibfnamefont {V.~E.}\ \bibnamefont
  {{Elfving}}},\ }\href {https://doi.org/10.1103/PhysRevA.103.052416}
  {\bibfield  {journal} {\bibinfo  {journal} {\pra}\ }\textbf {\bibinfo
  {volume} {103}},\ \bibinfo {eid} {052416} (\bibinfo {year} {2021})},\ \Eprint
  {https://arxiv.org/abs/2011.10395} {arXiv:2011.10395 [quant-ph]} \BibitemShut
  {NoStop}%
\bibitem [{\citenamefont {Paine}\ \emph {et~al.}(2021)\citenamefont {Paine},
  \citenamefont {Elfving},\ and\ \citenamefont {Kyriienko}}]{Paine2021}%
  \BibitemOpen
  \bibfield  {author} {\bibinfo {author} {\bibfnamefont {A.~E.}\ \bibnamefont
  {Paine}}, \bibinfo {author} {\bibfnamefont {V.~E.}\ \bibnamefont {Elfving}},\
  and\ \bibinfo {author} {\bibfnamefont {O.}~\bibnamefont {Kyriienko}},\ }\href
  {https://doi.org/10.48550/ARXIV.2108.03190} {\bibinfo {title} {Quantum
  quantile mechanics: Solving stochastic differential equations for generating
  time-series}} (\bibinfo {year} {2021})\BibitemShut {NoStop}%
\bibitem [{\citenamefont {Heim}\ \emph {et~al.}(2021)\citenamefont {Heim},
  \citenamefont {Ghosh}, \citenamefont {Kyriienko},\ and\ \citenamefont
  {Elfving}}]{Heim2021}%
  \BibitemOpen
  \bibfield  {author} {\bibinfo {author} {\bibfnamefont {N.}~\bibnamefont
  {Heim}}, \bibinfo {author} {\bibfnamefont {A.}~\bibnamefont {Ghosh}},
  \bibinfo {author} {\bibfnamefont {O.}~\bibnamefont {Kyriienko}},\ and\
  \bibinfo {author} {\bibfnamefont {V.~E.}\ \bibnamefont {Elfving}},\ }\href
  {https://doi.org/10.48550/ARXIV.2111.06376} {\bibinfo {title} {Quantum
  model-discovery}} (\bibinfo {year} {2021})\BibitemShut {NoStop}%
\bibitem [{pat()}]{patent}%
  \BibitemOpen
  \href@noop {} {\bibinfo {title} {{A patent application for the method
  described in this manuscript has been submitted by PASQAL SAS with Savvas
  Varsamopoulos, Evan Philip, and Vincent E. Elfving as
  inventors.}}}\BibitemShut {Stop}%
\bibitem [{\citenamefont {Schuld}\ \emph {et~al.}(2019)\citenamefont {Schuld},
  \citenamefont {Bergholm}, \citenamefont {Gogolin}, \citenamefont {Izaac},\
  and\ \citenamefont {Killoran}}]{Schuld2019a}%
  \BibitemOpen
  \bibfield  {author} {\bibinfo {author} {\bibfnamefont {M.}~\bibnamefont
  {Schuld}}, \bibinfo {author} {\bibfnamefont {V.}~\bibnamefont {Bergholm}},
  \bibinfo {author} {\bibfnamefont {C.}~\bibnamefont {Gogolin}}, \bibinfo
  {author} {\bibfnamefont {J.}~\bibnamefont {Izaac}},\ and\ \bibinfo {author}
  {\bibfnamefont {N.}~\bibnamefont {Killoran}},\ }\href
  {https://doi.org/10.1103/PhysRevA.99.032331} {\bibfield  {journal} {\bibinfo
  {journal} {Phys. Rev. A}\ }\textbf {\bibinfo {volume} {99}},\ \bibinfo
  {pages} {032331} (\bibinfo {year} {2019})}\BibitemShut {NoStop}%
\bibitem [{\citenamefont {Liao}\ \emph {et~al.}(2021)\citenamefont {Liao},
  \citenamefont {Hsieh},\ and\ \citenamefont {Ferrie}}]{Liao2021}%
  \BibitemOpen
  \bibfield  {author} {\bibinfo {author} {\bibfnamefont {Y.}~\bibnamefont
  {Liao}}, \bibinfo {author} {\bibfnamefont {M.-H.}\ \bibnamefont {Hsieh}},\
  and\ \bibinfo {author} {\bibfnamefont {C.}~\bibnamefont {Ferrie}},\
  }\href@noop {} {\bibinfo {title} {Quantum optimization for training quantum
  neural networks}} (\bibinfo {year} {2021}),\ \Eprint
  {https://arxiv.org/abs/2103.17047} {arXiv:2103.17047 [quant-ph]} \BibitemShut
  {NoStop}%
\bibitem [{\citenamefont {McClean}\ \emph {et~al.}(2018)\citenamefont
  {McClean}, \citenamefont {Boixo}, \citenamefont {Smelyanskiy}, \citenamefont
  {Babbush},\ and\ \citenamefont {Neven}}]{McClean2018}%
  \BibitemOpen
  \bibfield  {author} {\bibinfo {author} {\bibfnamefont {J.~R.}\ \bibnamefont
  {McClean}}, \bibinfo {author} {\bibfnamefont {S.}~\bibnamefont {Boixo}},
  \bibinfo {author} {\bibfnamefont {V.~N.}\ \bibnamefont {Smelyanskiy}},
  \bibinfo {author} {\bibfnamefont {R.}~\bibnamefont {Babbush}},\ and\ \bibinfo
  {author} {\bibfnamefont {H.}~\bibnamefont {Neven}},\ }\bibfield  {journal}
  {\bibinfo  {journal} {Nature Communications}\ }\textbf {\bibinfo {volume}
  {9}},\ \href {https://doi.org/10.1038/s41467-018-07090-4}
  {10.1038/s41467-018-07090-4} (\bibinfo {year} {2018})\BibitemShut {NoStop}%
\bibitem [{\citenamefont {Grant}\ \emph {et~al.}(2019)\citenamefont {Grant},
  \citenamefont {Wossnig}, \citenamefont {Ostaszewski},\ and\ \citenamefont
  {Benedetti}}]{Grant2019}%
  \BibitemOpen
  \bibfield  {author} {\bibinfo {author} {\bibfnamefont {E.}~\bibnamefont
  {Grant}}, \bibinfo {author} {\bibfnamefont {L.}~\bibnamefont {Wossnig}},
  \bibinfo {author} {\bibfnamefont {M.}~\bibnamefont {Ostaszewski}},\ and\
  \bibinfo {author} {\bibfnamefont {M.}~\bibnamefont {Benedetti}},\ }\href
  {https://doi.org/10.22331/q-2019-12-09-214} {\bibfield  {journal} {\bibinfo
  {journal} {{Quantum}}\ }\textbf {\bibinfo {volume} {3}},\ \bibinfo {pages}
  {214} (\bibinfo {year} {2019})}\BibitemShut {NoStop}%
\bibitem [{\citenamefont {Cerezo}\ \emph {et~al.}(2021)\citenamefont {Cerezo},
  \citenamefont {Sone}, \citenamefont {Volkoff}, \citenamefont {Cincio},\ and\
  \citenamefont {Coles}}]{Cerezo2021b}%
  \BibitemOpen
  \bibfield  {author} {\bibinfo {author} {\bibfnamefont {M.}~\bibnamefont
  {Cerezo}}, \bibinfo {author} {\bibfnamefont {A.}~\bibnamefont {Sone}},
  \bibinfo {author} {\bibfnamefont {T.}~\bibnamefont {Volkoff}}, \bibinfo
  {author} {\bibfnamefont {L.}~\bibnamefont {Cincio}},\ and\ \bibinfo {author}
  {\bibfnamefont {P.~J.}\ \bibnamefont {Coles}},\ }\bibfield  {journal}
  {\bibinfo  {journal} {Nature Communications}\ }\textbf {\bibinfo {volume}
  {12}},\ \href {https://doi.org/10.1038/s41467-021-21728-w}
  {10.1038/s41467-021-21728-w} (\bibinfo {year} {2021})\BibitemShut {NoStop}%
\bibitem [{\citenamefont {Wang}\ \emph {et~al.}(2021)\citenamefont {Wang},
  \citenamefont {Fontana}, \citenamefont {Cerezo}, \citenamefont {Sharma},
  \citenamefont {Sone}, \citenamefont {Cincio},\ and\ \citenamefont
  {Coles}}]{Wang2021}%
  \BibitemOpen
  \bibfield  {author} {\bibinfo {author} {\bibfnamefont {S.}~\bibnamefont
  {Wang}}, \bibinfo {author} {\bibfnamefont {E.}~\bibnamefont {Fontana}},
  \bibinfo {author} {\bibfnamefont {M.}~\bibnamefont {Cerezo}}, \bibinfo
  {author} {\bibfnamefont {K.}~\bibnamefont {Sharma}}, \bibinfo {author}
  {\bibfnamefont {A.}~\bibnamefont {Sone}}, \bibinfo {author} {\bibfnamefont
  {L.}~\bibnamefont {Cincio}},\ and\ \bibinfo {author} {\bibfnamefont {P.~J.}\
  \bibnamefont {Coles}},\ }\bibfield  {journal} {\bibinfo  {journal} {Nature
  Communications}\ }\textbf {\bibinfo {volume} {12}},\ \href
  {https://doi.org/10.1038/s41467-021-27045-6} {10.1038/s41467-021-27045-6}
  (\bibinfo {year} {2021})\BibitemShut {NoStop}%
\bibitem [{\citenamefont {Cerezo}\ and\ \citenamefont
  {Coles}(2021)}]{Cerezo2021c}%
  \BibitemOpen
  \bibfield  {author} {\bibinfo {author} {\bibfnamefont {M.}~\bibnamefont
  {Cerezo}}\ and\ \bibinfo {author} {\bibfnamefont {P.~J.}\ \bibnamefont
  {Coles}},\ }\href {https://doi.org/10.1088/2058-9565/abf51a} {\bibfield
  {journal} {\bibinfo  {journal} {Quantum Science and Technology}\ }\textbf
  {\bibinfo {volume} {6}},\ \bibinfo {pages} {035006} (\bibinfo {year}
  {2021})}\BibitemShut {NoStop}%
\bibitem [{\citenamefont {Arrasmith}\ \emph {et~al.}(2021)\citenamefont
  {Arrasmith}, \citenamefont {Cerezo}, \citenamefont {Czarnik}, \citenamefont
  {Cincio},\ and\ \citenamefont {Coles}}]{Arrasmith2021}%
  \BibitemOpen
  \bibfield  {author} {\bibinfo {author} {\bibfnamefont {A.}~\bibnamefont
  {Arrasmith}}, \bibinfo {author} {\bibfnamefont {M.}~\bibnamefont {Cerezo}},
  \bibinfo {author} {\bibfnamefont {P.}~\bibnamefont {Czarnik}}, \bibinfo
  {author} {\bibfnamefont {L.}~\bibnamefont {Cincio}},\ and\ \bibinfo {author}
  {\bibfnamefont {P.~J.}\ \bibnamefont {Coles}},\ }\href
  {https://doi.org/10.22331/q-2021-10-05-558} {\bibfield  {journal} {\bibinfo
  {journal} {{Quantum}}\ }\textbf {\bibinfo {volume} {5}},\ \bibinfo {pages}
  {558} (\bibinfo {year} {2021})}\BibitemShut {NoStop}%
\bibitem [{\citenamefont {Holmes}\ \emph {et~al.}(2021)\citenamefont {Holmes},
  \citenamefont {Arrasmith}, \citenamefont {Yan}, \citenamefont {Coles},
  \citenamefont {Albrecht},\ and\ \citenamefont {Sornborger}}]{Holmes2021}%
  \BibitemOpen
  \bibfield  {author} {\bibinfo {author} {\bibfnamefont {Z.}~\bibnamefont
  {Holmes}}, \bibinfo {author} {\bibfnamefont {A.}~\bibnamefont {Arrasmith}},
  \bibinfo {author} {\bibfnamefont {B.}~\bibnamefont {Yan}}, \bibinfo {author}
  {\bibfnamefont {P.~J.}\ \bibnamefont {Coles}}, \bibinfo {author}
  {\bibfnamefont {A.}~\bibnamefont {Albrecht}},\ and\ \bibinfo {author}
  {\bibfnamefont {A.~T.}\ \bibnamefont {Sornborger}},\ }\href
  {https://doi.org/10.1103/PhysRevLett.126.190501} {\bibfield  {journal}
  {\bibinfo  {journal} {Phys. Rev. Lett.}\ }\textbf {\bibinfo {volume} {126}},\
  \bibinfo {pages} {190501} (\bibinfo {year} {2021})}\BibitemShut {NoStop}%
\bibitem [{\citenamefont {Marrero}\ \emph {et~al.}(2021)\citenamefont
  {Marrero}, \citenamefont {Kieferov\'a},\ and\ \citenamefont
  {Wiebe}}]{Ortiz-Marrero2021}%
  \BibitemOpen
  \bibfield  {author} {\bibinfo {author} {\bibfnamefont {C.~O.}\ \bibnamefont
  {Marrero}}, \bibinfo {author} {\bibfnamefont {M.}~\bibnamefont
  {Kieferov\'a}},\ and\ \bibinfo {author} {\bibfnamefont {N.}~\bibnamefont
  {Wiebe}},\ }\href {https://doi.org/10.1103/PRXQuantum.2.040316} {\bibfield
  {journal} {\bibinfo  {journal} {PRX Quantum}\ }\textbf {\bibinfo {volume}
  {2}},\ \bibinfo {pages} {040316} (\bibinfo {year} {2021})}\BibitemShut
  {NoStop}%
\bibitem [{\citenamefont {Patti}\ \emph {et~al.}(2021)\citenamefont {Patti},
  \citenamefont {Najafi}, \citenamefont {Gao},\ and\ \citenamefont
  {Yelin}}]{Patti2021}%
  \BibitemOpen
  \bibfield  {author} {\bibinfo {author} {\bibfnamefont {T.~L.}\ \bibnamefont
  {Patti}}, \bibinfo {author} {\bibfnamefont {K.}~\bibnamefont {Najafi}},
  \bibinfo {author} {\bibfnamefont {X.}~\bibnamefont {Gao}},\ and\ \bibinfo
  {author} {\bibfnamefont {S.~F.}\ \bibnamefont {Yelin}},\ }\href
  {https://doi.org/10.1103/PhysRevResearch.3.033090} {\bibfield  {journal}
  {\bibinfo  {journal} {Phys. Rev. Research}\ }\textbf {\bibinfo {volume}
  {3}},\ \bibinfo {pages} {033090} (\bibinfo {year} {2021})}\BibitemShut
  {NoStop}%
\bibitem [{\citenamefont {Zhao}\ and\ \citenamefont {Gao}(2021)}]{Zhao2021b}%
  \BibitemOpen
  \bibfield  {author} {\bibinfo {author} {\bibfnamefont {C.}~\bibnamefont
  {Zhao}}\ and\ \bibinfo {author} {\bibfnamefont {X.-S.}\ \bibnamefont {Gao}},\
  }\href {https://doi.org/10.22331/q-2021-06-04-466} {\bibfield  {journal}
  {\bibinfo  {journal} {{Quantum}}\ }\textbf {\bibinfo {volume} {5}},\ \bibinfo
  {pages} {466} (\bibinfo {year} {2021})}\BibitemShut {NoStop}%
\bibitem [{\citenamefont {Abbas}\ \emph {et~al.}(2021)\citenamefont {Abbas},
  \citenamefont {Sutter}, \citenamefont {Zoufal}, \citenamefont {Lucchi},
  \citenamefont {Figalli},\ and\ \citenamefont {Woerner}}]{Abbas2021}%
  \BibitemOpen
  \bibfield  {author} {\bibinfo {author} {\bibfnamefont {A.}~\bibnamefont
  {Abbas}}, \bibinfo {author} {\bibfnamefont {D.}~\bibnamefont {Sutter}},
  \bibinfo {author} {\bibfnamefont {C.}~\bibnamefont {Zoufal}}, \bibinfo
  {author} {\bibfnamefont {A.}~\bibnamefont {Lucchi}}, \bibinfo {author}
  {\bibfnamefont {A.}~\bibnamefont {Figalli}},\ and\ \bibinfo {author}
  {\bibfnamefont {S.}~\bibnamefont {Woerner}},\ }\href
  {https://doi.org/10.1038/s43588-021-00084-1} {\bibfield  {journal} {\bibinfo
  {journal} {Nature Computational Science}\ }\textbf {\bibinfo {volume} {1}},\
  \bibinfo {pages} {6} (\bibinfo {year} {2021})}\BibitemShut {NoStop}%
\bibitem [{\citenamefont {Coles}(2021)}]{Coles2021}%
  \BibitemOpen
  \bibfield  {author} {\bibinfo {author} {\bibfnamefont {P.~J.}\ \bibnamefont
  {Coles}},\ }\href {https://doi.org/10.1038/s43588-021-00088-x} {\bibfield
  {journal} {\bibinfo  {journal} {Nature Computational Science}\ }\textbf
  {\bibinfo {volume} {1}},\ \bibinfo {pages} {6} (\bibinfo {year}
  {2021})}\BibitemShut {NoStop}%
\bibitem [{\citenamefont {Lewenstein}\ \emph {et~al.}(2021)\citenamefont
  {Lewenstein}, \citenamefont {Gratsea}, \citenamefont {Riera-Campeny},
  \citenamefont {Aloy}, \citenamefont {Kasper},\ and\ \citenamefont
  {Sanpera}}]{Lewenstein2021}%
  \BibitemOpen
  \bibfield  {author} {\bibinfo {author} {\bibfnamefont {M.}~\bibnamefont
  {Lewenstein}}, \bibinfo {author} {\bibfnamefont {A.}~\bibnamefont {Gratsea}},
  \bibinfo {author} {\bibfnamefont {A.}~\bibnamefont {Riera-Campeny}}, \bibinfo
  {author} {\bibfnamefont {A.}~\bibnamefont {Aloy}}, \bibinfo {author}
  {\bibfnamefont {V.}~\bibnamefont {Kasper}},\ and\ \bibinfo {author}
  {\bibfnamefont {A.}~\bibnamefont {Sanpera}},\ }\href
  {https://doi.org/10.1088/2058-9565/ac070f} {\bibfield  {journal} {\bibinfo
  {journal} {Quantum Science and Technology}\ }\textbf {\bibinfo {volume}
  {6}},\ \bibinfo {pages} {4} (\bibinfo {year} {2021})}\BibitemShut {NoStop}%
\bibitem [{\citenamefont {Wright}\ and\ \citenamefont
  {McMahon}(2019)}]{Wright2019}%
  \BibitemOpen
  \bibfield  {author} {\bibinfo {author} {\bibfnamefont {L.~G.}\ \bibnamefont
  {Wright}}\ and\ \bibinfo {author} {\bibfnamefont {P.~L.}\ \bibnamefont
  {McMahon}},\ }\href@noop {} {\bibinfo {title} {The capacity of quantum neural
  networks}} (\bibinfo {year} {2019}),\ \Eprint
  {https://arxiv.org/abs/1908.01364v1} {arXiv:1908.01364v1 [quant-ph]}
  \BibitemShut {NoStop}%
\bibitem [{\citenamefont {Kandala}\ \emph {et~al.}(2017)\citenamefont
  {Kandala}, \citenamefont {Mezzacapo}, \citenamefont {Temme}, \citenamefont
  {Takita}, \citenamefont {Brink}, \citenamefont {Chow},\ and\ \citenamefont
  {Gambetta}}]{Kandala2017}%
  \BibitemOpen
  \bibfield  {author} {\bibinfo {author} {\bibfnamefont {A.}~\bibnamefont
  {Kandala}}, \bibinfo {author} {\bibfnamefont {A.}~\bibnamefont {Mezzacapo}},
  \bibinfo {author} {\bibfnamefont {K.}~\bibnamefont {Temme}}, \bibinfo
  {author} {\bibfnamefont {M.}~\bibnamefont {Takita}}, \bibinfo {author}
  {\bibfnamefont {M.}~\bibnamefont {Brink}}, \bibinfo {author} {\bibfnamefont
  {J.~M.}\ \bibnamefont {Chow}},\ and\ \bibinfo {author} {\bibfnamefont
  {J.~M.}\ \bibnamefont {Gambetta}},\ }\href
  {https://doi.org/10.1038/nature23879} {\bibfield  {journal} {\bibinfo
  {journal} {Nature}\ }\textbf {\bibinfo {volume} {549}},\ \bibinfo {pages}
  {242} (\bibinfo {year} {2017})}\BibitemShut {NoStop}%
\bibitem [{\citenamefont {Kingma}\ and\ \citenamefont {Ba}(2014)}]{Kingma2014}%
  \BibitemOpen
  \bibfield  {author} {\bibinfo {author} {\bibfnamefont {D.~P.}\ \bibnamefont
  {Kingma}}\ and\ \bibinfo {author} {\bibfnamefont {J.}~\bibnamefont {Ba}},\
  }\href@noop {} {\bibinfo {title} {Adam: A method for stochastic
  optimization}} (\bibinfo {year} {2014}),\ \Eprint
  {https://arxiv.org/abs/1412.6980v9} {arXiv:1412.6980v9 [cs.LG]} \BibitemShut
  {NoStop}%
\bibitem [{\citenamefont {Garey}\ \emph {et~al.}(1976)\citenamefont {Garey},
  \citenamefont {Johnson},\ and\ \citenamefont {Stockmeyer}}]{Garey1976}%
  \BibitemOpen
  \bibfield  {author} {\bibinfo {author} {\bibfnamefont {M.}~\bibnamefont
  {Garey}}, \bibinfo {author} {\bibfnamefont {D.}~\bibnamefont {Johnson}},\
  and\ \bibinfo {author} {\bibfnamefont {L.}~\bibnamefont {Stockmeyer}},\
  }\href {https://doi.org/https://doi.org/10.1016/0304-3975(76)90059-1}
  {\bibfield  {journal} {\bibinfo  {journal} {Theoretical Computer Science}\
  }\textbf {\bibinfo {volume} {1}},\ \bibinfo {pages} {237} (\bibinfo {year}
  {1976})}\BibitemShut {NoStop}%
\bibitem [{\citenamefont {Cook}\ \emph {et~al.}(1995)\citenamefont {Cook},
  \citenamefont {Lovasz},\ and\ \citenamefont {Seymour}}]{Cook1995}%
  \BibitemOpen
  \bibfield  {author} {\bibinfo {author} {\bibfnamefont {W.~J.}\ \bibnamefont
  {Cook}}, \bibinfo {author} {\bibfnamefont {L.}~\bibnamefont {Lovasz}},\ and\
  \bibinfo {author} {\bibfnamefont {P.}~\bibnamefont {Seymour}},\ }\href
  {https://doi.org/https://doi.org/10.1090/dimacs/020} {\emph {\bibinfo {title}
  {Combinatorial Optimization}}},\ Vol.~\bibinfo {volume} {20}\ (\bibinfo
  {publisher} {American Mathematical Society},\ \bibinfo {year}
  {1995})\BibitemShut {NoStop}%
\bibitem [{\citenamefont {Berman}\ and\ \citenamefont
  {Karpinski}(1999)}]{Berman1999}%
  \BibitemOpen
  \bibfield  {author} {\bibinfo {author} {\bibfnamefont {P.}~\bibnamefont
  {Berman}}\ and\ \bibinfo {author} {\bibfnamefont {M.}~\bibnamefont
  {Karpinski}},\ }in\ \href@noop {} {\emph {\bibinfo {booktitle} {Automata,
  Languages and Programming}}},\ \bibinfo {editor} {edited by\ \bibinfo
  {editor} {\bibfnamefont {J.}~\bibnamefont {Wiedermann}}, \bibinfo {editor}
  {\bibfnamefont {P.}~\bibnamefont {van Emde~Boas}},\ and\ \bibinfo {editor}
  {\bibfnamefont {M.}~\bibnamefont {Nielsen}}}\ (\bibinfo  {publisher}
  {Springer Berlin Heidelberg},\ \bibinfo {address} {Berlin, Heidelberg},\
  \bibinfo {year} {1999})\ pp.\ \bibinfo {pages} {200--209}\BibitemShut
  {NoStop}%
\bibitem [{\citenamefont {Everitt}\ \emph {et~al.}(2011)\citenamefont
  {Everitt}, \citenamefont {Landau}, \citenamefont {Leese},\ and\ \citenamefont
  {Stahl}}]{Everitt2011}%
  \BibitemOpen
  \bibfield  {author} {\bibinfo {author} {\bibfnamefont {B.~S.}\ \bibnamefont
  {Everitt}}, \bibinfo {author} {\bibfnamefont {S.}~\bibnamefont {Landau}},
  \bibinfo {author} {\bibfnamefont {M.}~\bibnamefont {Leese}},\ and\ \bibinfo
  {author} {\bibfnamefont {D.}~\bibnamefont {Stahl}},\ }\href@noop {} {\emph
  {\bibinfo {title} {Cluster Analysis}}}\ (\bibinfo  {publisher} {Wiley},\
  \bibinfo {year} {2011})\BibitemShut {NoStop}%
\bibitem [{\citenamefont {Garey}\ \emph {et~al.}(2017)\citenamefont {Garey},
  \citenamefont {Johnson},\ and\ \citenamefont {Stockmeyer}}]{Altman2017}%
  \BibitemOpen
  \bibfield  {author} {\bibinfo {author} {\bibfnamefont {M.}~\bibnamefont
  {Garey}}, \bibinfo {author} {\bibfnamefont {D.}~\bibnamefont {Johnson}},\
  and\ \bibinfo {author} {\bibfnamefont {L.}~\bibnamefont {Stockmeyer}},\
  }\bibfield  {journal} {\bibinfo  {journal} {Nature Methods}\ }\textbf
  {\bibinfo {volume} {14}},\ \href {https://doi.org/10.1038/nmeth.4299}
  {10.1038/nmeth.4299} (\bibinfo {year} {2017})\BibitemShut {NoStop}%
\bibitem [{\citenamefont {Nowak}\ and\ \citenamefont
  {Tibshirani}(2008)}]{Nowak2008}%
  \BibitemOpen
  \bibfield  {author} {\bibinfo {author} {\bibfnamefont {G.}~\bibnamefont
  {Nowak}}\ and\ \bibinfo {author} {\bibfnamefont {R.}~\bibnamefont
  {Tibshirani}},\ }\bibfield  {journal} {\bibinfo  {journal} {Biostatistics}\
  }\textbf {\bibinfo {volume} {9}},\ \href
  {https://doi.org/10.1093/biostatistics/kxm046} {10.1093/biostatistics/kxm046}
  (\bibinfo {year} {2008})\BibitemShut {NoStop}%
\bibitem [{\citenamefont {Bengio}\ and\ \citenamefont
  {Bengio}(1999)}]{1999Bengio}%
  \BibitemOpen
  \bibfield  {author} {\bibinfo {author} {\bibfnamefont {Y.}~\bibnamefont
  {Bengio}}\ and\ \bibinfo {author} {\bibfnamefont {S.}~\bibnamefont
  {Bengio}},\ }in\ \href
  {https://proceedings.neurips.cc/paper/1999/file/e6384711491713d29bc63fc5eeb5ba4f-Paper.pdf}
  {\emph {\bibinfo {booktitle} {Advances in Neural Information Processing
  Systems}}},\ Vol.~\bibinfo {volume} {12},\ \bibinfo {editor} {edited by\
  \bibinfo {editor} {\bibfnamefont {S.}~\bibnamefont {Solla}}, \bibinfo
  {editor} {\bibfnamefont {T.}~\bibnamefont {Leen}},\ and\ \bibinfo {editor}
  {\bibfnamefont {K.}~\bibnamefont {M\"{u}ller}}}\ (\bibinfo  {publisher} {MIT
  Press},\ \bibinfo {year} {1999})\BibitemShut {NoStop}%
\bibitem [{\citenamefont {Schuld}\ and\ \citenamefont
  {Killoran}(2019)}]{Schuld2019b}%
  \BibitemOpen
  \bibfield  {author} {\bibinfo {author} {\bibfnamefont {M.}~\bibnamefont
  {Schuld}}\ and\ \bibinfo {author} {\bibfnamefont {N.}~\bibnamefont
  {Killoran}},\ }\href {https://doi.org/10.1103/PhysRevLett.122.040504}
  {\bibfield  {journal} {\bibinfo  {journal} {Phys. Rev. Lett.}\ }\textbf
  {\bibinfo {volume} {122}},\ \bibinfo {pages} {040504} (\bibinfo {year}
  {2019})}\BibitemShut {NoStop}%
\bibitem [{\citenamefont {Henry}\ \emph {et~al.}(2021)\citenamefont {Henry},
  \citenamefont {Thabet}, \citenamefont {Dalyac},\ and\ \citenamefont
  {Henriet}}]{Henry2021}%
  \BibitemOpen
  \bibfield  {author} {\bibinfo {author} {\bibfnamefont {L.-P.}\ \bibnamefont
  {Henry}}, \bibinfo {author} {\bibfnamefont {S.}~\bibnamefont {Thabet}},
  \bibinfo {author} {\bibfnamefont {C.}~\bibnamefont {Dalyac}},\ and\ \bibinfo
  {author} {\bibfnamefont {L.}~\bibnamefont {Henriet}},\ }\href
  {https://doi.org/10.1103/PhysRevA.104.032416} {\bibfield  {journal} {\bibinfo
   {journal} {Phys. Rev. A}\ }\textbf {\bibinfo {volume} {104}},\ \bibinfo
  {pages} {032416} (\bibinfo {year} {2021})}\BibitemShut {NoStop}%
\bibitem [{\citenamefont {Guilu~Long}(2008)}]{Long2008}%
  \BibitemOpen
  \bibfield  {author} {\bibinfo {author} {\bibfnamefont {Y.~L.}\ \bibnamefont
  {Guilu~Long}},\ }\href {https://doi.org/10.1007/s11704-008-0021-z} {\bibfield
   {journal} {\bibinfo  {journal} {Frontiers of Computer Science}\ }\textbf
  {\bibinfo {volume} {2}},\ \bibinfo {eid} {167} (\bibinfo {year}
  {2008})}\BibitemShut {NoStop}%
\bibitem [{\citenamefont {Childs}\ and\ \citenamefont
  {Wiebe}(2012)}]{Childs2012}%
  \BibitemOpen
  \bibfield  {author} {\bibinfo {author} {\bibfnamefont {A.~M.}\ \bibnamefont
  {Childs}}\ and\ \bibinfo {author} {\bibfnamefont {N.}~\bibnamefont {Wiebe}},\
  }\href@noop {} {\bibfield  {journal} {\bibinfo  {journal} {Quantum Info.
  Comput.}\ }\textbf {\bibinfo {volume} {12}},\ \bibinfo {pages} {901–924}
  (\bibinfo {year} {2012})}\BibitemShut {NoStop}%
\bibitem [{\citenamefont {Berry}\ \emph {et~al.}(2015)\citenamefont {Berry},
  \citenamefont {Childs}, \citenamefont {Cleve}, \citenamefont {Kothari},\ and\
  \citenamefont {Somma}}]{Berry2015}%
  \BibitemOpen
  \bibfield  {author} {\bibinfo {author} {\bibfnamefont {D.~W.}\ \bibnamefont
  {Berry}}, \bibinfo {author} {\bibfnamefont {A.~M.}\ \bibnamefont {Childs}},
  \bibinfo {author} {\bibfnamefont {R.}~\bibnamefont {Cleve}}, \bibinfo
  {author} {\bibfnamefont {R.}~\bibnamefont {Kothari}},\ and\ \bibinfo {author}
  {\bibfnamefont {R.~D.}\ \bibnamefont {Somma}},\ }\href
  {https://doi.org/10.1103/PhysRevLett.114.090502} {\bibfield  {journal}
  {\bibinfo  {journal} {Phys. Rev. Lett.}\ }\textbf {\bibinfo {volume} {114}},\
  \bibinfo {pages} {090502} (\bibinfo {year} {2015})}\BibitemShut {NoStop}%
\bibitem [{\citenamefont {Childs}\ \emph {et~al.}(2018)\citenamefont {Childs},
  \citenamefont {Maslov}, \citenamefont {Nam}, \citenamefont {Ross},\ and\
  \citenamefont {Su}}]{Childs2018}%
  \BibitemOpen
  \bibfield  {author} {\bibinfo {author} {\bibfnamefont {A.~M.}\ \bibnamefont
  {Childs}}, \bibinfo {author} {\bibfnamefont {D.}~\bibnamefont {Maslov}},
  \bibinfo {author} {\bibfnamefont {Y.}~\bibnamefont {Nam}}, \bibinfo {author}
  {\bibfnamefont {N.~J.}\ \bibnamefont {Ross}},\ and\ \bibinfo {author}
  {\bibfnamefont {Y.}~\bibnamefont {Su}},\ }\href
  {https://doi.org/10.1073/pnas.1801723115} {\bibfield  {journal} {\bibinfo
  {journal} {Proceedings of the National Academy of Sciences}\ }\textbf
  {\bibinfo {volume} {115}},\ \bibinfo {pages} {9456} (\bibinfo {year}
  {2018})},\ \Eprint
  {https://arxiv.org/abs/https://www.pnas.org/doi/pdf/10.1073/pnas.1801723115}
  {https://www.pnas.org/doi/pdf/10.1073/pnas.1801723115} \BibitemShut {NoStop}%
\bibitem [{\citenamefont {{Jerbi}}\ \emph {et~al.}(2021)\citenamefont
  {{Jerbi}}, \citenamefont {{Fiderer}}, \citenamefont {{Poulsen Nautrup}},
  \citenamefont {{K{\"u}bler}}, \citenamefont {{Briegel}},\ and\ \citenamefont
  {{Dunjko}}}]{Jerbi2021}%
  \BibitemOpen
  \bibfield  {author} {\bibinfo {author} {\bibfnamefont {S.}~\bibnamefont
  {{Jerbi}}}, \bibinfo {author} {\bibfnamefont {L.~J.}\ \bibnamefont
  {{Fiderer}}}, \bibinfo {author} {\bibfnamefont {H.}~\bibnamefont {{Poulsen
  Nautrup}}}, \bibinfo {author} {\bibfnamefont {J.~M.}\ \bibnamefont
  {{K{\"u}bler}}}, \bibinfo {author} {\bibfnamefont {H.~J.}\ \bibnamefont
  {{Briegel}}},\ and\ \bibinfo {author} {\bibfnamefont {V.}~\bibnamefont
  {{Dunjko}}},\ }\href@noop {} {\bibfield  {journal} {\bibinfo  {journal}
  {arXiv e-prints}\ ,\ \bibinfo {eid} {arXiv:2110.13162}} (\bibinfo {year}
  {2021})},\ \Eprint {https://arxiv.org/abs/2110.13162} {arXiv:2110.13162
  [quant-ph]} \BibitemShut {NoStop}%
\bibitem [{\citenamefont {{Goto}}\ \emph {et~al.}(2021)\citenamefont {{Goto}},
  \citenamefont {{Tran}},\ and\ \citenamefont {{Nakajima}}}]{Goto2021}%
  \BibitemOpen
  \bibfield  {author} {\bibinfo {author} {\bibfnamefont {T.}~\bibnamefont
  {{Goto}}}, \bibinfo {author} {\bibfnamefont {Q.~H.}\ \bibnamefont {{Tran}}},\
  and\ \bibinfo {author} {\bibfnamefont {K.}~\bibnamefont {{Nakajima}}},\
  }\href {https://doi.org/10.1103/PhysRevLett.127.090506} {\bibfield  {journal}
  {\bibinfo  {journal} {\prl}\ }\textbf {\bibinfo {volume} {127}},\ \bibinfo
  {eid} {090506} (\bibinfo {year} {2021})},\ \Eprint
  {https://arxiv.org/abs/2009.00298} {arXiv:2009.00298 [quant-ph]} \BibitemShut
  {NoStop}%
\bibitem [{\citenamefont {{Karniadakis}}\ \emph {et~al.}(2021)\citenamefont
  {{Karniadakis}}, \citenamefont {{Kevrekidis}}, \citenamefont {{Lu}},
  \citenamefont {{Perdikaris}}, \citenamefont {{Wang}},\ and\ \citenamefont
  {{Yang}}}]{pinn_review}%
  \BibitemOpen
  \bibfield  {author} {\bibinfo {author} {\bibfnamefont {G.~E.}\ \bibnamefont
  {{Karniadakis}}}, \bibinfo {author} {\bibfnamefont {I.~G.}\ \bibnamefont
  {{Kevrekidis}}}, \bibinfo {author} {\bibfnamefont {L.}~\bibnamefont {{Lu}}},
  \bibinfo {author} {\bibfnamefont {P.}~\bibnamefont {{Perdikaris}}}, \bibinfo
  {author} {\bibfnamefont {S.}~\bibnamefont {{Wang}}},\ and\ \bibinfo {author}
  {\bibfnamefont {L.}~\bibnamefont {{Yang}}},\ }\href
  {https://doi.org/10.1038/s42254-021-00314-5} {\bibfield  {journal} {\bibinfo
  {journal} {Nature Reviews Physics}\ }\textbf {\bibinfo {volume} {3}},\
  \bibinfo {pages} {422} (\bibinfo {year} {2021})}\BibitemShut {NoStop}%
\bibitem [{\citenamefont {{Perdomo}}\ \emph {et~al.}(2019)\citenamefont
  {{Perdomo}}, \citenamefont {{Leyton-Ortega}},\ and\ \citenamefont
  {{Perdomo-Ortiz}}}]{2019Perdomo}%
  \BibitemOpen
  \bibfield  {author} {\bibinfo {author} {\bibfnamefont {O.}~\bibnamefont
  {{Perdomo}}}, \bibinfo {author} {\bibfnamefont {V.}~\bibnamefont
  {{Leyton-Ortega}}},\ and\ \bibinfo {author} {\bibfnamefont {A.}~\bibnamefont
  {{Perdomo-Ortiz}}},\ }\href@noop {} {\bibfield  {journal} {\bibinfo
  {journal} {arXiv e-prints}\ ,\ \bibinfo {eid} {arXiv:1903.01940}} (\bibinfo
  {year} {2019})},\ \Eprint {https://arxiv.org/abs/1903.01940}
  {arXiv:1903.01940 [quant-ph]} \BibitemShut {NoStop}%
\end{thebibliography}%
\appendix

\section{Comparison to QAOA \label{app:QAOA}}
Recall that the goal of QEL is to find $\bm{x}_\mathrm{opt} \in X$ such that $f(\bm{x}_\mathrm{opt})$ is the extremal value of a general unknown $f: X \rightarrow \mathbb{R}$, given only a partial set $\Set{(\bm{x}_i, f(\bm{x}_i))}$. The goal of QAOA, in contrast, is to find $\bm{x}_\mathrm{opt} \in \Set{0,1}^{\otimes N}$ such that $f(\bm{x}_\mathrm{opt})$ is as close as possible to the maximum of $f$, 
where $ f(\bm{x}) = \sum_{i=1}^M C_i(\bm{x}) $ and each ``clause" $C_i\vcentcolon \Set{0,1}^{\otimes N} \rightarrow \Set{0, 1}$ is given. 

\begin{figure}[!htb]
    {
    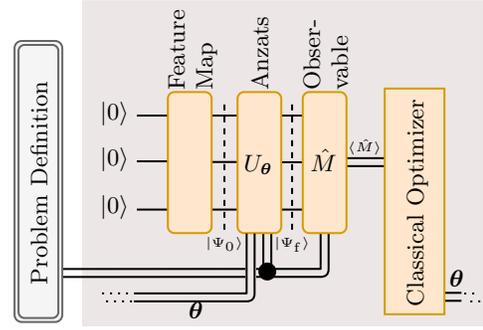
    \caption{QAOA algorithm in the same style and notation of Fig.~\ref{fig:StepI}. One epoch of QAOA is shown in the gray rectangle. Table~\ref{table:QAOA} provides a dictionary to compare to the notation used in \textcite{Farhi2014}.
    \label{fig:QAOA}}}
\end{figure}

We will now describe the QAOA algorithm using notation similar to Sec.~\ref{section:methodology} to make direct comparison easy. Fig.~\ref{fig:QAOA} shows the QAOA algorithm in the same style and notation of Fig.~\ref{fig:StepI} and Table~\ref{table:QAOA} provides a dictionary to compare to the notation used in \textcite{Farhi2014}.

Our algorithm assumes an unknown function that maps a discrete or continuous independent variable (input) $\bm{x} \in X$ to an associated dependent variable (output) $y = f(\bm{x}) \in \mathbb{R}$. A finite set of such pairs $\Set{(\bm{x}_i, y_i)}$ is known and is the training data for our algorithm. The goal of our algorithm is to find $\bm{x}_\mathrm{opt} \in X$ such that $f(\bm{x}_\mathrm{opt})$ is the extremal value of $f(x)$. 

The feature map in QAOA is a simple fixed circuit applying the Hadamard gate on each input qubit to convert the input wavefunction into $\ket{\Psi_0}$, an equal superpostion of all the bitstrings. Then a unitary operator $U_{\bm{\theta}}$ depending on $2p$ angles (``parameter" $p$ from \textcite{Farhi2014}) and the problem definition is applied to arrive at the final wavefunction $\ket{\Psi_\mathrm{f}}$. The expectation value of an operator $\hat{M}$, which is again designed according to the problem definition, is calculated with respect to the final wavefunction $\ket{\Psi_\mathrm{f}}$ and the value is used by a classical optimizer. The goal of the classical optimizer is to update ${\bm{\theta}}$ for the next epoch such that the expectation value of $\hat{M}$ is maximized over multiple iterations. 

Once the expectation value of $\hat{M}$ satisfies some predetermined criteria for maximization, $\hat{M}$ is disconnected from the circuit and $\ket{\Psi_\mathrm{f}}$ is measured in the $Z$-basis to obtain the optimal bitstring.

\begin{table}
\caption{Dictionary for comparison to QAOA.\label{table:QAOA}}
\begin{ruledtabular}
\begin{tabular}{c c} 
Fig.~\ref{fig:QAOA} & \textcite{Farhi2014} \\
\midrule
Problem definition & Objective function $C$ \\
$\ket{\Psi_0}$ & $\ket{s}$ \\
$U_{\bm{\theta}}$ & $U(B, \beta_p) U(C, \gamma_p) \ldots U(B, \beta_1) U(C, \gamma_1)$ \\
$\ket{\Psi_\mathrm{f}}$ & $\ket{\bm{\gamma},\bm{\beta}}$ \\
$\hat{M}$ & $C$\\
$\langle\hat{M}\rangle$ & $\bra{\bm{\gamma},\bm{\beta}}C\ket{\bm{\gamma},\bm{\beta}}$\\
$\bm{\theta} = \Set{\theta_1, \ldots, \theta_{2 p} }$ & $\Set{\gamma_1, \ldots, \gamma_p} \cup \Set{\beta_1, \ldots, \beta_p}$
\end{tabular}
\end{ruledtabular}
\end{table}

\section{Generalization \label{app:generalization}}
The description in Sec.~\ref{section:methodology} are edge cases of the following general description of the algorithm.

Suppose we have an unknown function 
\begin{equation}
\begin{aligned}
        f \vcentcolon \mathcal{X} \subset \mathbb{R}^n &\rightarrow \mathbb{R}\\
        \bm{x} \phantom{\subset \mathbb{R}^n} &\rightarrow f(\bm{x}),  
\end{aligned}
\end{equation}
where $\mathcal{X}$ is not necessarily path connected. We know some information about $f$ in the form of a set of conditions 
\begin{equation}\label{eq:conditions}
    \set{\mathcal{C}_i}_i.
\end{equation}
\begin{aside}
For example
\begin{equation}
    \Set{ f((0,0)) > 2 ,  \partial_{xx} f - \frac{1}{c}\partial_{tt} f \rvert_\text{interior} = 0, \ldots }.
\end{equation}
\end{aside}
\emph{\newline Requirement:} To implement this algorithm, conditions in \eqref{eq:conditions} should be enforceable through the construction in Eq.~\eqref{eq:f_approx} and the loss function in Eq.~\eqref{eq:loss}.

We are given a function 
\begin{equation}
\begin{aligned}
    L_\text{ext} \vcentcolon \left(\mathbb{R}^n\right)^{\mathbb{R}} \times \mathbb{R}^n &\rightarrow {\mathbb{R}}\\
    (g, \bm{x}) &\mapsto L_\text{ext}\left(g, \bm{x}\right).
\end{aligned}
\end{equation}
\begin{aside}
For example,
\begin{equation}
    L_\text{ext}(g, \bm{x}) = - g(\bm{x})
\end{equation}
or
\begin{equation}
    L_\text{ext}(g, \bm{x}) = \left\Vert\nabla g\rvert_{\bm{x}}\right\Vert + \tan \left\Vert\frac{2 \bm{x}}{\pi} \right\Vert.
\end{equation}
\end{aside}
Our goal is to find $\bm{x}_\text{ext} \in \mathcal{X}$ such that $L_\text{ext}\left(f, \bm{x}_\text{ext}\right)$ is the minimal value of $L_\text{ext}\left(f, \cdot \right)$ in $\mathcal{X}$. It is straightforward to generalize this to functions with multiple components.
\emph{\newline Requirement:} To fully utilize quantum parallelism in this algorithm, it must be possible to obtain $L_\text{ext}$ with a construction similar to Eq.~\eqref{eq:f_approx}.

\begin{figure}[!htb]
    {
    \scalebox{0.85}{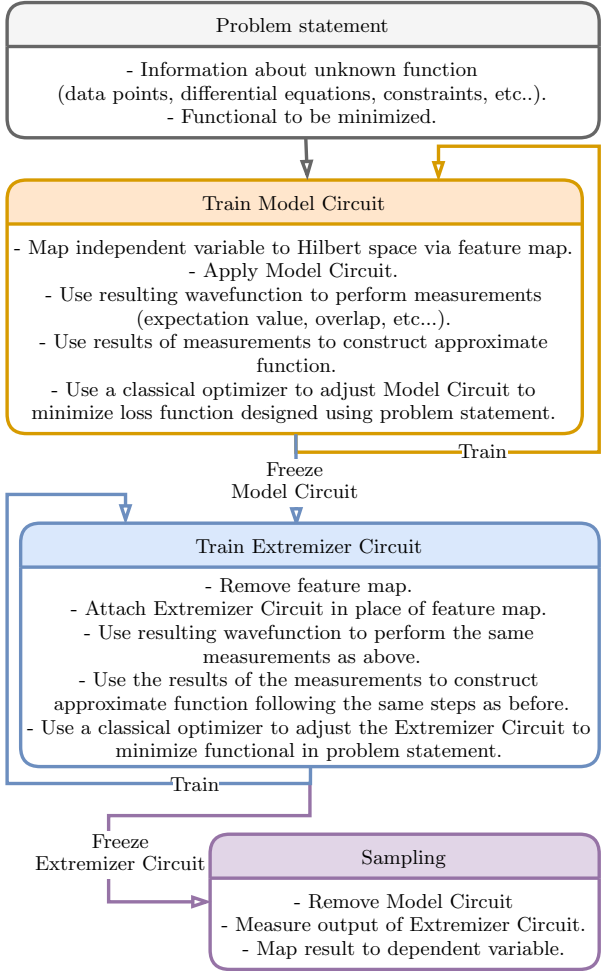}
    \caption{General workflow.} 
    \label{fig:QEL_general_workflow}}
\end{figure}

\subsection{Training the Quantum Model} 

\paragraph{} A quantum circuit parameterised by $\bm{\theta}$ is used to map $\bm{x} \in \mathcal{X}$ to a quantum state
\begin{equation}
    \Ket{\Psi_\text{M}\left(\bm{\theta}, \bm{x}\right)}.
\end{equation}
Retaining significant generality~\cite{Jerbi2021}, we will consider a quantum feature map $\psi$ which maps $\mathcal{X}$ to a Hilbert space $\mathcal{F}$ 
\begin{equation}
\begin{aligned}
    \psi \vcentcolon\mathcal{X} &\rightarrow \mathcal{F}\\
    \bm{x} &\mapsto \Ket{\psi(\bm{x})}
\end{aligned}
\end{equation}
and a Model Circuit $U_\text{M}({\bm{\theta}})$ that is applied to this to obtain the final state 
\begin{equation}
    \Ket{\Psi_\text{M}\left(\bm{\theta}, \bm{x}\right)} \vcentcolon= U_\text{M}({\bm{\theta}})\Ket{\psi(\bm{x})}.
\end{equation}
If the decomposition of $\mathcal{X}$ in terms of path connected pieces is
\begin{equation}
    \mathcal{X} = \bigcup\limits_{i=1}^{m} \mathcal{X}_{i}
\end{equation}
the Hilbert space $\mathcal{F}$ we use for our computation is chosen such that it is the Hilbert space direct sum of $m$ mutually orthogonal subspaces $\mathcal{F}_i$. That is
\begin{equation}
    \mathcal{F} = \bigoplus\limits_{i=1}^{m}  \mathcal{F}_i. 
\end{equation}
We also assume there are continuous feature maps
\begin{equation}
    \psi_i \vcentcolon\mathcal{X}_i \rightarrow \mathcal{F}_i
\end{equation}
and
\begin{equation}\label{eq:feature_map}
    \psi(\bm{x}) \vcentcolon= \psi_i(\bm{x}) 
\end{equation}
for $i$ such that $\bm{x} \in \mathcal{X}_i$.
\begin{aside}
For example, if
\begin{equation}
\begin{aligned}
    \mathcal{X} &= \Set{1, \ldots, 2^3}\times [0, 1] \\
    &= \bigcup\limits_{i=1}^{8} \Set{i}\times [0, 1] =\vcentcolon \bigcup\limits_{i=1}^{8} \mathcal{X}_{i} 
\end{aligned}
\end{equation}
one could use 
\begin{equation}
\begin{aligned}
    \mathcal{F} &= H_{s_{1/2}} \otimes  H_{s_{1/2}} \otimes H_{s_{1/2}} \otimes H_{s_{1/2}}\\
    &\cong \bigoplus\limits_{i=1}^{8}  \left(H_{s_0} \otimes H_{s_{1/2}}\right)
\end{aligned}
\end{equation}
with
\begin{equation}
\begin{aligned}
\psi_i \vcentcolon \Set{\mathtt{I}}\times [0, 1] &\rightarrow H_{s_0} \otimes H_{s_{1/2}}\\
\alpha &\mapsto \Ket{\mathtt{I}} \otimes R_y(\alpha \pi)\Ket{0} 
\end{aligned}
\end{equation}
and \begin{equation}
\begin{aligned}
\psi \vcentcolon &&\mathcal{X} &\rightarrow  \bigoplus\limits_{i=1}^{8}  \left(H_{s_0} \otimes H_{s_{1/2}}\right)\\
&&(i, \alpha) &\mapsto 0 + \Ket{\mathtt{I}}_i \otimes R_y(\alpha \pi)\Ket{0}_i + 0. 
\end{aligned}
\end{equation}
\end{aside}

\paragraph{} Next, $\Ket{\Psi_\text{M}}$ is mapped to $\mathbb{R}$ via a map 
\begin{equation}
    \mathcal{M} \vcentcolon\mathcal{F} \times \mathcal{F} \rightarrow \mathbb{R},
\end{equation}
which is accomplished via a quantum measurement involving states $\Ket{\Psi_\text{M}}$. The results of $\mathcal{M}$ are used to construct an approximation of $f$ parameterised by $\bm{\theta}$ and a new set of variables $\bm{a}$
\begin{equation}
\begin{aligned}
    \check{f}(\bm{\theta}, \bm{a}) \vcentcolon\mathcal{X} &\rightarrow \mathbb{R}\\
    \bm{x} &\mapsto \check{f}(\bm{x} ;\bm{\theta}, \bm{a}).
\end{aligned}
\end{equation}
Here
\begin{multline}
\label{eq:f_approx}
    \check{f}(\bm{x} ;\bm{\theta}, \bm{a}) \coloneqq \mathcal{A}\Big(\bigtimes_{i=1}^{n'} \mathcal{M}\Big(\Ket{\Psi_\text{M}\big(\bm{\theta}, \phi_i(\bm{x})\big)},\\
    \Ket{\Psi_\text{M}\big(\bm{\theta}, \phi'_i(\bm{x})\big)}\Big); \bm{a}\Big),
\end{multline}
where 
\begin{equation}
\begin{aligned} \label{eq:phiphimap}
    (\phi_i, \phi'_i) \vcentcolon\mathcal{X} &\rightarrow (\mathcal{X})^{2}\\
    \bm{x} &\mapsto \left(\phi_i(\bm{x}), \phi'_i(\bm{x})\right)
\end{aligned}
\end{equation}
and
\begin{equation}
\begin{aligned}
    \mathcal{A}(\bm{a}) \vcentcolon\mathbb{R}^{n'} &\rightarrow \mathbb{R}.
\end{aligned}
\end{equation}
Though we are not explicitly constraining $\mathcal{A}$, it has to be of appropriate structure (considering effect of interference, etc...) especially for the second part of the algorithm to be successful. For every $\bm{x} \in \mathcal{X}_j$, $\phi_i(\bm{x}) \in \mathcal{X}_k$ for some particular $k$, say $\phi_i(j)$. Either $\phi_i(\set{1, \ldots, m})$ has to be a bijection or $\phi_i(\mathcal{X})$ has to be a constant function. These conditions have to hold for $\phi'_i$ too.
\begin{aside}
$\check{f}$ could be constructed by~\cite{Goto2021}, for example :
\begin{itemize}
    \item The measurement of the expectation value of an observable $\hat{M}$
        \begin{equation}
            \mathcal{M}(\Ket{\Psi_\text{M}},\cdot) =  \Braket{\Psi_\text{M}|\hat{M}|\Psi_\text{M}}
        \end{equation}
        and
        \begin{equation}
        \begin{aligned}
             \check{f}(\bm{x} ;\bm{\theta}, \bm{a}) &=  a_0 + \\
            &a_1 \Braket{\Psi_\text{M}\left(\bm{\theta}, \bm{x}\right)|\hat{M}|\Psi_\text{M}\left(\bm{\theta}, \bm{x}\right)}.
        \end{aligned}
        \end{equation}
    \item The measurement of the overlap to determine a kernel
        \begin{equation}
        \begin{aligned}
             k(\bm{x},\bm{x}'&; \bm{\theta}) \\
             &\vcentcolon= \lvert\mathcal{M}(\ket{\Psi_\text{M}\left(\bm{\theta},\bm{x}\right)}, \ket{\Psi_\text{M}(\bm{\theta}, \bm{x}')})\rvert^2\\ 
            &\phantom{\vcentcolon}= \Big\lvert \Braket{\Psi_\text{M}\left(\bm{\theta}, \bm{x}\right) | \Psi_\text{M}(\bm{\theta}, \bm{x}')}\Big\rvert^2
        \end{aligned}
        \end{equation}
        and
        \begin{equation}
            \check{f}(\bm{x} ;\bm{\theta}, \bm{a}) =  a_0 + \sum_{i=1}^{n'} a_i k(\bm{x}, \phi'_i(\bm{x}); \bm{\theta}).
        \end{equation}
\end{itemize}
\end{aside}

This construction of $\check{f}$ could be used to enforce some of the conditions in \eqref{eq:conditions}.
\begin{aside}
For example, using an appropriately constrained UFA for $\Ket{\Psi_\text{M}\left(\bm{\theta}, \bm{x}\right)}$ and using $\mathcal{A}$ with the desired symmetries. Techniques already developed for PIML~\cite{pinn_review} and DQC~\cite{Kyriienko2021} could be used here.
\end{aside}

\paragraph{} A loss function
\begin{equation}
\begin{aligned} \label{eq:loss}
    L \vcentcolon \left(\mathbb{R}^n\right)^{\mathbb{R}} &\rightarrow \mathbb{R}\\
    \phantom{(}g &\mapsto L(g)
\end{aligned}
\end{equation}
is constructed beforehand using the conditions in \eqref{eq:conditions} such that as $L(g) \xrightarrow[]{} \min (L)$, the function $g$ satisfies all the conditions in \eqref{eq:conditions} (the conditions already enforced through the construction in Eq.~\eqref{eq:f_approx} can be ignored). That implies, as $L(g) \xrightarrow[]{} \min (L)$, the function $g$ satisfies all that is known about $f$ so it would be a ``best estimate" of $f$ we can obtain with the given information.
\begin{aside}
For example, if
\begin{multline}
    \Set{\mathcal{C}_i}_i = \\ 
    \Set{ f((0,0)) - 2 = 0,  \partial_{xx} f - \frac{1}{c}\partial_{tt} f \rvert_\text{interior} = 0}
\end{multline}
then a possible 
\begin{equation}
    L(g) = \Vert g((0,0)) - 2 \Vert + \sum_\text{interior} \left\Vert \partial_{xx} g - \frac{1}{c}\partial_{tt} g \right\Vert.
\end{equation}
Techniques used for the construction of the loss function in DQC~\cite{Kyriienko2021} and PIML~\cite{pinn_review} could be used here.
\end{aside}

\paragraph{} The primary objective of this step is to minimize $L(\check{f}(\bm{\theta}, \bm{a}))$ with respect to parameters $\bm{\theta}$ and $\bm{a}$ and thereby obtain an approximation of $f$. Note that $\check{f}(\bm{\theta}, \bm{a})$ could be constrained by its construction and it is not necessary that $\min (L(\check{f}(\cdot))) = \min (L)$. If the minimization is successful, 
\begin{equation}
    \check{f}(\bm{x}; \bm{\theta}_{\text{min}}, \bm{a}_{\text{min}}) \approx f(\bm{x}).
\end{equation}
and in general, we can obtain 
\begin{equation}
     L_\text{ext}(\check{f}(\bm{\theta}_{\text{min}}, \bm{a}_{\text{min}}), \bm{x}) \approx L_\text{ext}(f, \bm{x}).
\end{equation}
with a construction similar to Eq.~\eqref{eq:f_approx}.

\subsection{Finding the Extremizing Input} 

\paragraph{} Recall the definition of the feature map in Eq.~\eqref{eq:feature_map}. Defining
\begin{equation}
    \widetilde{\mathcal{X}} \vcentcolon= \bigtimes\limits_{i=1}^{m} \left(\mathbb{C}, \mathcal{X}_{i}\right),
\end{equation}
we may redefine
\begin{equation}
\begin{aligned}
    \psi \vcentcolon && \widetilde{\mathcal{X}} &\rightarrow \mathcal{F}\\
    &&\times_{i=1}^{m}(\lambda_i, \bm{x}_i) &\mapsto \sum_{i=1}^{m} \frac{\lambda_i}{\Vert\bm{\lambda}\Vert} \Ket{\psi_i(\bm{x}_i)}. 
\end{aligned}
\end{equation}
Note that if we write $\bm{x} \in \mathcal{X}_{j} \subset \mathcal{X}$ as 
\begin{equation}
    \left(\times_{i=1}^{j} (0, \bm{x}_i)\right) \times (1,\bm{x}) \left(\times_{i=j+1}^{m} (0, \bm{x}_i)\right) \in \widetilde{\mathcal{X}} ,
\end{equation}
this matches the old definition. 
\begin{aside}
For example, if
\begin{equation}
\begin{aligned}
    \mathcal{X} &= \Set{1, \ldots, 2^3}\times [0, 1],
\end{aligned}
\end{equation}
\begin{equation}
\begin{aligned}
    \mathcal{F} =  \bigoplus\limits_{i=1}^{8}  \left(H_{s_0} \otimes H_{s_{1/2}}\right)
\end{aligned}
\end{equation}
and \begin{equation}
\begin{aligned}
\psi \vcentcolon &&\mathcal{X} &\rightarrow  \bigoplus\limits_{i=1}^{8}  \left(H_{s_0} \otimes H_{s_{1/2}}\right)\\
&&(i, \alpha) &\mapsto 0 + \Ket{\mathtt{I}}_i \otimes R_y(\alpha \pi)\Ket{0}_i + 0. 
\end{aligned}
\end{equation}
we have the extension
\begin{equation}
\begin{aligned}
\psi \vcentcolon \bigtimes\limits_{i=1}^{8}\left(\mathbb{C}, \Set{1} \times [0,1]
\right)&\rightarrow  \bigoplus\limits_{i=1}^{8}  \left(H_{s_0} \otimes H_{s_{1/2}}\right)\\
\bigtimes\limits_{i=1}^{8} \left( \lambda_i, \alpha_i\right) &\mapsto \sum_{i=1}^{8} \frac{\lambda_i}{\Vert\bm{\lambda}\Vert} \big( \Ket{\mathtt{I}}_i \otimes \\
\span \span R_y(\alpha_i \pi)\Ket{0}_i\big). 
\end{aligned}
\end{equation}
\end{aside}

Next, we extend the definition of $\phi_i$ and $\phi'_i$ in Eq.~\eqref{eq:phiphimap}. They are identical, so we will drop the subscript and call it $\phi$. Recalling  $\phi(j)=i$ implies $\phi(\bm{x}) \in \mathcal{X}_i$ for every $\bm{x} \in \mathcal{X}_j$,
\begin{equation}
\begin{aligned}\label{eq:phiphiextensions}
    \phi \vcentcolon &&\widetilde{\mathcal{X}} &\rightarrow \bigtimes\limits_{\phi(j)=1}^{m} \left(\mathbb{C}, \mathcal{X}_{\phi(j)}\right)\\
    &&\times_{i=1}^{m}(\lambda_i, \bm{x}_i) &\mapsto  \bigtimes\limits_{\phi(j)=1}^{m} \big(\lambda_{\phi(j)},\phi(\bm{x}_j)\big),
\end{aligned}
\end{equation}
or, if $\phi$ is a constant function, it continues to map every input to the same constant. 

We now have a map
\begin{equation}
\begin{aligned}
    \psi\circ\phi \vcentcolon \widetilde{\mathcal{X}} &\rightarrow \mathcal{F}\\
    \tilde{\bm{x}}
    &\mapsto \Ket{\psi(\phi(\tilde{\bm{x}}))}.
\end{aligned}
\end{equation}
Assuming the decomposition $\tilde{\bm{x}} = \times_{i=1}^{m}(\lambda_i, \bm{x}_i)$, 
\begin{align}
    \Ket{\psi(\phi(\tilde{\bm{x}}))} &=   \Ket{\psi\left(\bigtimes\limits_{\phi(j)=1}^{m} \left(\lambda_{\phi(j)},\phi(\bm{x}_j)\right)\right)}\\
    &= \sum_{\phi(j)=1}^{m} \frac{\lambda_{\phi(j)}}{\Vert\bm{\lambda}\Vert} \Ket{\psi_{\phi(j)}\left(\phi(\bm{x}_j)\right)},\label{eq:state_decomp}
\end{align}
where $\bm{x}_j \in \mathcal{X}_j$. Recalling $\phi(j)=i$ implies $\phi(\bm{x}_j) \in \mathcal{X}_i$ and $\mathcal{F} = \bigoplus\limits_{i=1}^{m}  \mathcal{F}_i$, 
\begin{equation}
    \Ket{\psi_{\phi(j)}\left(\phi(\bm{x}_j)\right)} \in \mathcal{F}_{\phi(j)}
\end{equation}
so Eq.~\eqref{eq:state_decomp} is nothing but the decomposition of a vector in $\mathcal{F}$ in terms of its components in  mutually orthogonal subspaces $\mathcal{F}_i$.

\paragraph{} Say $\tilde{\bm{x}} \in \widetilde{\mathcal{X}}$ is parameterised by $\Set{\bm{\lambda}, \bm{\Phi}}$. We may generate the state
\begin{equation} \label{eq:Ext_circ}
\Ket{\psi\left(\phi_i\left(\tilde{\bm{x}}(\bm{\lambda}, \bm{\Phi})\right)\right)} \in \mathcal{F}
\end{equation} 
using a quantum circuit. Let us call this the Extremizer Circuit.  
\begin{aside}
For example, if
\begin{equation}
\begin{aligned}
    \mathcal{X} &= \Set{1, 2}\times [0, 1],
\end{aligned}
\end{equation}
\begin{equation}
\begin{aligned}
    \mathcal{F} &=  \left(H_{s_0} \otimes H_{s_{1/2}}\right) \oplus \left(H_{s_0} \otimes H_{s_{1/2}}\right),\\
\end{aligned}
\end{equation}
and
\begin{equation}
    \phi_i = \mathbb{I},
\end{equation}
\begin{equation}
\begin{aligned}
\psi &\vcentcolon \bigtimes\limits_{i=1}^{2}\left(\mathbb{C}, \Set{1} \times [0,1]
\right) &&\rightarrow &\bigoplus\limits_{i=1}^{2}  \left(H_{s_0} \otimes H_{s_{1/2}}\right)\\
\span \left( \left( \cos \left({{\lambda}/{2}}\right), \Phi_1 \right) , \left( \sin \left({{\lambda}/{2}}\right) , \Phi_2 \right) \right) \mapsto\span \span \span \\
\span \span \span \span \cos \left({{\lambda}/{2}}\right) \left( \Ket{\mathtt{I}} \otimes R_y(\Phi_1 \pi)\Ket{0} \right) +\\
\span \span \span \span  \sin \left({{\lambda}/{2}}\right) \left( \Ket{\mathtt{II}} \otimes R_y(\Phi_2 \pi)\Ket{0} \right),
\end{aligned}
\end{equation}
where $\Ket{\mathtt{I}}$ and $\Ket{\mathtt{II}}$ denote the basis of two distinct one dimensional Hilbert spaces.
We can implement this with conventional gate-based circuit with a change of basis since
\begin{equation}
    \bigoplus\limits_{i=1}^{2}  \left(H_{s_0} \otimes H_{s_{1/2}}\right) \cong \left(H_{s_{1/2}} \otimes H_{s_{1/2}}\right).
\end{equation}
\end{aside}

\paragraph{} 
Recall $\mathcal{A}(\Set{\mathcal{M(\cdot)}}_i; \bm{a})$ in Eq.~\eqref{eq:f_approx} is defined on all of $\left(\mathcal{F} \times \mathcal{F}\right)^{n'}$. So the map $\left(\mathcal{F} \times \mathcal{F}\right)^{n'} \rightarrow \mathbb{R}$ given by 
\begin{multline}\label{eq:output_operator}
    \mathcal{A}\Big(\bigtimes_{i=1}^{n'}\mathcal{M}\big(U_\text{M}({\bm{\theta}_{\text{min}}})\ket{\alpha_i},
    U_\text{M}({\bm{\theta}_{\text{min}}})\ket{\alpha'_i}\big);\\ \bm{a}_{\text{min}}\Big)
\end{multline}
is well defined for all $\left(\ket{\alpha_i},\ket{\alpha'_i}\right) \in \left(\mathcal{F} \times \mathcal{F}\right)^{n'}$. We already have the machinery corresponding to this; the trained Model Circuit $U_\text{M}({\bm{\theta}_{\text{min}}})$ can be applied to any $\left(\ket{\alpha_i},\ket{\alpha'_i}\right) \in \left(\mathcal{F} \times \mathcal{F}\right)^{n'}$ and the same steps used during the training of the Quantum Model can be applied. If 
\begin{equation}
    \left(\ket{\alpha_i},\ket{\alpha'_i}\right)=\Big(\Ket{\psi\big(\phi_i(\bm{x})\big)}, \Ket{\psi\big(\phi'_i(\bm{x})\big)} \Big)
\end{equation}
for $\bm{x}\in \mathcal{X}$, Eq.~\eqref{eq:output_operator} yields
\begin{equation}
    \check{f}(\bm{x}; \bm{\theta}_{\text{min}}, \bm{a}_{\text{min}}).
\end{equation}

The Extremizer Circuit in Eq.~\eqref{eq:Ext_circ} allows us to obtain
\begin{equation}
\begin{aligned}\label{eq:extremizer_out}
\psi \circ \left(\bigtimes_{i=1}^{n'}\left(\phi_i, \phi'_i\right)\right) \vcentcolon\span\\ \span \mathbb{R}^{\dim{\bm{\lambda}}} \times \mathbb{R}^{\dim{\bm{\Phi}}} &\rightarrow& \mathcal{F}^{2 n'}\\
    \span \left(\bm{\lambda},\bm{\Phi}\right)
    &\mapsto& \bigtimes_{i=1}^{n'} \Big(\Ket{\psi\left(\phi_i\left(\tilde{\bm{x}} \left(\bm{\lambda},\bm{\Phi}\right)\right)\right)},\\ 
    \span \span \span \Ket{\psi\left(\phi'_i\left(\tilde{\bm{x}} \left(\bm{\lambda},\bm{\Phi}\right)\right)\right)}\Big) .
\end{aligned}
\end{equation}
Combining Eq.~\eqref{eq:output_operator} with Eq.~\eqref{eq:extremizer_out} allows us to obtain 
\begin{equation}
    \tilde{f} \vcentcolon \mathbb{R}^{\dim{\bm{\lambda}}} \times \mathbb{R}^{\dim{\bm{\Phi}}} \rightarrow \mathbb{R}.
\end{equation}
This $\tilde{f}$ is an extension of the approximation $\check{f}(\bm{\theta}_{\text{min}}, \bm{a}_{\text{min}})$ of $f$ over a path connected domain, if $\bm{\Phi}$ parameterizes all $\bm{x} \in \mathcal{X}$.

Following the same steps, $L_\text{ext}$ can be also be extended over the same domain using the previously constructed circuit for $L_\text{ext}$.

\paragraph{} When $\tilde{\bm{x}} \in \tilde{\mathcal{X}}$, an appropriately constructed $L_\text{ext}$ will be able to yield the sum weighted by $\bm{\lambda}$ of the loss from the contributing $\bm{x}_i \in \mathcal{X}$ that constitute $\tilde{\bm{x}}$.
\emph{\newline Requirement:} To implement this algorithm, it has to be possible to meet this condition.\newline
Hence, we simultaneously search $m$ points of the domain (from the $m$ distinct pieces). Note that we do not need to modify the quantum circuit or the subsequent meansurement steps we used in the previous step to obtain the approximation of $f$, we just need to remove the map $\phi$ that loads the input during the training of the Quantum Model and replace it with the Extremizer Circuit. This lets us utilize the parallelism of quantum superposition during extremal learning.

Furthermore, if $\bm{\Phi}$ parameterizes all $\bm{x} \in \mathcal{X}$ we can search all of $\mathcal{X}$ as though it were a continuous domain. Note that we did not have to construct an interpolation of the disconnected domain $\mathcal{X}$ of the function $f$ we were trying to optimize; we simply leverage the ``inherent interpolation" through quantum superposition from the circuit we trained to construct the Quantum Model.

We may now optimize
\begin{equation}
    L_{\text{ext}}\left(\tilde{f}\left(\bm{\lambda},\bm{\Phi}\right),\tilde{\bm{x}} \left(\bm{\lambda},\bm{\Phi}\right)\right)
\end{equation}
with respect to the continuous variables $\left(\bm{\lambda},\bm{\Phi}\right)$. If $\mathcal{X}$ is path connected, $\dim{\bm{\lambda}} = 0$ and if $\mathcal{X}$ is discrete, $\dim{\bm{\Phi}} = 0$. Under these edge cases, the generalization described in this section reduces to those in Sec.~\ref{section:methodology}.

\paragraph{} Note that there is no obvious way to map $\tilde{\bm{x}} \in \tilde{\mathcal{X}}$ to an input value. However, since 
\begin{equation}
    \mathcal{F} = \bigoplus\limits_{i=1}^{m}  \mathcal{F}_i, 
\end{equation}
where all $\mathcal{F}_i$ are mutually perpendicular, there is always an operator such that its measurement will send a general vector in $\mathcal{F}$ to a vector in one of the $\mathcal{F}_i$.

Let
\begin{equation}
    \Ket{\psi\left(\tilde{\bm{x}} \left(\bm{\lambda}_\text{ext},\bm{\Phi}_\text{ext}\right)\right)} = \sum_{i=1}^{m} \frac{\lambda_i}{\Vert\bm{\lambda}_\text{ext}\Vert} \Ket{\psi_i(\bm{x}_i)}.
\end{equation}
The coefficient $\frac{\lambda_i}{\Vert\bm{\lambda}_\text{ext}\Vert}$ is proportional to the probability of obtaining a state in each of the perpendicular subspaces of the Hilbert space after measurement.  The expectation is that the construction of $\Ket{\psi\left(\tilde{\bm{x}} \left(\bm{\lambda},\bm{\Phi}\right)\right)}$ makes it most likely for this projection to correspond to the $\mathcal{X}_i$ containining the extremal input, provided optimization was successful.

\paragraph{} After the measurement, we get a wavefunction of the form $\Ket{\psi(\bm{x}_\text{ext})}$ for some $\bm{x}_\text{ext} \in \mathcal{X}$. This is the same encoding that was initially employed  on the input to map it to the Hilbert space. We can map it back to the input variable by performing measurements on $\Ket{\psi(\bm{x}_\text{ext})}$.

\subsection{Example}

Consider $f(x,n)\vcentcolon \mathbb{R} \times \Set{1 ,2 ,3 ,4} \rightarrow \mathbb{R}$  defined by
\begin{equation}
\begin{aligned}\label{eq:mixed_fn}
f(x, 1) &= - x^2 \sin(2 x+2)\\
f(x, 2) &= x^3 \sin(2 x) - 0.2\\
f(x, 3) &= \sin^2(2 x-0.5) - 0.6\\
f(x, 4) &= - x \sin(2 x + 2)/2.
\end{aligned}
\end{equation}
The minima of $f$ under the additional condition $x \in [ -1 , 1 ]$ is at $(x,n)=(0.25,3)$. We use this example to demonstrate the capacity of the QEL algorithm to work with a mixture of discrete and continuous variables. 

As a training dataset, we give only values away from the minimum, to avoid potential biases during our modeling and optimization process. We used $21$ points for each of the values of $n$.  

\begin{figure}[!htb]
    
    \includegraphics[width=0.9\linewidth]{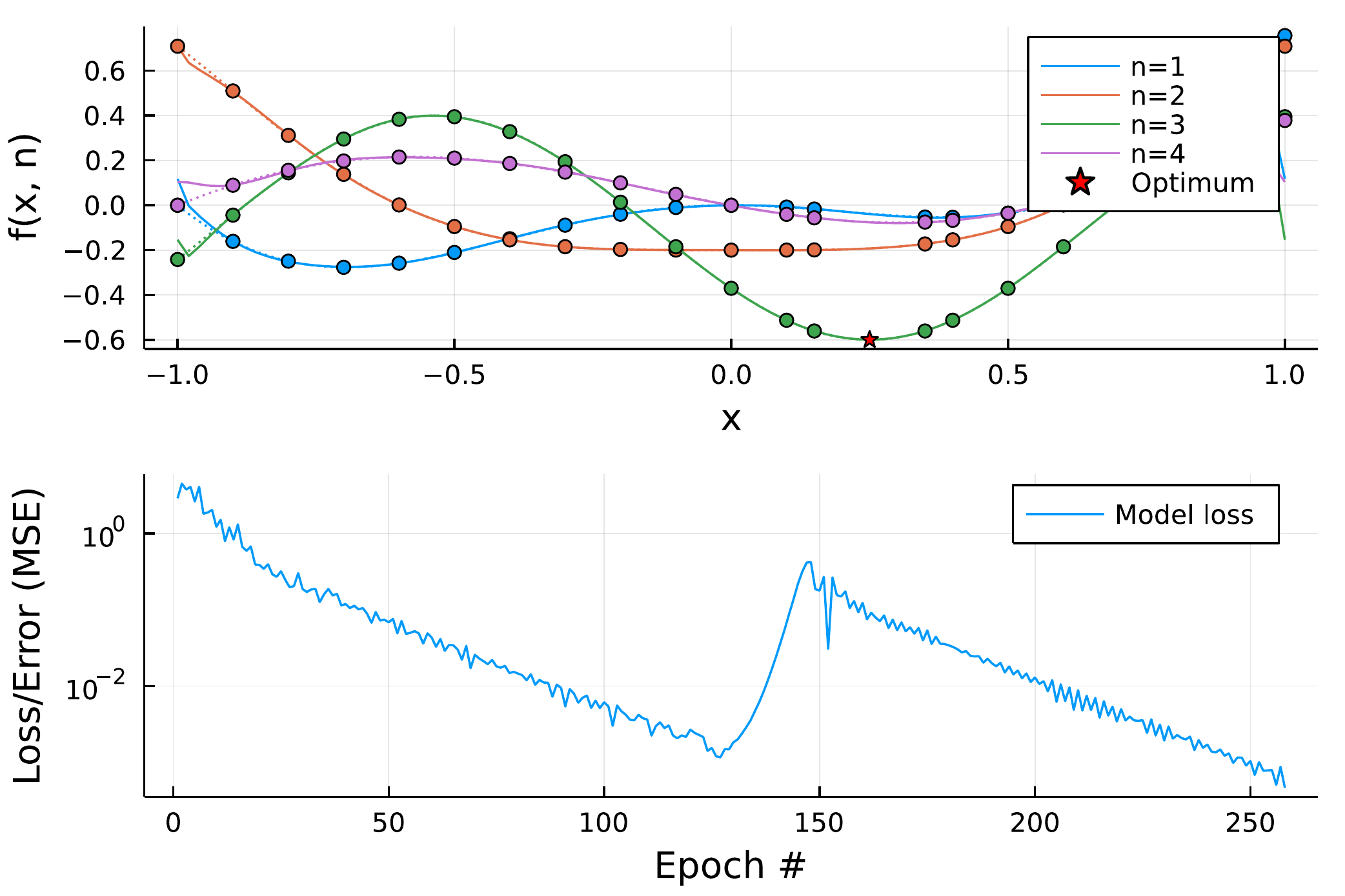}
    \caption{The top plot shows the original function (dotted line), the training data  selected from this function (circle markers), the trained QNN model (solid line), and the true optimum (star). The bottom plot shows the model loss (MSE) as a function of epoch number in the ADAM~\cite{Kingma2014} optimization. We used a learning rate of $0.5$ and 258 epochs to train the model.}
    \label{fig:qcl_mixed}
\end{figure}

First, to fit the data, the QCL-type model \cite{Mitarai2018} we use is a QNN circuit with 5 qubits. We initialize the QNN in the all-zero state. To encode $x$, we apply a feature map with $\hat{R}_y(2j\arccos(x))$, where $j$ is the qubit index starting from 1 and $\hat{R}_y$ is a single-qubit Pauli-Y rotation, to the first 3 qubits. We use simple digital encoding on the last 2 qubits to represent $n$. Next, we apply a Hardware Efficient Ansatz (HEA)~\cite{Kandala2017} of depth 10 and finally measure the total magnetization $\hat{M}=\sum_j \hat{\sigma}_Z^j$ such that the quantum model is $f(x, n)=\langle\hat{M}\rangle$. The result of the modeling stage is found in Figure \ref{fig:qcl_mixed}.

Extremization of the trained model was performed using an extremizer circuit with $\hat{R}_y(2j\arccos(x))$, where $j$ is the qubit index starting from 1 and $\hat{R}_y$ is a single-qubit Pauli-Y rotation, on the first 3 qubits. The last 2 qubits use the circuit in \textcite{2019Perdomo} to create superpositions of the representation of the discrete variable $n$. Initially, we started from $x=0$ and an equal superposition of all values of $n$. The four parameters of the extremizer circuit were optimized using ADAM~\cite{Kingma2014} optimization with a learning rate of $0.01$ and $100$ epochs. Measurements were performed on the resulting state to ascertain the minima of $f$. 

\begin{table}
\caption{Probability distribution of value of $n$\label{table:mixed}}
\begin{ruledtabular}
\begin{tabular}{c c} 
$n$ & Probability \\
\midrule
1 & 0.0417 \\
2 & 0.0114 \\
3 & 0.8689 \\
4 & 0.078 \\
\end{tabular}
\end{ruledtabular}
\end{table}
The value of $n$ is probabilistic; the result is summarized in Table~\ref{table:mixed}. The value of $x$ can be obtained exactly by inverting the feature map. The final result is as shown in Table~\ref{table:mixed_result}
\begin{table}
\caption{Result of application of QEL to obtain the minima of a function $f(x,n)$ with one discrete variable and one continuous variable defined in \eqref{eq:mixed_fn}.\label{table:mixed_result}}
\begin{ruledtabular}
\begin{tabular}{c c c} 
Variable & QEL Result & Exact Result \\
\midrule
$x$ & 0.249 & 0.250\\
$n$ & 3 & 3\\
\end{tabular}
\end{ruledtabular}
\end{table}

\section{A Use Case}
The novelty of QEL in the context extremal learning involving discrete variables is evident, considering extremal learning involving discrete variables have not been discussed in the literature before. We provide an example from product design here to clarify the utility of extremal learning even in cases where discrete variables are not involved, inherited from its incorporation of DQC. 

If the driving power source has a fixed amplitude, there is a particular driving frequency called the resonant frequency at which the steady-state amplitude of a driven RLC oscillator is maximum. The commercial applications of electronic RLC oscillators are plentiful, ranging from television receivers to induction heaters. 

It is easy to see why it may be desirable to know the resonant frequency of the RLC circuit (for instance, to maximize efficiency). Consider an application where a device desires to always drive an internal RLC oscillator at resonant frequency. Being an  RLC oscillator, we know it is governed by the equation 
\begin{equation} \label{eq:RLC}
L \frac{\partial^2}{\partial t^2} I(t) + R \frac{\partial}{\partial t} I(t) + \frac{1}{C} I(t) = \omega V_0 \cos (\omega t).
\end{equation}
We have full control over the driving power source and we can measure the output $I$ and the internal temperature $T$ of the device. As the device operates and generates heat, however, it slowly alters the values $R$, $L$ and $C$ of the circuit components and changes the resonant frequency. We cannot directly control the internal temperature of the device nor measure the value of the circuit components. Assuming the value of the device components $R$, $L$ and $C$ only depends on the internal temperature of the device, this is an ideal problem to demonstrate the utility of extremal learning.

In this case, our unknown function is $I(t)$. Extremal learning could leverage DQC to solve Eq.~\eqref{eq:RLC} governing the RLC circuit (assume, for the moment, it is not easily solvable analytically). Our goal is not, however, to optimize the unknown function $I$ as a function of time $t$ but driving frequency $\omega$. Of course, this is also possible within the framework by treating both time $t$ and the parameter driving frequency $\omega$ in Eq.~\eqref{eq:RLC} as variables and solving for $I(t; \omega)$. In this case, the ``data" regarding the unknown function (or, in terms of our generalized framework in Appendix~\ref{app:generalization}, the ``condition" in \eqref{eq:conditions}) is simply the differential equation in Eq.~\eqref{eq:RLC}. Once the model is learned, we may extremize it and find the resonant frequency.

There is still more to this problem. Eq.~\eqref{eq:RLC} alone does not take into account the variation of the resonant frequency as a function of the internal temperature of the device. Our framework allows us to incorporate this information too, by adding the output of the circuit at various temperatures and frequencies as additional data points. With sufficient data points and a circuit with sufficient expressivity, it would be possible to model $I(t; \omega, T)$ and then extremize it to find the resonant frequency at any temperature $T$.    

Notice the role of the different independent variables/parameters in this example.
\begin{itemize}
    \item The variable $t$ is related to the the unknown function $I$ through a differential equation. The differential equation is the ``data" used to learn the dependence on this variable.
    \item The relation between parameter $T$ and the unknown function $I$ is unknown and is learned from experimental data, as opposed to being dictated by an equation. We can measure $T$, but not directly control it. The dependence on this variable is learned from experimental data.
    \item The parameter $\omega$ is related to the the unknown function $I$ through a differential equation, but it is a parameter in the equation, not a differentiating variable. We can adjust $\omega$ and our goal is to find the value of $\omega$ that extremizes the amplitude of $I(t)$ at a given temperature $T$.
\end{itemize}

The QEL framework is flexible enough to tackle the problem even in a case like this where information about the system is available in mixed form. Moreover, one or several of the parameters could be discrete.

\end{document}